\documentclass[a4paper,fleqn,usenatbib]{mnras}

  \usepackage{graphicx}
  \usepackage{mathtools}
  \usepackage{color}
  \usepackage{thumbpdf}
  \usepackage{float}
  \usepackage{newtxtext,newtxmath}
  \usepackage[T1]{fontenc}
  \usepackage{ae,aecompl}
              
  \graphicspath{{.}{../Figures/}}
  \DeclareGraphicsExtensions{.png,.pdf}

  \newcommand{\eg}{{\it e.g.}}
  \newcommand{\ie}{{\it i.e.}}
  \newcommand{\etc}{{\it etc.}}

  \newcommand{\ncode}[1]{{\tt #1}}
  \newcommand{\sms}[1]{{\mbox{{\scriptsize #1}}}}
  \newcommand{\citext}[1]{{\it #1}}
  \newcommand{\citengl}[1]{``{\it #1}''}
  \newcommand{\seppar}{\vspace{10pt}}

  \newcommand{\mmic}{\;\mu\rm m}
  \newcommand{\mic}{$\mu\rm m$}
  \newcommand{\dd}{{\rm d}}
  \newcommand{\ddiff}{{\;\rm d}}
  \newcommand{\E}[1]{\times10^{#1}}

  \newcommand{\vect}[1]{\vec{#1}}
  \newcommand{\mat}[1]{\mathbf{#1}}

  \renewcommand{\refeq}[1]{Eq.~(\ref{#1})}
  \newcommand{\refeqs}[1]{Eqs.~(\ref{#1})}
  \newcommand{\refeqp}[1]{(Eq.~\ref{#1})}
  \newcommand{\refeqnp}[1]{Eq.~\ref{#1}}
  
  \newcommand{\reftab}[1]{Table~\ref{#1}}
  \newcommand{\reffig}[1]{Figure~\ref{#1}}
  \newcommand{\reffigs}[1]{Figures.~\ref{#1}}
  \newcommand{\refsec}[1]{Section~\ref{#1}}
  \newcommand{\refsecs}[1]{Sections~\ref{#1}}
  
  \newcommand{\refapp}[1]{Appendix~\ref{#1}}
  \newcommand{\refapps}[1]{Appendices~\ref{#1}}

  \newcommand{\MBB}{{\sl MBB}}
  \newcommand{\BBQ}{{\sl BBQ}}
  \newcommand{\BEMBB}{{\sl BEMBB}}
  \newcommand{\deltaU}{{\sl deltaU}}
  \newcommand{\powU}{{\sl powerU}}
  \newcommand{\Star}{{\sl starBB}}
  \newcommand{\radio}{{\sl radio}}
  \newcommand{\HB}{\ncode{HerBIE}}
  \newcommand{\minpack}{\ncode{MINPACK}}
  \newcommand{\Uavm}{\bar{U}}
  \newcommand{\Uav}{$\Uavm$}

  \newcommand{\hi}{H$\,${\sc i}}
  \newcommand{\hii}{H$\,${\sc ii}}
  \newcommand{\hmol}{H$_2$}

  \newcommand{\spitz}{{\it Spitzer}}
  
  \newcommand{\wise}{{\it WISE}}

  \newcommand{\hersc}{{\it Herschel}}
  \newcommand{\planck}{{\it Planck}}

  \newcommand{\akari}{{\it Akari}}
  \newcommand{\spica}{{\it SPICA}}
  \newcommand{\cobe}{{\it COBE}}
  
  \newcommand{\IRACi}{IRAC$_\sms{3.6$\mu m$}$}

  \newcommand{\akariii}{IRC$_\sms{3.2$\mu m$}$}
  \newcommand{\akariiii}{IRC$_\sms{4.1$\mu m$}$}
  \newcommand{\akariiv}{IRC$_\sms{7.0$\mu m$}$}
  \newcommand{\akariv}{IRC$_\sms{9.0$\mu m$}$}
  \newcommand{\akarivi}{IRC$_\sms{11$\mu m$}$}
  \newcommand{\akarivii}{IRC$_\sms{15$\mu m$}$}
  \newcommand{\akariviii}{IRC$_\sms{18$\mu m$}$}
  \newcommand{\wisei}{\wise$_\sms{3.4$\mu m$}$}
  \newcommand{\wiseii}{\wise$_\sms{4.6$\mu m$}$}
  \newcommand{\wiseiii}{\wise$_\sms{11.6$\mu m$}$}
  \newcommand{\wiseiv}{\wise$_\sms{22.1$\mu m$}$}

  \newcommand{\PACSii}{PACS$_\sms{100$\mu m$}$}
  \newcommand{\PACSiii}{PACS$_\sms{160$\mu m$}$}
  \newcommand{\SPIREi}{SPIRE$_\sms{250$\mu m$}$}
  \newcommand{\SPIREii}{SPIRE$_\sms{350$\mu m$}$}
  \newcommand{\SPIREiii}{SPIRE$_\sms{500$\mu m$}$}

  \newcommand{\DIRBEviii}{DIRBE$_\sms{100$\mu m$}$}
  \newcommand{\DIRBEix}{DIRBE$_\sms{140$\mu m$}$}
  \newcommand{\DIRBEx}{DIRBE$_\sms{240$\mu m$}$}
  \newcommand{\HFIi}{HFI$_\sms{350$\mu m$}$}
  \newcommand{\HFIii}{HFI$_\sms{550$\mu m$}$}
  \newcommand{\HFIiii}{HFI$_\sms{850$\mu m$}$}
  
\def\apjl{ApJ}%
\def\apjs{ApJS}%
\def\ao{Appl.~Opt.}%
\def\aap{A\&A}%
  \defcitealias{kelly12}{K12}
  \defcitealias{galliano18b}{Paper~II}
  \newcommand{\citeprep}[1]{\citeauthor{#1} {\it in prep.}}
  
  \newcommand{\citepprep}[1]{(\citeauthor{#1}, {\it in prep.})}

  \newcounter{textlistctr}
  \newcommand{\thetextlist}{, }
  \newcommand{\textlist}[1]
              {\setcounter{textlistctr}{1}
               \renewcommand{\thetextlist}
               {{\it (\roman{textlistctr})}\stepcounter{textlistctr}}#1
                }

  \title[Hierarchical Bayesian SED modelling]{A Dust spectral energy 
         distribution model with hierarchical Bayesian
         inference. I. Formalism \&\ benchmarking} 
  \author[F.~Galliano]{Fr\'ed\'eric Galliano\thanks{frederic.galliano@cea.fr}
         \\
         IRFU, CEA, Universit\'e Paris-Saclay, F-91191 Gif-sur-Yvette, 
         France \\
         Universit\'e Paris-Diderot, AIM, Sorbonne Paris Cit\'e, CEA, CNRS,
         F-91191 Gif-sur-Yvette, France \\
         }
  \date{Accepted 2018 January 17. Received 2018 January 17; in original form 2017 October 6}
  \pubyear{2017}

\begin{document}
  \label{firstpage}
  \pagerange{\pageref{firstpage}--\pageref{lastpage}}
  \maketitle

\begin{abstract}
  This article presents a new dust SED model, named \HB,
  aimed at eliminating the noise-induced correlations and large scatter 
  obtained when performing least-squares fits.
  The originality of this code is to apply the hierarchical Bayesian 
  approach to full dust models, including realistic optical properties, 
  stochastic heating and the mixing of physical conditions in the observed 
  regions.
  We test the performances of our model by applying it to synthetic 
  observations.
  We explore the impact on the recovered parameters of several effects:
  signal-to-noise ratio, SED shape, sample size, the presence of 
  intrinsic correlations, the wavelength coverage and the use of 
  different SED model components.
  We show that this method is very efficient:
  the recovered parameters are consistently distributed 
  around their true values.
  We do not find any clear bias, even for the most degenerate 
  parameters, or with extreme signal-to-noise ratios.
\end{abstract}


  \section{Introduction}

Dust grains are an important, but elusive, component of the interstellar medium 
(ISM).
Their ubiquity makes them a potential asset to understanding galaxy evolution, 
but their inherent complexity impedes our ability to use them to their full 
potential.
As a diagnostic tool, dust can be used to estimate gas masses, without 
suffering from the phase-bias that gas lines are subject to.
For this reason, dust is a popular dark gas tracer 
\citep[\eg][]{grenier05,leroy11,planck-collaboration11b}.
However, our poor knowledge of the grain properties puts caution on the 
accuracy of these measurements.
It is true that the grain cross-section per Hydrogen atom can be empirically 
calibrated in the ultraviolet \citep[UV; \eg][]{fitzpatrick07}, for extinction 
studies, or in the infrared \citep[IR; \eg][]{planck-collaboration14b}, 
for emission studies.
The problem is that these empirical laws lack an underlying physical model 
describing how dust properties vary in different environments.

As a modelling ingredient, dust provides the major source of heating of the gas 
in photo-dissociation regions \citep[PDRs; \eg][]{hollenbach97}, through the
photoelectric effect\footnote{Except in extremely low-metallicity systems, where X-rays could be the dominant heating source \citep{lebouteiller17}.} \citep{bakes94,weingartner01b,kimura16}.
The main discrepancies between different PDR codes actually originates in
the diversity of assumptions about the grain physics \citep{rollig07}.
In addition, grains are the catalysts for numerous chemical reactions, 
including the formation of \hmol, the most abundant molecule in the universe.
Accounting for the stochastic temperature fluctuations of small grains 
has lead to a significant revision of \hmol\ formation rates 
\citep{le-bourlot12,bron14}.
Overall, detailed modelling of the PDR and molecular lines in star forming 
regions shows a qualitative agreement between the derived dust and gas physical 
conditions \citep[\eg][]{wu15,chevance16}, although the quantitative comparison 
can be more difficult to assess \citepprep{wu18}.

The current census on interstellar dust relies mainly on observations of the 
Galaxy.
Most dust models are designed to reproduce the extinction, emission and 
depletion pattern of the diffuse ISM
\citep{zubko04,draine07,compiegne11,jones17}.
Some also take into account polarization constraints \citep{siebenmorgen14,guillet17}.
The dust properties in other environments are more sketchy.
For instance, there is clear evidence that the grain far-IR/submm emissivity 
increases by a factor of $\simeq 2-4$, from the diffuse ISM to denser 
environments \citep[\eg][]{stepnik03,planck-collaboration11c,roy13}.
This increase can be explained by mantle accretion and coagulation
\citep{kohler12,kohler15}.
However, since the accretion of mantles in dense regions and their recycling 
back in the diffuse ISM is a hysteresis process \citep{jones17}, 
parameterizing the dust properties as a sole function of the density of the 
medium, $n$, and the UV flux, $G_0$, may be too simplisitc.
Finally, the grain properties of other galaxies can exhibit significant 
differences \citep[\eg\ in the Magellanic clouds:][]{gordon03,galliano11,galametz16,chastenet17}.
In particular, the evolution of the dust-to-gas mass ratio with metallicty is 
a subject of debate 
\citep{lisenfeld98,draine07b,galliano08a,galametz11,remy-ruyer14,de-vis17}.
It appears to be non-linear in nearby galaxies \citep{remy-ruyer14}, 
consistent with theoretical models \citep{asano13,zhukovska14}.
On the contrary, studies in absorption on more distant systems suggest a 
constant dust-to-metal mass ratio down to extremely low-metallicity systems 
\citep{zafar13,de-cia13}.
\seppar

One of the ways to tackle these open questions consists in studying how the
observed grain properties vary with the physical conditions they are 
experiencing, in a large range of environments.
A fundamental step in this process is the accurate retrieval of the parameters 
and their intercorrelations.
The recent IR/submm space observatories \spitz, \akari, \hersc\ and \planck\ 
have provided us with invaluable data on the spectral energy distributions 
(SED) of a wide variety of objects.
However, despite a wealth of good quality observations, with complete 
spectral coverage, and calibration uncertainties down to only a few percents 
\citep{planck-collaboration14a}, we are still facing limitations using basic 
analysis techniques.
Among these limitations, there is a series of noise-induced false correlations 
between derived parameters, when performing least-squares fits
\citep[hereafter $\chi^2$;][]{shetty09}.

A very efficient way to treat these degeneracies was proposed by 
\citet{kelly12}.
These authors designed a hierarchical Bayesian model (hereafter HB), 
accounting for both noise and correlated calibration uncertainties.
We will explain in detail, in the present article, how an HB model works.
For now, we just need to note that it deals with two classes of 
parameters:\textlist{\thetextlist~the dust parameters of each source (mass, 
temperature, \etc); and
\thetextlist~a set of {\it hyperparameters} controlling the distribution of 
these dust parameters.}It is called a {\it multi-level} approach, for that 
reason.
A least-squares fit is a single-level model, as it only deals with the dust 
parameters.
In a single-level approach, the parameter's probability distribution of each 
source is independent of the other sources in the sample.
Introducing hyperparameters allows one to sample a single large probability 
distribution for the parameters of all the sources.
This way, the information about the distribution of parameters, among the different sources, has an impact on the likelihoods of individual sources.

\citet{kelly12} were able to show that the HB approach could correct the false 
negative 
correlation between the temperature, $T$, and the emissivity index, $\beta$, 
of modified black bodies (hereafter MBB).
We emphasize that it is the multi-level nature of the HB approach that could 
lead to such an improvement.
Indeed, several non-hierarchical Bayesian codes had been developed to interpret 
dust SEDs \citep{da-cunha08,paradis10,serra11,planck-collaboration11d}, but those could not treat the 
degeneracies, due to their single-level nature.
Following \citet{kelly12}, several articles presented HB codes for single MBBs 
\citep{juvela13}, or the linear combination of two MBBs, and MBBs with 
parameterized emissivity \citep{veneziani13}.
Until now, we were lacking an HB code working with full dust models, 
including:\textlist{\thetextlist~several types of grains (silicates 
and carbonaceous),
\thetextlist~with realistic optical properties, 
\thetextlist~a size distribution, and
\thetextlist~accounting for the stochastic heating of small grains,
\thetextlist~with the possibility to combine several components to account for 
the mixing of physical conditions in the observed regions.}Such
a model would allow us to apply state-of-the-art dust models to IR 
observations, extracting the physical parameters in an optimal way, consistent with the various known sources of uncertainties.
This tool could provide unique constraints on dust evolution.
This article discusses some efforts developed towards that goal.
We present here a new hierarchical Bayesian dust SED model, named \HB\ 
(\citext{HiERarchical Bayesian Inference for dust Emission}).
\seppar

The paper is ordered as follows.
In \refsec{sec:phys}, we present the different physical components that can be 
combined to fit the data.
\refsec{sec:HB} attempts at giving a comprehensive view on the statistical 
hypotheses of the HB model and its numerical implementation.
In \refsecs{sec:refgrid}-\ref{sec:othergrid}, we extensively discuss the tests 
we have conducted in order to assess the performances of the code.
\refsec{sec:refgrid} presents a systematic analyses of the code performances, 
varying signal-to-noise ratios, SED shape and sample size.
\refsec{sec:vargrid} explores additional effects: wavelength coverage; the 
presence of intrinsic correlations; the addition of external parameters to the 
model.
\refsec{sec:othergrid} briefly demonstrates the code performances using 
different physical models.
Finally, \refsec{sec:concl} summarizes our results.
We devote \refapps{app:template}-\ref{app:CPU} to technical explanations that 
would otherwise alter the flow of the discussion.

This is the first paper in a series of two articles.
Paper~II \citepprep{galliano18b} will address the robustness of our model.

  \section{The Physical model}
  \label{sec:phys}

    \subsection{Notation conventions}
    \label{sec:conv}

\HB\ is designed to solve a generic problem: the fit of a linear combination of  SED model components to a set of $n$ sources, $s_i$, observed through $m$ frequencies, $\nu_j$.
The sources, $s_i$ ($i=1,\ldots,n$), can be pixels in an image or individual objects (whole galaxies, star forming regions, \etc) in a sample.
The frequencies, $\nu_j$ ($j=1,\ldots,m$), can be photometric bands or frequency elements in a spectrum.
The total physical model is controlled by $q$ parameters, $x_k$ 
($k=1,\ldots,q$).

We note $\vect{y}$, vectors of elements $y_i$, and $\mat{Y}$, matrices of 
elements $Y_{i,j}$.
We note $|\mat{Y}|$ the determinant of matrix $\mat{Y}$.
Probability density functions (PDF) are written $p(\ldots)$;
the name of the variable between parentheses defines the actual density we are refering to.

Due to the inherent non-linearity of dust models, the derived PDF of several parameters is significantly skewed.
Although this is natural, it makes quoting mean values and uncertainties complicated, as they do not coincide with the median nor the mode of the distribution.
For these parameters, we rather consider the natural logarithm (noted $\ln$), 
in order to obtain a more symmetric distribution.

We use the adjective \citext{specific} to denote any quantity \citext{per unit 
mass}.

  \subsection{Individual SED model components}
  \label{sec:SEDcomp}

Fitting an observed SED requires choosing the number and the nature of the 
physical components.
Each case has to be adapted, depending on the assumed properties of the studied 
object (\eg\ diffuse cloud, \hii\ region, whole galaxy, \etc) and on the 
wavelength coverage.
In this section, we describe the different components that our code currently 
implements.
This list is meant to grow with future developments.

    \subsubsection{Equilibrium grains (\BBQ)}
    \label{sec:BBQ}
  
This component accounts for the emission of a collection of identical large 
grains at thermal equilibrium with a uniform radiation field.
It is particularly useful to model \hii\ regions where strong solid-state 
emission bands can be observed (\eg\ \citeprep{wu18}).
This approximation would not be valid for smaller grains as they would be 
stochastically heated \citep[\eg][]{draine03c}.
We designate this component as \BBQ.

The parameters controlling this component are simply the mass ($M_i$) and 
temperature ($T_i$): 
$\vect{x}_i=(\ln M_i,\ln T_i)$.
The monochromatic luminosity of source $s_i$, at frequency $\nu_j$ is expressed as:
\begin{equation}
  L_\nu^\sms{mod}(\nu_j,\vect{x}_i) 
    = M_i\times\frac{3\pi}{a\rho}Q_\sms{abs}(a,\nu_j)
                      \times B_\nu(T_i,\nu_j),
\end{equation}
where $a$ is the grain radius, $B_\nu$ is the Planck function, and $Q_\sms{abs}$ is the absorption efficiency of the chosen material, with mass density $\rho$.
We have a large database of optical properties to choose the $Q_\sms{abs}$ from, depending on the spectral features we want to model.
We arbitrarily set $a=30\;\rm nm$\footnote{Grains with 
$a\lesssim10\;\rm nm$ are usually stochastically heated, and cannot be modelled 
with a single temperature; grains with $a\gtrsim0.1\mmic$ have a size dependent 
equilibrium temperature, due to their grey UV extinction.}, as $Q_\sms{abs}/a$ 
is almost perfectly independent of $a$ in the IR-mm range.

  \subsubsection{Modified black body (\MBB)}
  \label{sec:MBB}

This component is similar to \BBQ, in terms of physical assumptions.
The only difference is that instead of considering a realistic $Q_\sms{abs}$, measured in the laboratory, we make the common assumption that it is a power-law of the frequency: $Q_\sms{abs}\propto\nu^\beta$,
the emissivity index $\beta$ being a free parameter.
This approximation is the most widely used dust model, as it is supposed to provide constraints both on the physical conditions experienced by the grains (through the temperature) and on their intrinsic properties (through $\beta$).
We designate this component as \MBB.

The parameters controlling this component are $\vect{x}_i=(\ln M_i,\ln T_i, \beta_i)$.
The monochromatic luminosity of source $s_i$, at frequency $\nu_j$ is:
\begin{equation}
  L_\nu^\sms{mod}(\nu_j,\vect{x}_i)
    = M_i\times 4\pi\kappa(\nu_0)
      \left(\frac{\nu_j}{\nu_0}\right)^{\beta_i}\times B_\nu(T_i,\nu_j).
\end{equation}
The opacity at reference wavelength $\lambda_0=c/\nu_0=160\mmic$ is
fixed to $\kappa(\nu_0)=1.4\;\rm m^2\, kg^{-1}$, as it corresponds to the \citet[][BARE-GR-S]{zubko04} Galactic dust model opacity.
Although \citet{hildebrand83} recommended normalizing the opacity in the submm regime, recent laboratory studies \citep[\eg][]{coupeaud11} tend to show that dust analogs have less dispersion in the far-IR.
This is the reason why we choose $\lambda_0=160\mmic$ rather than $850\mmic$.

  \subsubsection{Broken emissivity modified black body (\BEMBB)}
  \label{sec:BEMBB}

This component, introduced by \citet{gordon14}, is a refinement of \MBB, accounting for a possible change of the emissivity slope at long wavelengths, similar to what is observed in the laboratory \citep[\eg][]{coupeaud11}.
There are now two emissivity indices: $\beta_1$ for $\nu>\nu_b$
and $\beta_2$ for $\nu\le\nu_b$.
In practice, this component could be used to fit an observed SED, finely sampled in the far-IR/submm range.
We designate this component as \BEMBB.

The parameters controlling this component are 
$\vect{x}_i=(\ln M_i,\ln T_i, \beta_{1,i}, \beta_{2,i}, \nu_{b,i})$.
The monochromatic luminosity of source $s_i$, at frequency $\nu_j$ is:
\begin{equation}
  L_\nu^\sms{mod}(\nu_j,\vect{x}_i)
      = 4\pi\kappa(\nu_j,\beta_{1,i},\beta_{2,i},\nu_{b,i})\times 
        B_\nu(\nu_j,T_i),
\end{equation}
where the \citext{broken emissivity} is parameterized as:
\begin{equation}
  \kappa(\nu,\beta_{1,i},\beta_{2,i},\nu_{b,i})
    = \left\{
      \begin{array}{ll}
        \displaystyle\kappa(\nu_0)\left(\frac{\nu}{\nu_0}
                                         \right)^{\beta_{1,i}} 
          & \nu > \nu_{b,i} \\
        \displaystyle\kappa(\nu_0)
          \left(\frac{\nu}{\nu_0}\right)^{\beta_{2,i}} 
          \left(\frac{\nu_{b,i}}{\nu_0}\right)^{\beta_{1,i}-\beta_{2,i}}
          & \nu \le \nu_{b,i}.
      \end{array}
      \right.
\end{equation}
\citet{gordon14} \citengl{calibrate} the opacity at reference wavelength
$\kappa(\nu_0)$ on the diffuse ISM of the Milky Way.
Here, we test this component with the same $\kappa(\nu_0)$ value as the \MBB,
for simplicity.

  \subsubsection{Uniformly illuminated dust mixture (\deltaU)}
  \label{sec:deltaU}

This component represents a full ISM dust mixture, heated by a uniform interstellar radiation field (ISRF) with intensity $U$. 
This mixture is made of grains of different compositions (silicates, amorphous carbon, PAHs, \etc) each having a different size distribution.
Several dust mixtures and their size distributions are implemented into our code: the BARE-GR-S model of \citet{zubko04}, the model of \citet{compiegne11}, the AC model of \citet{galliano11} and the THEMIS model \citep{jones17}.
We assume that the radiation field has the spectral shape of the solar neighborhood \citep[with mean intensity noted $J_\nu^\odot(\nu)$;][]{mathis83} scaled by the parameter $U$: $J_\nu(\nu)=U\times J_\nu^\odot(\nu)$.
The mean intensity is normalized so that $U=1$ corresponds to:
\begin{equation}
  \int_{c/8\,\mu m}^{c/0.0912\,\mu m} 4\pi J_\nu^\odot(\nu)\ddiff\nu = 2.2\E{-5}\;\rm W\,m^{-2}.
\end{equation}

The parameters controlling this component are the mass ($M_i$), the radiation field intensity ($U_i$), the PAH mass fraction ($q^\sms{PAH}_i$) and the 
fraction of charged PAHs ($f^+_i$):
$\vect{x}_i=(\ln M_i,\ln U_i,q^\sms{PAH}_i,f^+_i)$.
The last two parameters allow us enough flexibility to fit the complex mid-IR range: $q^\sms{PAH}$ controls the relative strength of the aromatic features, while $f^+$ controls mainly the ratio between C--C and C--H bands \citep[\eg][]{allamandola99,li01,galliano08b}.
The monochromatic luminosity of source $s_i$, at frequency $\nu_j$, is:
\begin{eqnarray}
  L_\nu^\sms{mod}(\nu_j,\vect{x}_i) & = & M_i\times (
    q^\sms{PAH}_i\times f_i^+\times l_\nu^{\sms{PAH}^+}(U_i,\nu_j) 
   \nonumber\\
   & + & q^\sms{PAH}_i\times (1-f_i^+)\times l_\nu^{\sms{PAH}^0}(U_i,\nu_j)
   \nonumber\\
   & + & (1-q^\sms{PAH}_i)\times l_\nu^\sms{non-PAH dust}(U_i,\nu_j)),
   \label{eq:deltaU}
\end{eqnarray}
where $l_\nu^\sms{X}$ is the specific monochromatic luminosity of component X.
The emission spectrum has been computed with the \citet{guhathakurta89} stochastic heating method and has been integrated over the size distribution.
We designate this component as \deltaU.

  \subsubsection{Non-uniformly illuminated dust mixture (\powU)}
  \label{sec:powerU}

The application of the previous component is often problematic as it does not allow us to account for the likely mixing of excitation conditions along the line of sight and in the instrumental beam.
To account for non-uniform illumination, we adopt the prescription of \citet{dale01}, assuming that the dust mass is distributed in different radiation field intensities following a power-law:
\begin{equation}
  \frac{\dd M}{\dd U}=\mathcal{N}\times U^{-\alpha} \;\;\;\mbox{with}\;\;\; 
  U_-\le U\le U_-+\Delta U,
  \label{eq:dale}
\end{equation}
where the normalization factor depends on the value of $\alpha$:
\begin{equation}
  \mathcal{N} = \left\{
  \begin{array}{ll}
    \displaystyle
    \frac{(1-\alpha)}{(U_-+\Delta U)^{1-\alpha}
                      -(U_-)^{1-\alpha}} 
      & \mbox{if $\alpha>1$} \\
    \displaystyle
    \frac{1}{\ln(U_-+\Delta U)-\ln(U_-)}
      & \mbox{if $\alpha=1$.}
  \end{array} \right.
\end{equation}
The three parameters of this power-law $U_-$, $\Delta U$ and $\alpha$ are free 
to vary to fit the shape of the observed SED.
Since these three parameters are sometimes degenerate, we often discuss the 
more stable average starlight intensity, defined as:
\begin{equation}
  \Uavm
  =
  \frac{1}{M}
  \int_{U_-}^{U_-+\Delta U} U\times \frac{\dd M}{\dd U}\ddiff U \\
\end{equation}
which also depends on the value of $\alpha$:
\begin{equation}
  \Uavm = 
  \left\{
  \begin{array}{ll}
    \displaystyle
    \frac{1-\alpha}{2-\alpha}
    \frac{\left(U_-+\Delta U\right)^{2-\alpha}-U_-^{2-\alpha}}
         {\left(U_-+\Delta U\right)^{1-\alpha}-U_-^{1-\alpha}}
    & \mbox{ if } \alpha\neq1 \;\&\; \alpha\neq2 \\
    & \\
    \displaystyle
    \frac{\Delta U}{\ln\left(U_-+\Delta U\right)-\ln U_-}
    & \mbox{ if } \alpha = 1 \\
    & \\
    \displaystyle
    \frac{\ln\left(U_-+\Delta U\right)-\ln U_-}
         {U_-^{-1}-\left(U_-+\Delta U\right)^{-1}}
    & \mbox{ if } \alpha = 2, \\
  \end{array}
  \right.
  \label{eq:avU}
\end{equation}

This component is simply the \deltaU\ component integrated over the distribution of \refeq{eq:dale}.
The parameters controlling this component are thus 
$\vect{x}_i=(\ln M_i,\ln U_{-,i},\ln\Delta U_i,\alpha_i,q^\sms{PAH}_i,f_i^+)$.
The monochromatic luminosity of source $s_i$, at frequency $\nu_j$ is:
\begin{equation}
L_\nu^\sms{mod}(\nu_j,\vect{x}_i)
      = M_i\times\mathcal{N}
        \int_{U_{-,i}}^{U_{-,i}+\Delta U_i}
               l_\nu(U_i,q^\sms{PAH}_i,f^+_i,\nu_j)\times
               U^{-\alpha_i}
               \ddiff U,
\end{equation}    
where $M_i\times l_\nu(U_i,q^\sms{PAH}_i,f^+_i,\nu_j)$ is the \deltaU\ component \refeqp{eq:deltaU}.
We designate this component as \powU.

  \subsubsection{Near-IR emission by stellar populations (\Star)}
  \label{sec:starBB}

Direct or scattered starlight can significantly contaminate the emission in the near-IR range.
Thus, to properly model the emission by PAHs and small grains, one needs to account for the stellar continuum.
This component is simply modelled as a $T_\star=50\,000\;\rm K$ black body, 
the only free parameter being its bolometric luminosity 
$\vect{x}_i=(\ln L^\star_i)$.
The monochromatic luminosity of source $s_i$, at frequency $\nu_j$ is:
\begin{equation}
  L_\nu^\sms{mod}(\nu_j,\vect{x}_i)
   = L^\star_i\times\frac{\pi}{\sigma T_\star^4} 
                                 B_\nu(T_\star,\nu_j),
\end{equation}
where $\sigma$ is the Stefan-Boltzmann constant and $B_\nu$ is the Planck function.
This parameterization is however too rough to provide any reliable constraint on the actual stellar populations.
In addition, it is only realistic in the near-IR regime.
We designate this component as \Star.

  \subsubsection{Free-free and synchrotron continua (\radio)}
  \label{sec:radio}

In addition to dust emission, submillimeter bands also contain radio continuum.
We therefore add the possibility to model this component, assuming it is the linear combination of two power-laws representing the free-free ($F_\nu\propto\nu^{-0.1}$) and synchrotron continua ($F_\nu\propto\nu^{-\alpha_s}$; $0.7\lesssim\alpha_s\lesssim0.9$).
The free parameters are the $\nu L_\nu$ at $\lambda_1=1\;\rm cm$, $L_{1,i}$,
the fraction of free-free at $\lambda_1$, $f_{\sms{FF},i}$, and the synchrotron index, $\alpha_{s,i}$:
$\vect{x}_i=(\ln L_{1,i},f_{\sms{FF},i},\alpha_{s,i})$.

The monochromatic luminosity of source $s_i$, at frequency $\nu_j$, is:
\begin{eqnarray}
  L_\nu^\sms{mod}(\nu_j,\vect{x}_i)
   & = & \frac{L_{1,i}}{\nu_1}\times 
  \left(
  f_{\sms{FF},i}\times\left(\frac{\nu}{\nu_1}\right)^{-0.1}\right. \nonumber\\
  & + &\left.(1-f_{\sms{FF},i})\times\left(\frac{\nu}{\nu_1}\right)^{-\alpha_{s,i}}\right),
\end{eqnarray}
with $\nu_1=c/\lambda_1$.
We designate this component as \radio.

  \subsection{Linear combination of individual components}
  \label{sec:SEDcomb}  

We can linearly combine any of the previous components to fit an observed SED.
Each parameter can be let free to vary (limits can be specified), or fixed to 
an arbitrary value.
Alternatively, we can tie two parameters together (\eg\ the $\beta$ of two 
\MBB\ components or the $q^\sms{PAH}$ of a \deltaU\ and \powU\ components).

From a numerical point of view, we have pre-computed large grids of models for a large range of all the parameters, and performed synthetic photometry for a wide list of instrumental filters.
Our code then simply interpolates these grids of templates to evaluate the SED model for a combination of parameters.
The model grids and the interpolation method are described in \refapp{app:template}.

  \section{Hierarchical Bayesian inference applied to SED fitting}
  \label{sec:HB}

Most of the formalism discussed in this section is the generalization to more complex dust models of the pioneering work of \citet{kelly12}.

  \subsection{The Non-hierarchial Bayesian point of view}
  \label{sec:bayes}

Let's first start by laying down the standard Bayesian formalism.
If, for now, we omit systematic uncertainties, we can simply express the 
observed SED of a single source as the sum of our emission model and a random deviation due to the noise:
\begin{equation}
  L_\nu^\sms{obs}(\nu_j) = L_\nu^\sms{mod}(\nu_j,\vect{x}) 
                         + \epsilon(\nu_j)\times \sigma_\nu^\sms{noise}(\nu_j),
  \label{eq:Lnu}               
\end{equation}
where $\epsilon(\nu)$ is a random variable with $\langle\epsilon\rangle=0$ and 
$\sigma(\epsilon)=1$.
The probability density function of the noise $p(\epsilon(\nu_j))$ can be 
arbitrary (gaussian, Student's $t$, \etc).
We can reverse \refeq{eq:Lnu} to make explicit its dependence on the parameters 
$\vect{x}$:
\begin{equation}
  \epsilon(\nu_j,\vect{x}) 
    = \frac{L_\nu^\sms{obs}(\nu_j)-L_\nu^\sms{mod}(\nu_j,\vect{x})}
           {\sigma_\nu^\sms{noise}(\nu_j)}.
  \label{eq:epsilon}
\end{equation}
In the absence of correlations, the likelihood can be expressed as a conditional 
PDF:
\begin{equation}
  p(\vect{L}_\nu^\sms{obs}|\vect{x}) 
    = \prod_{j=1}^m p(\epsilon(\nu_j,\vect{x})),
  \label{eq:LH}
\end{equation}
noting 
$\vect{L}_\nu^\sms{obs}=(L_\nu^\sms{obs}(\nu_1),\ldots,L_\nu^\sms{obs}(\nu_m))$, 
the SED vector containing the emission at each waveband.
This is a well-kown expression.
For instance, if we assume gaussian errors, finding the maximum of this 
likelihood is equivalent to minimizing the chi-squared.

From a Bayesian point view, we are more interested in the PDF of the 
parameters, knowing the observations, 
$p(\vect{x}|\vect{L}_\nu^\sms{obs})$, rather than in the distribution of 
\refeq{eq:LH}.
Conveniently, Bayes' theorem states that:
\begin{equation}
  p(\vect{x}|\vect{L}_\nu^\sms{obs}) 
    = \frac{p(\vect{L}_\nu^\sms{obs}|\vect{x})
      \times p(\vect{x})}{p(\vect{L}_\nu^\sms{obs})}
      \propto p(\vect{L}_\nu^\sms{obs}|\vect{x})\times p(\vect{x}).
  \label{eq:bayes}
\end{equation}
In the equation above, $p(\vect{x})$ is called the {\it prior} distribution.
It represents the intrinsic distribution of the parameters.
We usually do not have an accurate knowledge of this distribution.
The term $p(\vect{L}_\nu^\sms{obs})$ is a constant, since it is independent of 
the parameters.
It enters as the normalization factor in the second part of \refeq{eq:bayes}.

The left-hand term of \refeq{eq:bayes} is called the {\it posterior} distribution.
Standard Bayesian techniques consist of sampling it, \ie\ mapping it in the space of parameters $\vect{x}$.
An assumption has to be made on the prior.
It is common to assume it is constant or slowly varying over the interval range covered by the parameters.
From the knowledge of the distribution of \refeq{eq:bayes}, one can estimate parameter averages, standard deviations, estimate confidence intervals, test hypotheses, \etc.

  \subsection{The Hierarchical model}
  \label{sec:hierarchical}
 
 \subsubsection{The Posterior distribution}
 
The hierarchical Bayesian method is built upon the previous formalism, with the difference that the prior distribution is now inferred from the data.
This is achieved by parameterizing the shape and position of the new prior distribution with {\it hyperparameters} (which control the distribution of the parameters).
It is common to assume a unimodal prior (\eg\ multivariate gaussian or Student's $t$), where the hyperparameters are:\textlist{\thetextlist~the average of the parameter vector, $\vect{\mu}$, and \thetextlist~their covariance matrix, $\mat{\Sigma}$.}This 
approach is relevant only when analyzing a sample of $n>1$ sources.
\refeq{eq:bayes} now becomes, for one source $s_i$:
\begin{equation}
  p(\vect{x}_i|\vect{L}_{\nu,i}^\sms{obs},\vect{\mu},\mat{\Sigma})
  \propto p(\vect{L}_{\nu,i}^\sms{obs}|\vect{x}_i)
    \times p(\vect{x}_i|\vect{\mu},\mat{\Sigma}),
\end{equation}
where $p(\vect{x}_i|\vect{\mu},\mat{\Sigma})$ is the new prior parameterized by  $\vect{\mu}$ and $\mat{\Sigma}$.
The posterior distribution of the parameters and their hyperparameters is then:
\begin{equation}
  \begin{multlined}
    p(\vect{x}_1,\ldots,\vect{x}_n,\vect{\mu},\mat{\Sigma}
     |\vect{L}_{\nu,1}^\sms{obs},\ldots,\vect{L}_{\nu,n}^\sms{obs})
  \\
    \propto 
    \prod_{i=1}^n
      p(\vect{x}_i|\vect{L}_{\nu,i}^\sms{obs},\vect{\mu},\mat{\Sigma})
      \times p(\vect{\mu})\times p(\mat{\Sigma}),
  \end{multlined}
  \label{eq:HB}
\end{equation}
where $p(\vect{\mu})$ and $p(\mat{\Sigma})$ are the prior distributions of the hyperparameters.
They will be described in \refsec{sec:prior}.
Note that, in \refeq{eq:HB}, there is only one common set of hyperparameters for all of the sources.

\begin{figure}
  \includegraphics[width=\linewidth]{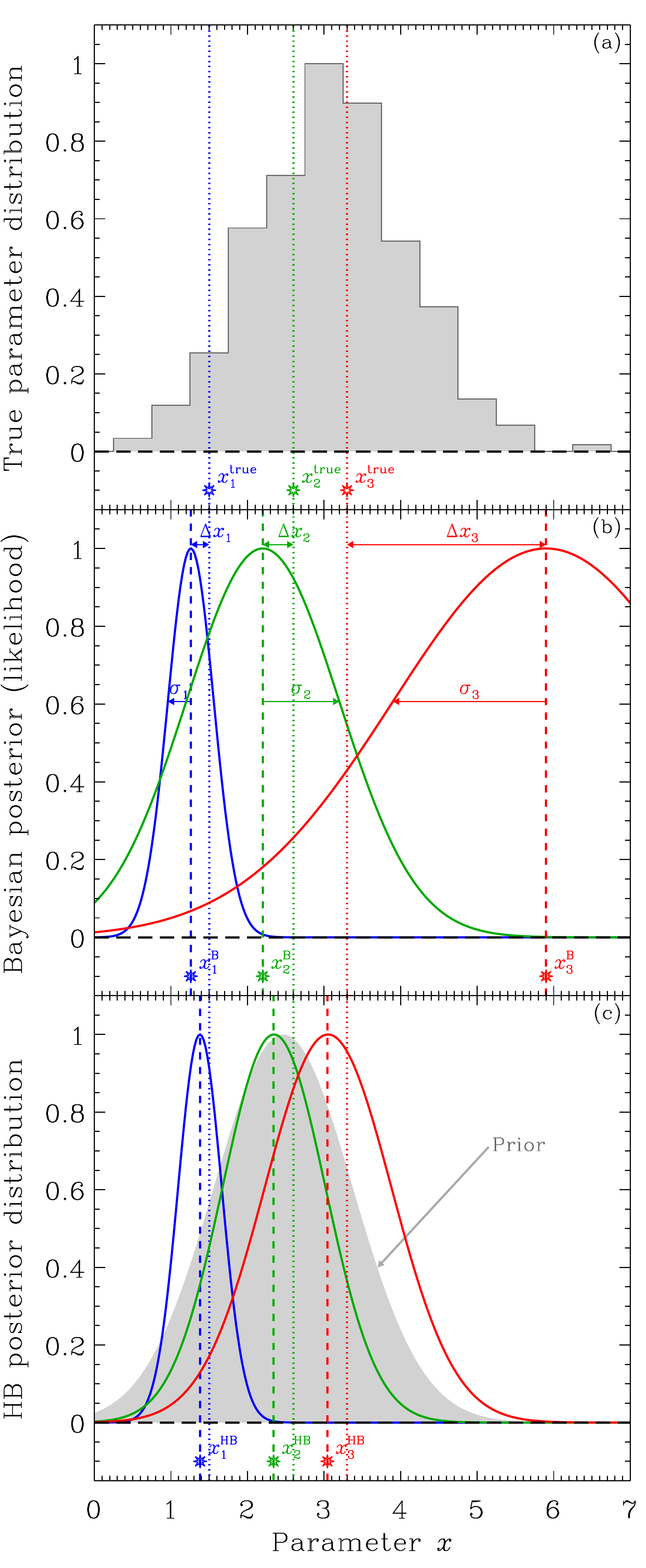}
  \caption{{\sl Illustration of the HB method.}
           Panel (a) shows the actual distribution of parameter $x$
           (in grey) and three values of these parameters 
           ($x_{1,2,3}^\sms{true}$; in colors; dotted lines) 
           that correspond to the observations.
           Panel (b) shows the likelihood of the fit (color curves)
           when introducing noise ($\sigma_{1,2,3}$).
           The mean value of $x$ derived from the likelihood 
           $x_{1,2,3}^\sms{B}$ is displaced from the true value by 
           $\Delta x_{1,2,3}=|x_{1,2,3}^\sms{B}-x_{1,2,3}^\sms{true}|$.
           Panel (c) shows the inferred prior (in grey), the posterior 
           distributions (color curves) and the derived mean values
           ($x_{1,2,3}^\sms{HB}$).}
  \label{fig:demoHB}
\end{figure}
\reffig{fig:demoHB} illustrates how the HB method works.
To make it simple, we consider a model with only one parameter $x$,  
such as, for example, the intensity of a single line fit.
We simulate, in panel (a), what could be the actual distribution of parameters (in grey).
We have drawn three values (color dotted lines) from this distribution, representing what we would get if we were observing a line in three different locations in a cloud: $x_{1,2,3}^\sms{true}$.
In panel (b), we introduce noise and show the likelihoods (color curves).
We have assigned an arbitrary uncertainty to the three parameters:
$\sigma_{1,2,3}$.
We have drawn a random deviation according to this uncertainty:
$\Delta x_{1,2,3}$.
If we were implementing a non-hierarchical Bayesian method or a least-squares fit, our estimate of the parameters would be the vertical dashed lines at the mean/mode of each likelihood: $x_{1,2,3}^\sms{B}$.
In panel (c), we show the inferred mean prior distribution (in grey).
The color curves represent the HB posterior, \ie\ the product of the prior and the likelihood.
The mean values that we would derive from the HB analysis are the $x_{1,2,3}^\sms{HB}$, which are closer to the true values than the $x_{1,2,3}^\sms{B}$.
The reason is that the multiplication by the prior reduces the dispersion of the values due to the noise.
The posterior number 1 (in blue) has a high signal-to-noise ratio.
Its likelihood is narrower than the prior. 
Thus, it is only weakly modified by the prior, and $x_1^\sms{B}\simeq x_1^\sms{HB}$.
In contrast to this, the posterior number 3 (in red) deviates more, since it has a low signal-to-noise ratio. 
The prior thus has a major effect on this distribution and brings 
$x_3^\sms{HB}$ closer to $x_3^\sms{true}$.

One of the subtleties of this process is the inference of the prior.
In our simple example, its hyperparameters are the mean, $\bar{x}$, and standard deviation, $\sigma_x$.
The full posterior distribution \refeqp{eq:HB} is thus a 5-dimension PDF depending here on $x_1$, $x_2$, $x_3$, $\bar{x}$ and $\sigma_x$.
The most likely hyperparameters will be the values of $\bar{x}$ and $\sigma_x$ for which the posterior is maximum, in the 5-dimension space.
We can also estimate averages of the hyperparameters $\langle\bar{x}\rangle$
and $\langle\sigma_x\rangle$, marginalizing the posterior over the rest of the variables. 
In that sense, the representation of panel (c) is a simplification.
It actually shows unidimensional cuts of the prior and the posteriors, fixing $\bar{x}$ to $\langle\bar{x}\rangle$ and $\sigma_x$ to $\langle\sigma_x\rangle$.

  \subsubsection{Introducing systematic uncertainties}
  \label{sec:delta}

Instrumental calibration uncertainties, on top of being partially correlated between wavebands, are assumed to be fully correlated between sources.
\citet{kelly12} treat these calibration errors as nuisance parameters.
Following their formalism, we can rewrite \refeq{eq:epsilon}:
\begin{equation}
  \epsilon(\nu_j,\vect{x},\delta_j)
  = \frac{L_\nu^\sms{obs}(\nu_j)
         -L_\nu^\sms{mod}(\nu_j,\vect{x})\times(1+\delta_j)}
         {\sigma_\nu^\sms{noise}(\nu_j)},
  \label{eq:epsdelta}
\end{equation}
where we have introduced the calibration offset $\delta_j$ at frequency 
$\nu_j$.
The posterior distribution of \refeq{eq:HB} becomes:
\begin{equation}
  \begin{multlined}
    p(\vect{x}_1,\ldots,\vect{x}_n,\vect{\mu},\mat{\Sigma},\vect{\delta}
     |\vect{L}_{\nu,1}^\sms{obs},\ldots,\vect{L}_{\nu,n}^\sms{obs})
  \\
    \propto 
    \prod_{i=1}^n\prod_{j=1}^m
      p(L_{\nu,i,j}^\sms{obs}|\vect{x}_i,\delta_j)
    \times p(\vect{x}_i|\vect{\mu},\mat{\Sigma})
      \times p(\vect{\mu})\times p(\mat{\Sigma})\times p(\vect{\delta}),
  \end{multlined}
  \label{eq:HBdelta}
\end{equation}
where the new likelihood, $p(L_{\nu,i,j}^\sms{obs}|\vect{x}_i,\delta_j)$, is simply $p(\epsilon(\nu_j,\vect{x}_i,\delta_j))$.
We have also introduced the prior distribution of $\vect{\delta}$, $p(\vect{\delta})$.
Its mean is $\langle\delta_j\rangle=0$ and its covariance matrix, $V_\sms{cal}$, is made of the calibration uncertainties of the different wavebands and their correlations.

  \subsubsection{The Noise distribution}
  \label{sec:noise}

The noise is assumed to be uncorrelated between wavelengths and between sources.
Our model let us choose between different distributions for the variable, depending on the actual distribution of the noise measured on the data.
We currently can choose between the three following types of noise.
\begin{description}
  \item[Normal noise:]
    this is the default case.
    The monochromatic likelihood of source $s_i$, at frequency $\nu_j$, 
    is in this case:
    \begin{equation}
      p(L_{\nu,i,j}^\sms{obs}|\vect{x}_i,\delta_j) 
      \propto
      \exp\left[-\frac{1}{2}
                \left(\epsilon(\nu_j,\vect{x}_i,\delta_j)\right)^2\right].
      \label{eq:LH1}
    \end{equation}
  \item[Robust noise:] 
    in case of outliers, we can assume that the 
    statistical errors follow a Student's $t$ distribution with $f=3$ degrees 
    of freedom.
    It has broader wings than a gaussian PDF with the same $\sigma$.
    The monochromatic likelihood of source $s_i$, at frequency $\nu_j$, 
    is then:
    \begin{equation}
      p(L_{\nu,i,j}^\sms{obs}|\vect{x}_i,\delta_j) 
      \propto
      \left[1+\frac{1}{f}
      \left(\epsilon(\nu_j,\vect{x}_i,\delta_j)\right)^2\right]^{-(f+1)/2}.
      \label{eq:LH2}
    \end{equation}
  \item[Asymmetric noise:]
    background galaxies or Galactic cirruses can skew the noise 
    distribution towards high fluxes.
    In this case, we use a split-normal PDF \citep{villani06}:
    \begin{equation}
      p(L_{\nu,i,j}^\sms{obs}|\vect{x}_i,\delta_j) 
      \propto
      \left\{\begin{array}{l}
      \displaystyle
      \exp\left[-\frac{1}{2}
        \left(\frac{L_{\nu,i,j}^\sms{obs}-L_{\nu,i,j}^\sms{mod}
                    \times(1+\delta_j)-\mu_{i,j}}
                   {\lambda_{i,j}}\right)^2\right]
      \\
        \;\;\;\;\;\mbox{if } 
         L_{\nu,i,j}^\sms{obs}-L_{\nu,i,j}^\sms{mod}\times(1+\delta_j) 
         \le \mu_{i,j} \\
      \\
      \displaystyle
      \exp\left[-\frac{1}{2}
        \left(\frac{L_{\nu,i,j}^\sms{obs}-L_{\nu,i,j}^\sms{mod}
                    \times(1+\delta_j)-\mu_{i,j}}
                   {\lambda_{i,j}\tau_{i,j}}\right)^2\right]
      \\
        \;\;\;\;\;\mbox{if } 
         L_{\nu,i,j}^\sms{obs}-L_{\nu,i,j}^\sms{mod}\times(1+\delta_j) 
         > \mu_{i,j} \\
       \end{array}\right.
       \label{eq:LH3}
    \end{equation}
    where the position parameter, $\mu_{i,j}$, the scale parameter, 
    $\lambda_{i,j}$, and the shape parameter, $\tau_{i,j}$, are derived from 
    the mean (0), standard deviation ($\sigma_{\nu,i,j}^\sms{noise}$) and 
    skewness of the noise (\refapp{app:split}).
\end{description}

  \subsubsection{The Prior distributions}
  \label{sec:prior}

We follow \citet{kelly12}, and assume a $g=8$ degrees of freedom multivariate Student's $t$ distribution, for the distribution of our parameters:
\begin{equation}
  p(\vect{x}_i|\vect{\mu},\mat{\Sigma})
  \propto \frac{1}{\sqrt{|\mat{\Sigma}|}}\times
  \left(1+\frac{1}{g}(\vect{x}_i-\vect{\mu})^T\mat{\Sigma}^{-1}(\vect{x}_i-\vect{\mu})\right)^{-(g+q)/2},
  \label{eq:prior}
\end{equation}
where $q$ is the number of parameters (\refsec{sec:conv}).
Notice that the factor $1/\sqrt{|\mat{\Sigma}|}$, in front of the exponential, is not a normalization constant here,
since we are sampling the distribution as a function of the elements of 
$\mat{\Sigma}$.

We assume a uniform prior on $\vect{\mu}$.
For the $q\times q$ covariance matrix, $\mat{\Sigma}$, we use the \citet{barnard00} {\it separation
strategy}, decomposing it as: $\mat{\Sigma}=\mat{S}\mat{R}\mat{S}$, where 
$\mat{S}$ is the diagonal matrix of standard deviations,
and $\mat{R}$ is the correlation matrix.
We place wide, independent normal priors on the diagonal elements of $\ln\mat{S}$, centered on the standard-deviations of the least-squares best fit parameters (\refapp{app:chi2}), $\ln S_{k,k}^{\chi^2}$, with $\sigma(\ln S_{k,k})=10$:
\begin{equation}
  p(\mat{S}) =
  \prod_{k=1}^q 
  \frac{1}{S_{k,k}}\frac{1}{\sqrt{2\pi}\sigma(\ln S_{k,k})}
  \exp\left[-\frac{1}{2}\left(\frac{\ln S_{k,k}-\ln S_{k,k}^{\chi^2}}{\sigma(\ln S_{k,k})}\right)^2\right].
\end{equation}
The prior on $\mat{R}$ is chosen to make sure that each correlation coefficient is uniformly distributed between $-1$ and $1$ and $\mat{R}$ is positive 
definite.
The formalism developed by \citet{barnard00} assumes that $p(\mat{\Sigma}|\mat{S})$ is distributed according to an inverse Wishart distribution, with $h=q+1$ degrees of freedom.
The resulting prior on $\mat{R}$ is then:
\begin{eqnarray}
  p(\mat{R}) 
    & \propto & |\mat{R}|^{(h-1)(q-1)/2-1}\left(\prod_{k=1}^q|\mat{R}_{(kk)}|\right)^{-h/2} \\
    & \propto & |\mat{R}|^{q(q-1)/2-1}\left(\prod_{k=1}^q|\mat{R}_{(kk)}|\right)^{-(q+1)/2}
  \label{eq:priorcorr}
\end{eqnarray}
where $\mat{R}_{(kk)}$ is the principal submatrix of order $k$, i.e.\
the matrix of elements $R_{k_1,k_2}$, with $k_1=1,\ldots,k$ and $k_2=1,\ldots,k$.
In the end, the prior distribution of the hyperparameters is:
\begin{equation}
  p(\vect{\mu})\times p(\mat{\Sigma}) \propto p(\mat{S})\times p(\mat{R}).
\end{equation}

Finally, the prior on the calibration offsets, $\vect{\delta}$, is designed to reflect the absolute calibration uncertainties recommended by each instrument's team, with $m\times m$ covariance matrix, $\mat{V}_\sms{cal}$.
Since the calibration factor $(1+\vect{\delta})$ can not be zero, we draw the variable 
$\vect{\delta^\prime}=\ln(1+\vect{\delta})$.
Noting that $\delta_j\ll 1$, we can assume that $\delta_j^\prime\simeq\delta_j$, and thus that $V_\sms{cal}$ can be used as the covariance matrix of $\vect{\delta}^\prime$.
Similarly to the statistical error, we consider several types of distribution.
The first case is a multivariate normal distribution:
\begin{equation}
  p(\vect{\delta^\prime})
  \propto
  \exp\left(-\frac{1}{2}\vect{\delta^\prime}^T\mat{V}_\sms{cal}^{-1}
             \vect{\delta^\prime}\right).
  \label{eq:priorcal1}
\end{equation}
The second case is a more robust multivariate Student's $t$ distribution with $f=3$ degrees of freedom:
\begin{equation}
  p(\vect{\delta^\prime})
  \propto
  \left(1+\frac{1}{f}\vect{\delta^\prime}^T\mat{V}_\sms{cal}^{-1}
                     \vect{\delta^\prime}\right)^{-(f+m)/2}
  \label{eq:priorcal2}
\end{equation}

  \subsection{The Numerical implementation}

Sampling the distribution of \refeq{eq:HBdelta} is a numerical challenge, as its number of dimensions is:
\begin{equation}
  N_\sms{dim} = \underbrace{n\times q}_\sms{parameters}
                +
                \underbrace{2\times q+q\times(q-1)/2}_\sms{hyperparameters}
                +
                \underbrace{m}_\sms{calibration}.
  \label{eq:Ndim}
\end{equation}
For a typical sample with $n=100$ sources, observed through $m=11$ photometric filters, modelled with $q=7$ free parameters, the dimension is $N_\sms{dim}=767$.
It is thus impractical to map the posterior on a regular cartesian grid of parameters.

  \subsubsection{The Metropolis-Hastings move with ancillarity/sufficiency 
                 interweaving strategy}
  \label{sec:sampler}

The most popular way to sample the posterior distribution is to use a Markov Chain Monte Carlo (MCMC).
This class of algorithm allows one to randomly draw variables from the posterior distribution.
An MCMC is essentially a sequence of values of the set of parameters and hyperparameters. 
The number density of values in the chain scales with the probability density,
\ie\ more models are computed around the maximum likelihood, and few or none in regions of the parameter space where the solution is unlikely.
It makes this method efficient, as the SED model (which can be numerically intensive) is computed only for relevant combinations of the parameters.
MCMCs also make post-processing simple, as one can easily marginalize over any parameter, estimate moments, test hypotheses, \etc\ without having to perform multidimensional integrals.

The MCMC sampler we have developed applies Gibbs sampling within the Metropolis-Hastings algorithm \citep[MH; \eg][]{geman84,gelman04,press07}.
This particular method consists of drawing each parameter, one by one, from their unidimensional conditional posterior distribution, fixing all the other parameters to their current value in the chain.
However, as noted by \citet{kelly11}, calibration uncertainties introduce correlations within the MCMC, and thus require running very long chains to ensure convergence.
The reason is that there is a degeneracy between the values of the calibration offsets, $\vect{\delta}$, and the SED model parameters, $\vect{x}$.
To address this problem, \citet{kelly11} demonstrated that the {\it ancillarity-sufficiency interweaving strategy} \citep[ASIS;][]{yu11} 
could reduce the autocorrelation of the MCMC and thus obtain convergence towards the posterior distribution with a shorter chain.
In simple terms, ASIS consits in inserting an extra step at each MCMC iteration.
In this new step, we draw values of the parameters in a direction where 
$(1+\delta_j)\times L_\nu^\sms{mod}(\vect{x}_i,\nu_j)$ is constant, in order to
decouple $\vect{\delta}$ and $\vect{x}$.
Our code applies ASIS to all model parameters.
    
At first, we set the initial values of the parameters, $\vect{x}_i$, of the MCMC to the best fit values given by the least-squares fit (\refsec{app:chi2}).
For the hyperparameters, we set $\vect{\mu}$ and the diagonal elements of $\mat{S}$ to the mean and standard deviation of the least-squares parameters.
The correlation coefficients (the non-diagonal elements of $\mat{R}$)
are set to 0.
Finally, the initial calibration offsets, $\vect{\delta}$ are set to $\vect{0}$.
We then iterate the following steps $N_\sms{MCMC}$ times.
\begin{enumerate}
  \item We draw the calibration offsets, for each frequency $\nu_j$, from:
    \begin{equation}
      \begin{multlined}
        p(\delta_j|\vect{\delta}_{j^\prime\ne j},
          \vect{x}_1,\ldots\vect{x}_n,
          L_{\nu,1,j}^\sms{obs},\ldots,L_{\nu,n,j}^\sms{obs})
      \\
      \propto
        p(\vect{\delta})
        \times
        \prod_{i=1}^n p(L_{\nu,i,j}^\sms{obs}|\vect{x}_i,\delta_j).
      \end{multlined}
      \label{eq:postdelta}
    \end{equation}
    If $\mat{V_\sms{cal}}$ is not diagonal (correlated calibration 
    uncertainties), 
    then all of the values of $\vect{\delta}$ contribute to every frequency, 
    $\nu_j$.
    Otherwise, only $\delta_j$ contributes.
  \item For each source, $s_i$, we draw each parameter, $x_{i,k}$, from:
    \begin{equation}
      \begin{multlined}
        p(x_{i,k}|\vect{\delta},\vect{x}_{i,k^\prime\ne k},\vect{\mu},
          \mat{\Sigma},\vect{L}_{\nu,i}^\sms{obs})
      \\
      \propto
        p(\vect{x}_i|\vect{\mu},\mat{\Sigma})
        \times
        \prod_{j=1}^m p(L_{\nu,i,j}^\sms{obs}|\vect{x}_i,\delta_j).
      \end{multlined}
      \label{eq:postpar}
    \end{equation}
  \item We implement the 
    {\it component-wise interweaving strategy} \citep[CIS;][]{yu11}.
    To do so, we iterate the following steps by looping on each 
    parameter of index $k^\prime$.
    \begin{enumerate}
      \item For each source, $s_i$, and each frequency, $\nu_j$, we compute the 
        new following variable:
        \begin{equation}
          \tilde{\delta}_{i,j} 
            = (1+\delta_j)\times L_\nu^\sms{mod}(\vect{x}_i,\nu_j).
          \label{eq:asis}
        \end{equation}
        In practice, $L_\nu^\sms{mod}(\vect{x}_i,\nu_j)$ can simply be 
        the model component which is controlled by the parameter we are 
        looping over.
        If the parameter is tied to another one, then its component needs to be
        added.
      \item Then, for one given source, $s_{i^\prime}$, we draw a new value of 
        the physical parameter we are looping over, 
        keeping $\vect{\tilde{\delta}}_{i^\prime}$ constant:
        \begin{equation}
          \begin{multlined}
            p(x_{i^\prime,k^\prime}|\vect{\tilde{\delta}}_{i^\prime},
                           x_{i^\prime,k\ne k^\prime},\vect{\mu},\mat{\Sigma})
          \propto
          \\
            p(\vect{x}_{i^\prime}|\vect{\mu},\mat{\Sigma})
            \times
            p\left(\vect{\delta^\prime}
                 =\ln\frac{\vect{\tilde{\delta}}_{i^\prime}}
             {\vect{L}_\nu^\sms{mod}(\vect{x}_{i^\prime})}\right),
          \end{multlined}
          \label{eq:asisdist}
        \end{equation}
        where the second term of the right-hand side is simply the distribution
        of \refeq{eq:priorcal1}, replacing each $\delta^\prime_j$ by
        $\ln(\tilde{\delta}_{i^\prime,j}
        /L_\nu^\sms{mod}(\vect{x}_{i^\prime},\nu_j))$.
        In practice, we select a different $i^\prime$ at each MCMC cycle, to
        improve statistical mixing.
        Notice that the likelihood does not appear, as 
        we sample in a direction where it is constant.
        That is the key of the success of ASIS.
      \item We then compute the values of the parameters of the remaining 
        sources, $x_{i\ne i^\prime,k^\prime}$.
        This is achieved by inverting \refeq{eq:asis} for an arbitrarily 
        chosen frequency $\nu_{j^\prime}$ (we change $j^\prime$ at each 
        MCMC cycle) and eliminating $\delta_j$:
        \begin{equation}
          x_{i\ne i^\prime,k^\prime} 
          = L_{\nu,i,j^\prime,k^\prime}^\sms{inv}
            \left(\frac{\tilde{\delta}_{i,j^\prime}}
                       {\tilde{\delta}_{i^\prime,j^\prime}}\times
                  L_\nu^\sms{mod}(\vect{x}_{i^\prime},\nu_{j^\prime})\right),
        \end{equation}
        where $L_{\nu,i,j,k}^\sms{inv}$ is the model inverse function, 
        \ie\ it is the value of parameter $x_k$ corresponding to
        the monochromatic luminosity in the argument, at frequency $\nu_j$,
        fixing the other parameters, $x_{k\ne k^\prime}$.
        Its numerical implementation is discussed in 
        \refapp{app:tempinv}.
      \item We update the calibration offsets, for each frequency, 
        $\nu_j$, by solving \refeq{eq:asis}:
        \begin{equation}
          \delta_j = \frac{\tilde{\delta}_{i^\prime,j}}
                          {L_\nu^\sms{mod}(\vect{x}_{i^\prime},\nu_j)}-1.
        \end{equation}
    \end{enumerate}
  \item The prior on $\vect{\mu}$ being uniform, we draw the $\mu_k$ elements,
    one by one, from:
    \begin{equation}
      p(\mu_k|\mu_{k^\prime\ne k},\vect{x}_1,\ldots,\vect{x}_n,\mat{\Sigma})
      \propto
      \prod_{i=1}^np(\vect{x}_i|\vect{\mu},\mat{\Sigma}).
      \label{eq:postmu}
    \end{equation}
  \item The standard deviations, $S_{k,k}$, are drawn, one by one, from:
    \begin{equation}
      p(S_{k,k}|S_{k^\prime\ne k,k^\prime\ne k},
        \vect{x}_1,\ldots,\vect{x}_n,\mat{R})
      \propto
      p(\mat{S})\times\prod_{i=1}^n p(\vect{x}_i|\vect{\mu},\mat{\Sigma}).
      \label{eq:postS}
    \end{equation}
  \item Finally, the elements of the correlation matrix, $\mat{R}$, are drawn,   
    one by one, from:
    \begin{equation}
      p(R_{k_1,k_2}|R_{k_1^\prime\ne k_1,k_2^\prime\ne k_2},\vect{x}_i,\mat{S})
      \propto
      p(\mat{R})\times\prod_{i=1}^n p(\vect{x}_i|\vect{\mu},\mat{\Sigma}).
      \label{eq:postR}
    \end{equation}
\end{enumerate}

  \subsubsection{Assessing optimal sampling}
  
We derive several quantities in order to assess the convergence of the MCMC.
The {\it autocorrelation function} (ACF) of a parameter, $x$, is, for a 
lag\footnote{The lag is the difference between two {\it times} (two steps) in 
the MCMC.}, $k$:
\begin{equation}
 \rho_x(k) = \frac{\displaystyle {\rm cov}_t(x^{(t)},x^{(t+k)})}
                {\displaystyle\sqrt{{\rm var}_t(x^{(t)})
                                    {\rm var}_t(x^{(t+k)})}},
  \label{eq:ACF}                                    
\end{equation}
where $x^{(t)}$ is the value of parameter $x$ at the step $t$ of the MCMC.
The notations ${\rm var}_t$ and ${\rm cov}_t$ indicate that the variance and covariance are taken over the index $t$.
We numerically evaluate it, using the FFT of the MCMC.
From the ACF, we can estimate the {\it integrated autocorrelation time} 
\citep[{\eg}][]{foreman-mackey13}:
\begin{equation}
  \tau_\sms{int}(x) = 1 + 2\sum_{k=1}^{N_\sms{MCMC}}\rho_x(k),
  \label{eq:tint}
\end{equation}
following the method of \citet{sokal96}.
It quantifies the length after which the drawings are truly independent.
The {\it effective sample size}, defined as:
\begin{equation}
  N_\sms{eff}(x) = \frac{N_\sms{MCMC}}{\tau_\sms{int}(x)},
  \label{eq:Neff}
\end{equation}
provides an estimate of the number of effective independent draws.
We let our MCMC run until $N_\sms{eff}>30$, at least, for each parameter.

  \subsubsection{Derived quantities}
  \label{sec:derived}
  
Once the MCMC has been computed, we estimate a few quantities 
characterizing the parameter distribution.
For each parameter and hyperparameter, $y$, we derive its mean and standard 
deviation, $\langle y\rangle$ and $\sigma_y$, marginalizing over the other 
parameters.
This is technically achieved by taking the average and standard deviation of 
the MCMC of the parameter.
Similarly, we can estimate any other quantity (correlation coefficients, degree 
of confidence, \etc), by computing this quantity at each MCMC step, and 
estimating the average and standard deviation of the resulting distribution of 
values.

  \section{Effects of the Noise, Sample Size and SED shape}
  \label{sec:refgrid}
  
In this section and in \refsecs{sec:vargrid} and \ref{sec:othergrid}, we dissect several tests aimed at assessing the performance of our HB code.
These tests are all performed on simulated data, so that one can compare the results of the model to the true values of the parameters.
In the present article, we exclusively simulate data with the same assumptions as in the HB model.
These assumptions are the followings.
\begin{description}
  \item[The noise:] 
    both the distribution (normal, asymmetric, \etc; 
    see \refsec{sec:noise}) and the amplitude of the noise have an effect 
    on the results.
    In this paper, we exclusively use normal, uncorrelated noise and assume 
    that we perfectly know its amplitude.
  \item[The physical model:]
    we simulate SEDs with the same combination of components 
    (see \refsec{sec:SEDcomp}) as in the HB method.
    In \refsecs{sec:refgrid} and \ref{sec:vargrid}, we use only the combination 
    of the \powU\ (\refsec{sec:powerU}) and \Star\ (\refsec{sec:starBB}) 
    components.
    This combination is indeed one of the most relevant, when modelling the 
    near-IR-to-submm emission of interstellar regions.
    We use the grain properties of the AC model of \citet{galliano11}.
    We demonstrate the model performances with several other components in 
    \refsec{sec:othergrid}.
  \item[The prior distribution:]
    the assumed shape of the prior \refeqp{eq:prior} also has an impact on
    low signal-to-noise ratio sources.
    In what follows, we draw parameter values from the same distribution as in
    \refeq{eq:prior}.
\end{description}

  \subsection{The Simulation grid}
  \label{sec:refgridesc}
  
We start by studying the combined effects of signal-to-noise ratio, sample size and SED shape on the performance of the HB method.
To do so, we simulate a grid of SED samples, varying these three quantities.

\begin{table} 
  \centering
  \begin{tabular}{lccc}
    \hline\hline
      & Cold & Warm & Hot \\
    \hline
      $\mu[\ln M]$ & \multicolumn{3}{c}{0} \\
      $\mu[\ln U_-]$ & $\ln 0.3$ & $\ln 10$ & $\ln 100$ \\
      $\mu[\ln\Delta U]$ & $\ln 10$ & $\ln 300$ & $\ln 10^4$ \\
      $\mu[\alpha]$ & \multicolumn{3}{c}{1.8} \\
      $\mu[q^\sms{PAH}]$ & 0.06 & 0.04 & 0.02 \\
      $\mu[f^+]$ & \multicolumn{3}{c}{0.5} \\
      $\mu[\ln L^\star]$ 
        & $\ln 10^3$ & $\ln 10^4$ & $\ln 10^5$ \\
    \hline
      $S[\ln M]$ & \multicolumn{3}{c}{0.5} \\
      $S[\ln U_-]$ & \multicolumn{3}{c}{0.4} \\
      $S[\ln\Delta U]$ & \multicolumn{3}{c}{0.5} \\
      $S[\alpha]$ & \multicolumn{3}{c}{0.3} \\
      $S[q^\sms{PAH}]$ & \multicolumn{3}{c}{0.01} \\
      $S[f^+]$ & \multicolumn{3}{c}{0.1} \\
      $S[\ln L^\star]$ & \multicolumn{3}{c}{0.1} \\
    \hline
      $R[\mbox{all}]$ & \multicolumn{3}{c}{0} \\
    \hline
  \end{tabular}
  \caption{{\sl True hyperparameters for the three SED shapes of 
            \refsec{sec:refgridesc}.}
           These values are the elements of $\vect{\mu}$, $\mat{S}$ and 
           $\mat{R}$ of the distribution in \refeq{eq:prior}.
           They correspond to the parameters described in \refsecs{sec:powerU}
           and \ref{sec:starBB}.
           $M$ is in $\rm M_\odot$ and $L^\star$ in $\rm L_\odot$.}
  \label{tab:prioref}
\end{table} 
\begin{figure} 
  \includegraphics[width=\linewidth]{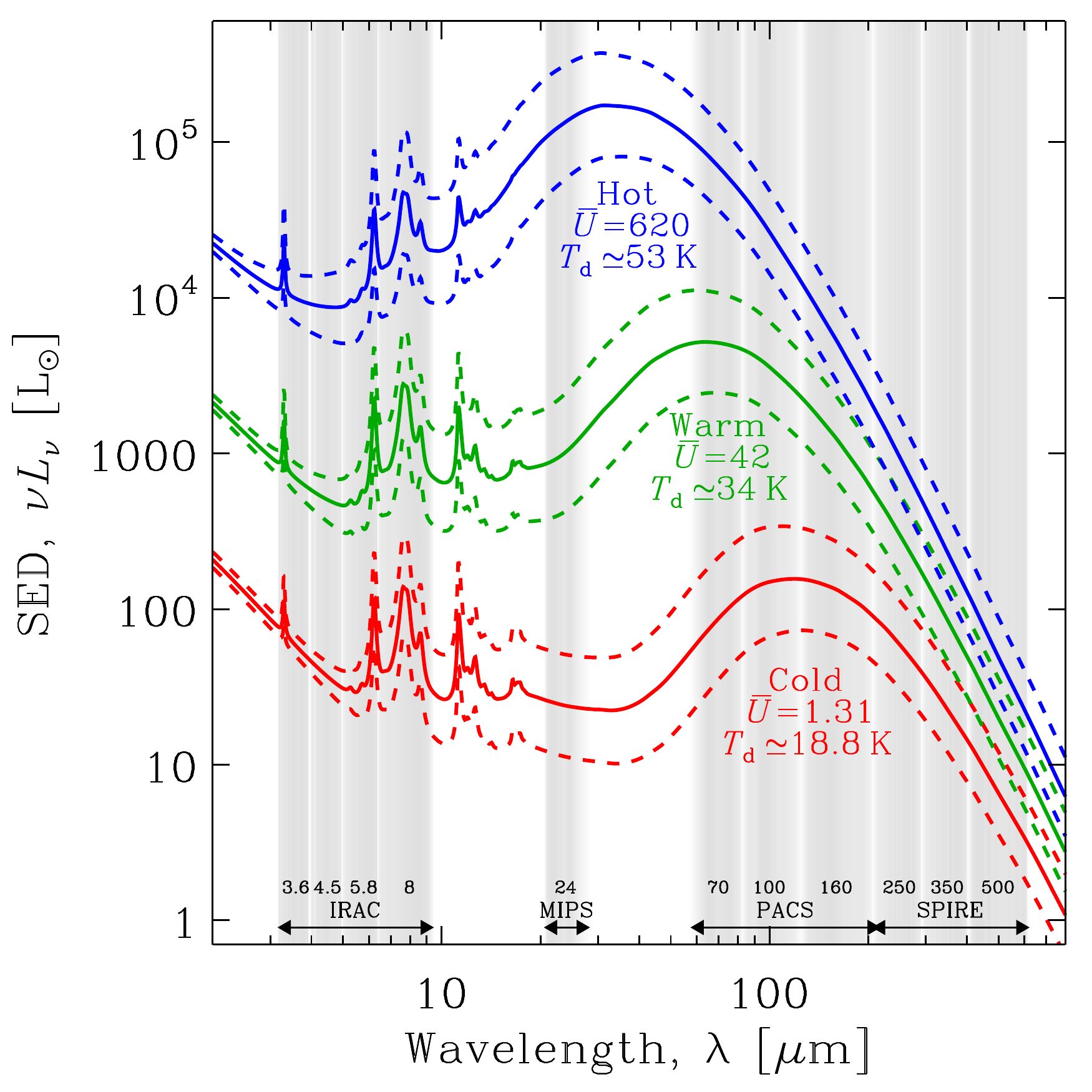}
  \caption{{\sl Classes of simulated SEDs.}
           The solid lines present the mean of the simulated SEDs of 
           \reftab{tab:prioref}.
           The dashed lines show the $\pm1\sigma$ ranges.
           For each class, we quote the average starlight intensity, 
           \Uav, of the mean, and the approximate equilibrium
           grain temperature, $T_\sms{d}$.
           The latter is derived from
           $T_\sms{d}\simeq\Uavm^{1/6}\times18\;\rm K$.
           We have displayed the transmission of each filter used (grey 
           densities) and labeled its nominal wavelength, in microns.}
  \label{fig:SEDshapes}
\end{figure}
We simulate three different classes of SED shapes, labeled {\it cold}, 
{\it warm} and {\it hot}.
The physical parameters of each simulation are drawn from the distribution of \refeq{eq:prior} with the hyperparameters listed in \reftab{tab:prioref}.
These SEDs are shown in \reffig{fig:SEDshapes}.
We assume that these SEDs are observed through a typical collection of photometric filters: the four \spitz/IRAC bands (3.6, 4.5, 5.8 and 8~\mic), the 
\spitz/MIPS band at 24~\mic, the three \hersc/PACS bands (70, 100 and 160~\mic) and the three \hersc/SPIRE bands (250, 350 and 500~\mic).
For each SED shape, we generate three samples of $n=10$, 100 and 1000 sources.
  
We add random deviations to the simulated SED samples.
These deviations are divided in the two following categories.
\begin{description}
  \item[Calibration offsets,] $\delta_j$, are drawn from the distribution of 
    \refeq{eq:priorcal1}, keeping the same values for each 
    source in the sample.
    However, we draw different offsets for each sample, in order to 
    average out biases that could result from a particular realization.
  \item[Noise deviations,] $\epsilon_{i,j}$, are drawn from a normal 
    distribution \refeqp{eq:LH1}.
    These variables are independent.
    We assume that the absolute noise level is the same in each source of a 
    given sample.
    This is to mimic observations of spatially resolved regions, where the 
    RMS per waveband is roughly constant.
    We set the noise uncertainty at a frequency, $\nu_j$, proportional to 
    the simulated median of the monochromatic luminosities of all the sources, 
    $s_i$:
    \begin{equation}
      \sigma_{\nu,j} = \frac{\displaystyle
        {\rm med}\left(L_\nu^\sms{mod}(\vect{x}_i,\nu_j)\right)}
        {f_\sms{S/N}},
      \label{eq:SovN}
    \end{equation}
    For each SED shape and sample size, we simulate three realizations of the 
    noise with median signal-to-noise ratios, $f_\sms{S/N}=0.3$, 3 and 30.
    Samples with $f_\sms{S/N}=0.3$ are dominated by the noise, while
    samples with $f_\sms{S/N}=30$ are dominated by the calibration errors.
\end{description}
In total, we have $3^3=27$ simulations.

  \subsection{Dissection of a model's results}
  \label{sec:example}
  
To start, we analyze in details the central run in the simulation grid (warm SED, with $n=100$ and $f_\sms{S/N}=3$), in order to demonstrate how the model works, on a concrete example.

\subsubsection{The MCMC}
\label{sec:MCMC}

\begin{figure*} 
  \includegraphics[width=\textwidth]{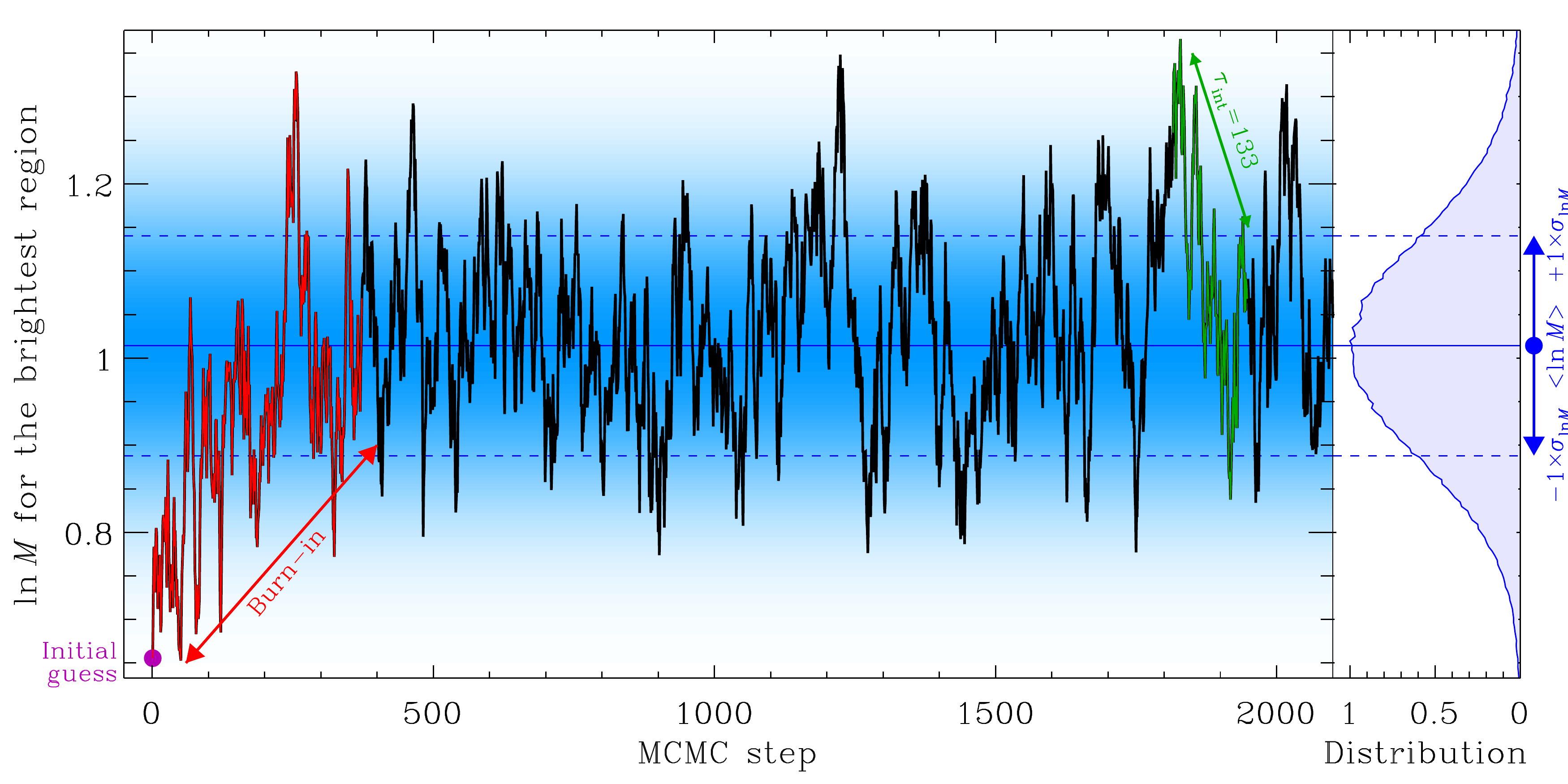}
  \caption{{\sl Markov Chain Monte Carlo of a parameter.}
           The parameter is $\ln M$ for the brightest source in the central 
           simulation of \refsec{sec:refgridesc}.
           The right panel shows the corresponding distribution of the 
           parameter, and its derived average and standard-deviation.
           We highlight the burn-in phase (in red) and a typical 
           autocorrelation time (in green).
           For clarity, we show only the first 2000 steps.}
  \label{fig:dissect_mcmc}
  \includegraphics[width=\textwidth]{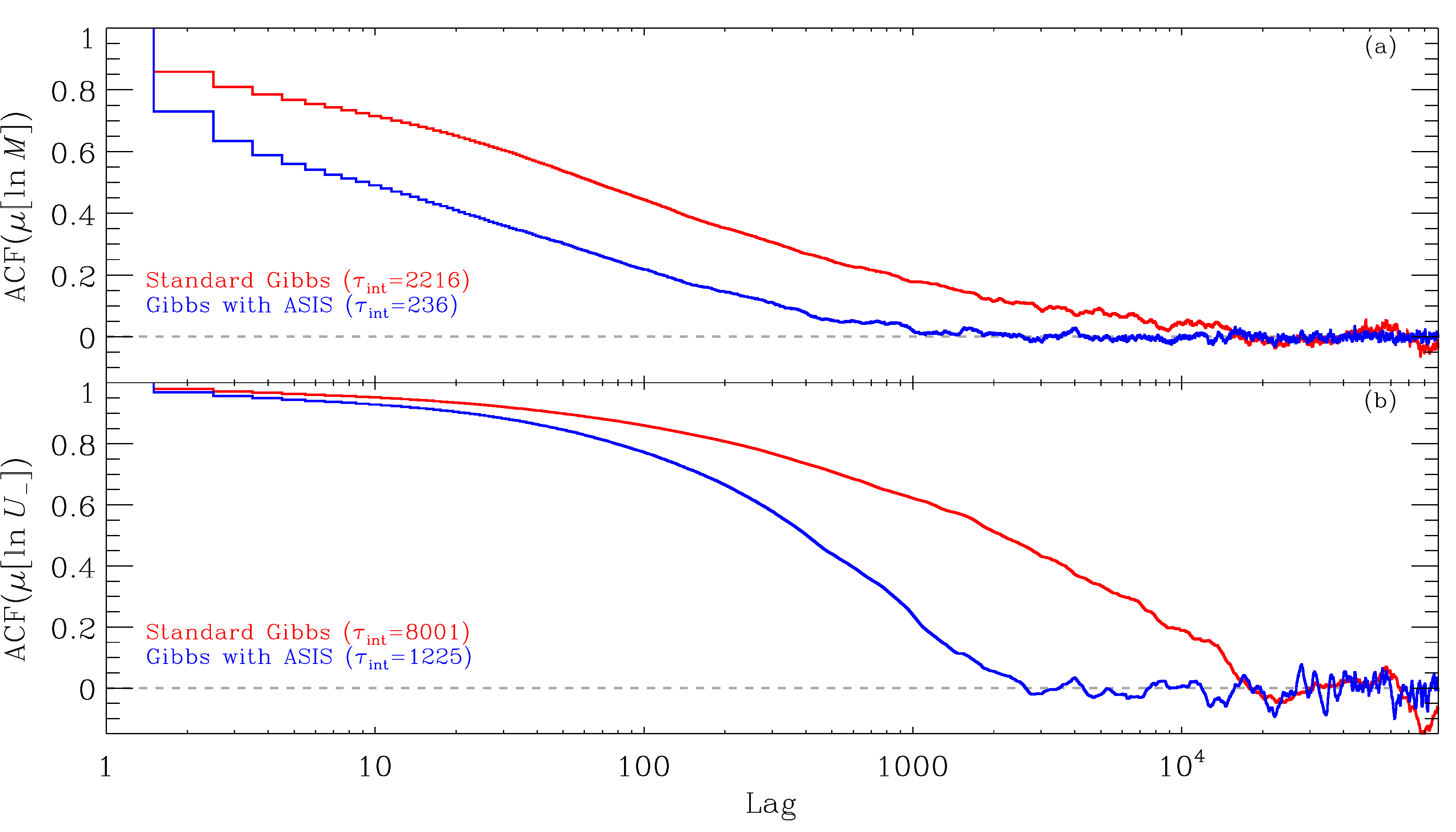}
  \caption{{\sl Autcorrelation functions \refeqp{eq:ACF} for two 
           hyperparameters.}
           The two hyperparameters are the averages of $\ln M$ and $\ln U_-$
           (\refsec{sec:powerU}), for the central simulation of 
           \refsec{sec:refgridesc}.
           The blue curves correspond to the ACFs obtained with our full 
           method, including ASIS (\refsec{sec:sampler}).
           The red curves correspond to the same ACFs, but switching off ASIS.}
  \label{fig:dissect_acf}
\end{figure*}
In \reffig{fig:dissect_mcmc}, we show the first 2000 steps of the MCMC of the
parameter $\ln M$, for the brigthest source in the sample.
The distribution of values sampled by the chain, shown in the right panel of 
\reffig{fig:dissect_mcmc}, is the marginal posterior distribution of the 
parameter.
Notice that the fluctuations of the chain are not independent.
There are structures, like the one highlighted in green, having a length of the order of the integrated autocorrelation time ($\tau_\sms{int}$; \refeqnp{eq:tint}).
We have chosen an example with a particularly short autocorrelation time, for clarity.
However, $\tau_\sms{int}$ can reach up to $\simeq10^5$ for some parameters (see \refsec{sec:anaref}).
We have also highlighted the {\it burn-in} phase, in red.
This phase corresponds to the time spent by the MCMC to walk from the initial condition to the region of relevant likelihood.
It is advised to exclude this part of the chain from the analysis.
Using the least-squares best fit parameters as initial conditions (\refsec{sec:sampler}), we have not witnessed any particularly long burn-in. 
It is usually of the order of $\tau_\sms{int}$.

The autocorrelation of the chain can be quantified.
\reffig{fig:dissect_acf} shows the ACF \refeqp{eq:ACF} for two hyperparameters of the central simulation.
These ACFs all have the same qualitative behaviour.
They start around 1, at small lags.
They then drop towards 0 in a time that is comparable to $\tau_\sms{int}$.
Finally, for large lags, they oscillate around 0 with a small amplitude.
In \reffig{fig:dissect_acf}, we compare the ACFs and integrated autocorrelation times obtained with:\textlist{\thetextlist~our full method including ASIS (\refsec{sec:sampler}), in blue;
and \thetextlist~standard Gibbs sampling, switching off ASIS, in red.}We
can see that, using ASIS, $\tau_\sms{int}$ is reduced by a factor of $\simeq 6-9$, in this particular case.
It illustrates that the implementation of ASIS can help reach convergence with a significantly shorter chain.

\subsubsection{The SEDs}

\begin{figure} 
  \includegraphics[width=\linewidth]{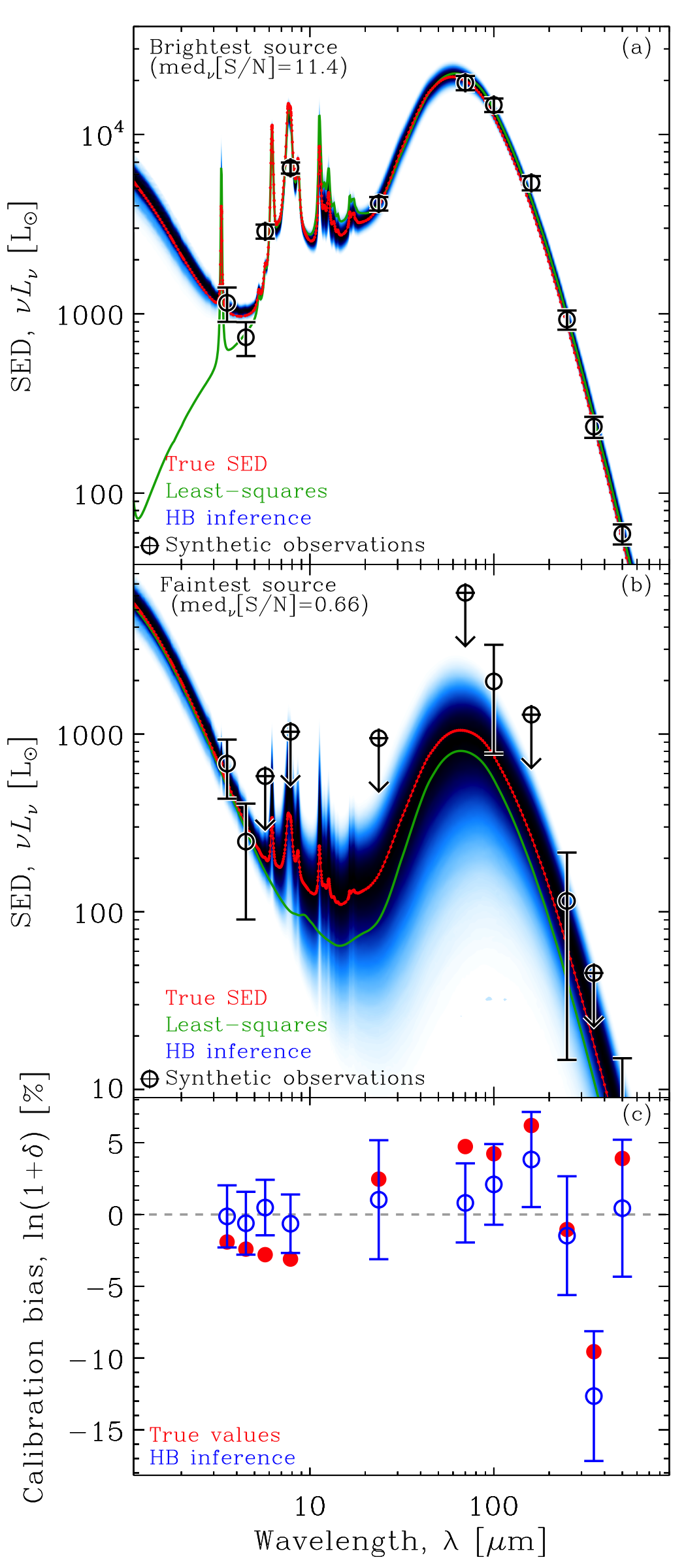}
  \caption{{\sl SED fits of two sources.}
           The two sources are the brightest (panel a) and the faintest 
           (panel b) of the central simulation of \refsec{sec:refgridesc}.
           The circle with error bars are the synthetic observations.
           Upper limits are quoted at $3\sigma$.
           The blue-to-black density shows the HB probability distribution of 
           the SED.
           The red line shows the true SED (without noise).
           The green lines are the least-squares fit, for comparison.
           For each SED, we quote the median (over frequencies) of the 
           signal-to-noise ratio (${\rm med}_\nu[\rm S/N]$).
           Panel (c) shows the calibration offsets $\vect{\delta}$ (common to 
           both SEDs).
           The red dots are the true offsets, and the blue circle with error 
           bars are the inferred values.}
  \label{fig:dissect_sed}
\end{figure}
\reffig{fig:dissect_sed} shows examples of SED fits for the central simulation.
Panel (a) shows the SED of the brightest pixel and panel (b), the faintest.
The SED probability density is simply the distribution of SED models, computed with the values of the parameters, at each step in the MCMC.
Obviously, the SED density is more dispersed for the low signal-to-noise ratio source.
We see that, in both cases, the HB SED density is in better agreement with the true SED (in red) than the least-squares fit (in green).
We also notice that the PAH fraction of the $\chi^2$, in panel (b), has been clearly underestimated, while the HB model is close to its true value.
Panel (c) shows the calibration offsets ($\vect{\delta}$).
The simulated offsets are shown in red.
We emphasize that they are common to all the sources in the sample.
We can see that the inferred values of $\vect{\delta}$ (in blue) are consistent with the true values.

\subsubsection{The Derived parameters}
\label{sec:dissect_par}

\begin{figure} 
  \includegraphics[width=\linewidth]{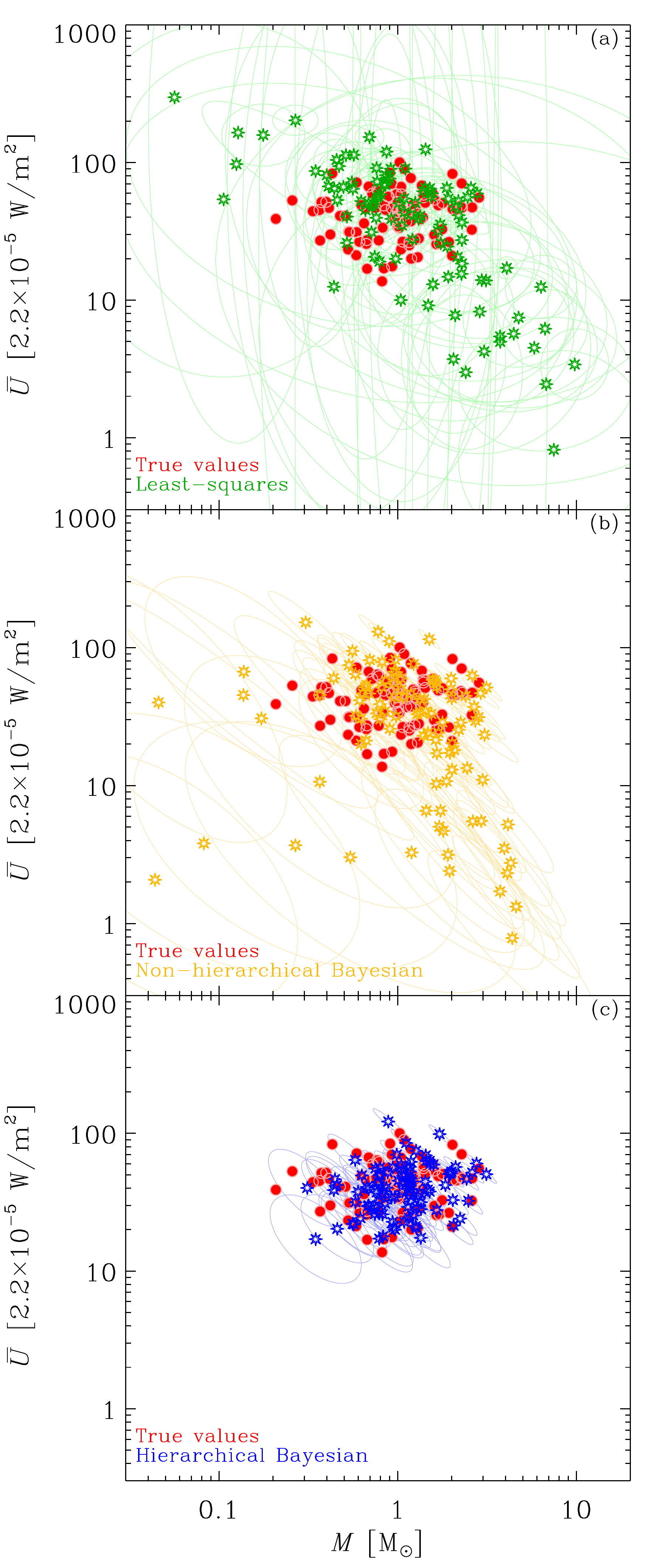}
  \caption{{\sl Efficiency of different methods at recovering parameters.}
           In each panel, we show the true (in red) and the inferred
           values (stars with error ellipses) of the mass, $M$, and average 
           starlight intensity, \Uav, for the central simulation 
           of \refsec{sec:refgridesc}.
           Panel (a) shows the least-squares results (\refapp{app:chi2}; 
           in green).
           Panel (b) shows the non-hierarchical Bayesian inference (in orange).
           Panel (c) shows the HB results (in blue).}
  \label{fig:dissect_correl}
\end{figure}
\reffig{fig:dissect_correl} compares the performances of different methods, applied to the central simulation.
It shows the derived correlation between two of the main parameters: the dust mass, $M$, and the average starlight intensity, \Uav\ \refeqp{eq:avU}.
The true values are shown in red, in each panel.
Notice that there is no intrinsic correlation between the two parameters.
However, the least-squares fit, in panel (a), shows a clear negative correlation, with a significant scatter.
This is the equivalent of the noise induced $\beta-T$ negative correlation, for modified black bodies, demonstrated by \citet{shetty09}.
In panel (b), we display the non-hierarchical Bayesian results.
The stars and error ellipses show the means and covariances of the posterior distribution \refeqp{eq:bayes}, including the calibration offsets in the likelihood and its prior distribution.
The inferred values show a significant scatter and a negative correlation between the two parameters.
Notice that the uncertainties are very different between panels (a) and (b).
The Bayesian error ellipses are more rigorous, as they are directly derived from the actual shape of the likelihood, and not from a parabolic approximation, in the $\chi^2$ case (\refapp{app:chi2}).
Panel (b) demonstrates that, although the non-hierarchical Bayesian method provides an accurate description of the likelihoods of each source, it does not help to significantly reduce the scatter and eliminate false correlations.
Finally, panel (c) shows the HB results (in blue).
The scatter is considerably reduced, compared to the previous cases.
In addition, there is no more false correlation between the parameters.
The sizes of the error ellipses have also been greatly reduced, especially for the lowest signal-to-noise ratio sources. 
However, notice that these uncertainties are still consistent with the true values.

\begin{figure} 
  \includegraphics[width=\linewidth]{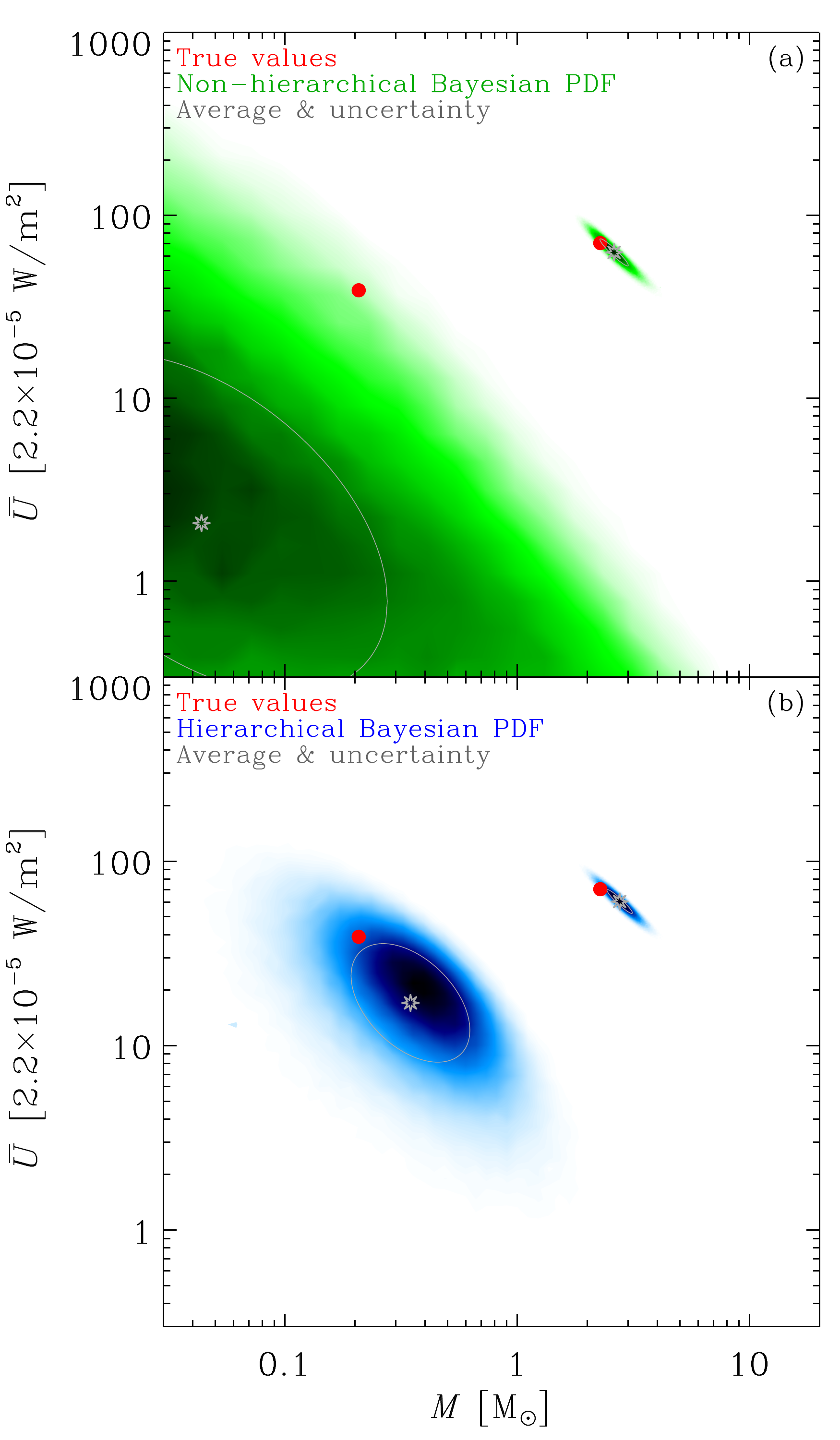}
  \caption{{\sl Posterior distributions of two sources.}
           These are the brightest and faintest sources of the central 
           simulation of \refsec{sec:refgridesc}.
           We show the same correlation between $M$ and \Uav\ as 
           in \reffig{fig:dissect_correl}, and keep the same dynamic range.
           Panel (a) shows the non-hierarchical Bayesian values 
           (green density), corresponding
           to panel (b) of \reffig{fig:dissect_correl}.
           Panel (b) shows the HB values (blue density), corresponding to 
           panel (c) of \reffig{fig:dissect_correl}.
           In each panel, the grey stars are the averages of the parameters
           over the posterior, and the grey ellipses are their covariances.
           The true values are shown in red, in both panels.}
  \label{fig:dissect_likelihood}
\end{figure}
Panel (c) of \reffig{fig:dissect_correl} demonstrates, on a concrete case, the effect that panel (c) of \reffig{fig:demoHB} was trying to illustrate: the reduction of the dispersion of low signal-to-noise ratio sources by the prior.
Panel (a) of \reffig{fig:dissect_likelihood} shows the non-hierarchical Bayesian posterior distribution (green density), for the brightest and faintest sources of the simulation, in the same parameter space as 
\reffig{fig:dissect_correl}.
The PDF of the faintest source clearly extends out of the range of the figure, as its SED has a median signal-to-noise ratio of only 0.66 (panel b of \reffig{fig:dissect_sed}).
Panel (b) of \reffig{fig:dissect_likelihood} shows the HB posteriors, for the same sources.
Comparing the two panels, we see that the PDF of the brigthest source is almost not modified by the prior. 
On the contrary, the PDF of the faintest source is brought back towards its true value.
Comparing this PDF to its mean value and error ellipse, we see it is noticeably skewed.

  \subsection{The Role of the prior distribution}
  \label{sec:prioref}
  
\begin{figure} 
  \includegraphics[width=\linewidth]{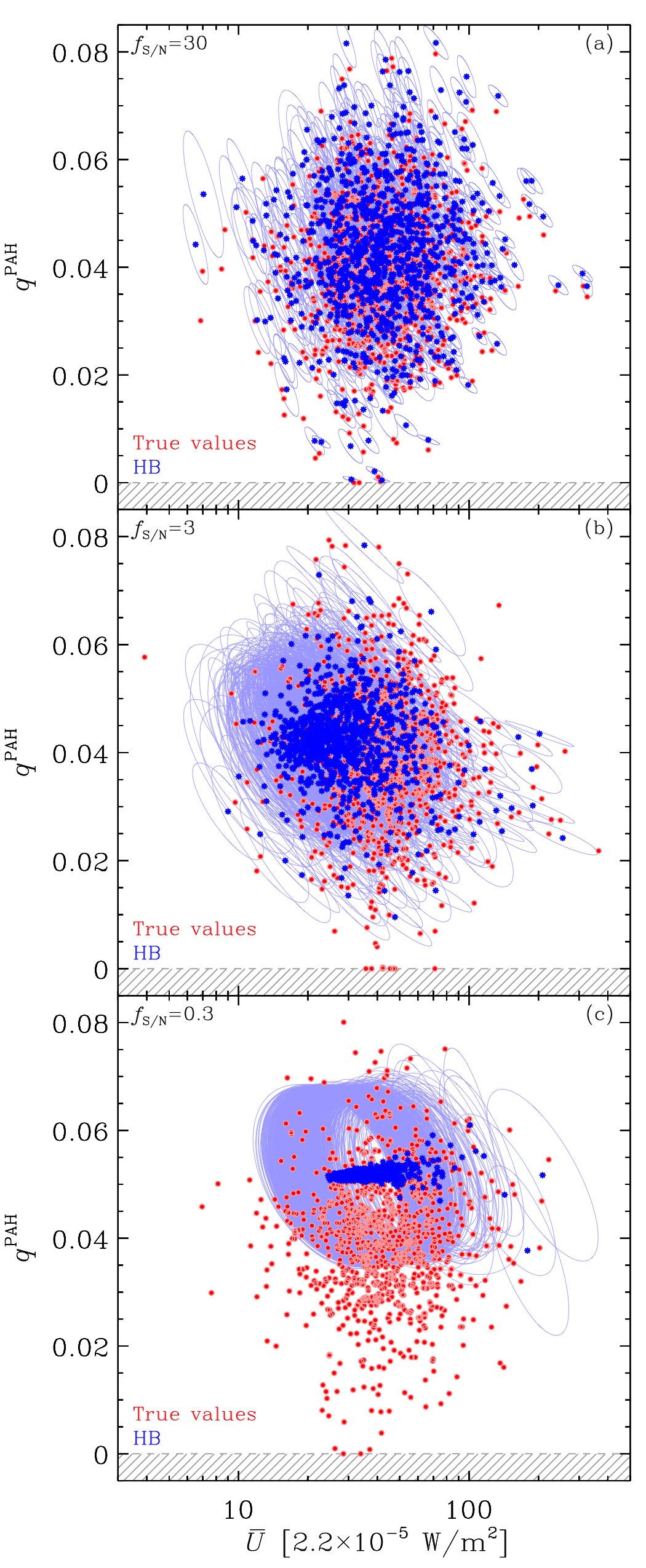}
  \caption{{\sl Demonstration of the role of the prior.}
           The three panels show the results for the simulations of
           \refsec{sec:refgridesc}, with a warm SED, $n=1000$ sources, and
           with median signal-to-noise ratios $f_\sms{S/N}=30$ (panel a),
           $f_\sms{S/N}=3$ (panel b), and $f_\sms{S/N}=0.3$ (panel c).
           The red dots show the true values of the average starlight 
           intensity, \Uav, and of the PAH mass fraction, $q^\sms{PAH}$.
           The blue stars and their error ellispes are the HB inference.} 
  \label{fig:refgrid_correl_SovN}
\end{figure}
To illustrate the role of the prior distribution, 
\reffig{fig:refgrid_correl_SovN} 
shows the results of the HB code on three of the simulations of 
\refsec{sec:refgridesc}: a warm SED, with $n=1000$ sources.
The three panels show the relation between the average starlight intensity, 
\Uav, and the PAH mass fraction, $q^\sms{PAH}$.
In panel (a), the median signal-to-noise ratio \refeqp{eq:SovN} is 
high ($f_\sms{S/N}=30$).
As a consequence, the parameters of each source are well constrained.
The uncertainty ellipses have a characteristic size much smaller than the width 
of the distribution of parameters.
The typical uncertainty in $q^\sms{PAH}$ is indeed
$\sigma_{q^\sms{PAH}}\simeq2\E{-3}$, while the standard deviation of the 
distribution along $q^\sms{PAH}$ is $S[q^\sms{PAH}]\simeq0.01$.
The prior distribution is thus rather flat compared to the likelihood of an 
individual source.
Therefore, the multiplication of the likelihood by the prior \refeqp{eq:HB}
does not have a significant effect.
As a result, the posterior distribution is close to the non-hierarchical case 
(\eg\ \reffig{fig:dissect_likelihood}).

In contrast, when the median signal-to-noise ratio decreases (panels b and c of 
\reffig{fig:refgrid_correl_SovN}), the width of the parameter distribution is 
unchanged ($S[q^\sms{PAH}]\simeq0.01$ for each panel), but the uncertainty on 
the parameters of each source increases.
In panel (b), with $f_\sms{S/N}=3$, the two quantities are roughly equal.
The multiplication of the likelihood by the prior thus has an effect on the 
posterior distributions.
In particular, the sources at low \Uav, have a lower signal-to-noise ratio.
Their mean values (blue stars) tend to delineate the shape of the prior 
distribution.
The size of their uncertainty ellipses is also reduced by the prior.
In panel (c), with $f_\sms{S/N}=0.3$, the signal-to-noise ratio is so low that 
the individual likelihoods are much larger than the prior.
As a result, the posterior distribution is very close to the prior 
distribution.
Consequently, all the mean values of $q^\sms{PAH}$ (blue stars) are almost
perfectly equal to the prior's average 
($\langle q^\sms{PAH}_i\rangle\simeq\mu[q^\sms{PAH}]$).
The uncertainty ellipses of the posterior distribution have the width of the
prior distribution ($\sigma_{q^\sms{PAH}_i}\simeq S[q^\sms{PAH}]$).
However, notice that along the horizontal axis, the parameter \Uav, which is 
better constrained, still exhibits an intrinsic distribution of values.
If we were to decrease even more the signal-to-noise ratio, the blue stars would all collapse onto one single point in the panel with coordinates
$(\mu[\Uavm],\mu[q^\sms{PAH}])$.

The HB method is particularly useful in cases like panel (c) of 
\reffig{fig:refgrid_correl_SovN}.
It is in such a case that it provides results significantly better than 
non-hierarchical Bayesian and least-squares methods.
In panel (c), the low signal-to-noise ratio prevents performing any relevant 
analysis of individual sources.
However, the fact that the HB method deals with the whole probability 
distribution of the sample, allows us to recover average properties. 
In other words, any constraint, even with an extremely low signal-to-noise 
ratio, is taken into account in the HB approach.

  \subsection{Systematic analysis of the model's performances}
  \label{sec:anaref}

After scrutinizing select model's results, let us now study the performances of the HB method over the whole simulation grid of \refsec{sec:refgridesc}.
In particular, we need to understand how close the derived parameters are from their true values.
We are also interested in knowing how the HB method improves the results, compared to the $\chi^2$ fit.
For that purpose, we define the following metric, for each parameter and hyperparameter, $y$:
\begin{equation}
  D[y] = \left\{
  \begin{array}{ll}
    \displaystyle
    \frac{\displaystyle \langle y^\sms{HB}\rangle-y^\sms{true}}
         {\displaystyle \sigma_y^\sms{HB}}
     & \mbox{for the HB case,} \\
    \displaystyle
    \frac{\displaystyle y_{\chi^2}-y^\sms{true}}
         {\displaystyle \sigma_y^\sms{HB}}
     & \mbox{for the $\chi^2$ case,}
   \end{array}
   \right.
   \label{eq:D}
\end{equation}
where $\langle y^\sms{HB}\rangle$ and $\sigma_y^\sms{HB}$ are the mean and standard deviation over the MCMC of $y$, $y_{\chi^2}$, the least-squares value, and $y^\sms{true}$, the true value. 
With this definition, we can study the relative deviation of the HB values, from their true values: a value $|D[y]| \le N_\sigma$ means that the HB value is consistent within $N_\sigma$.
In addition, we can directly compare the HB and $\chi^2$ deviations, as they have the same denominator.

\subsubsection{Performances for the hyperparameters}
\label{sec:refhyp}

\begin{figure*} 
  \includegraphics[width=\linewidth]{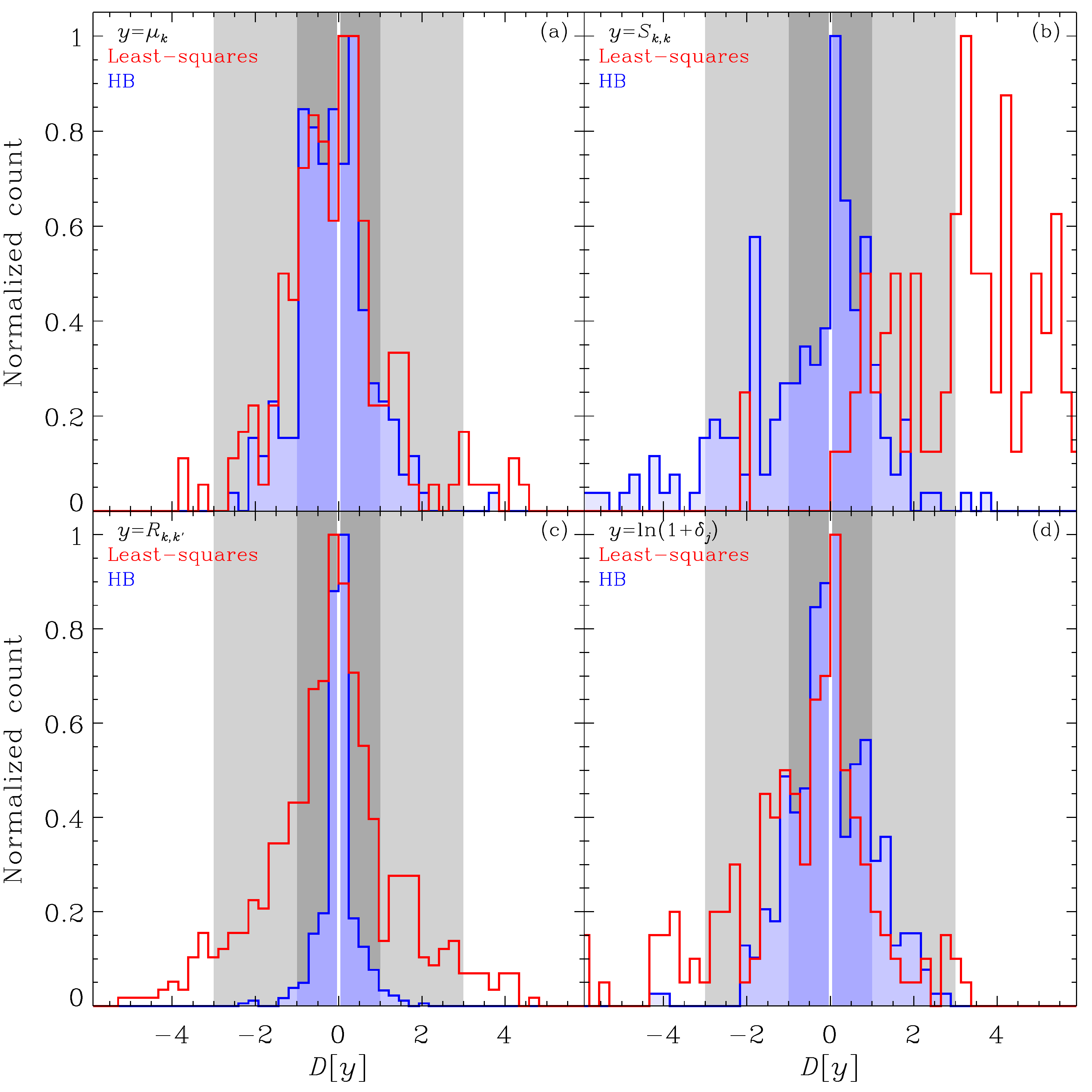}
  \caption{{\sl Recovery performances for the hyperparameters and calibration 
            offsets.}
           The four panels show the distribution of the
           relative deviation $D[y]$ \refeqp{eq:D}, for the hyperparameters
           of all the simulations of \refsec{sec:refgridesc}.
           We have separated the hyperparameters by type: panel (a)
           shows the distribution for all the means, $\mu_k$;
           panel (b), all the standard deviations, $S_{k,k}$;
           panel (c), all the correlation coefficients, 
           $R_{k,k^\prime}$;
           and panel (d), all the calibration offsets, 
           $\ln(1+\delta_j)$.
           For the $\chi^2$ method, we quote the mean relative residuals, 
           as the calobration offsets.
           The blue histograms correspond to the HB values, while the
           red histograms represent the $\chi^2$ results.
           We have highlighted the $1\sigma$ (in dark grey) and 
           the $3\sigma$ (in light grey) ranges.}
  \label{fig:refgrid_hyp_all}
\end{figure*}
\reffig{fig:refgrid_hyp_all} compares the distributions of $D[y]$ for all of the 
hyperparameters of the simulation grid (\refsec{sec:refgridesc}).
Panel~(a) shows the distribution of the recovered means, $\langle\mu_k\rangle$
(\refsec{sec:prior}).
Since there are 27 models, with 7 parameters per model (\reftab{tab:prioref}), 
this distribution contains a total of $7\times27=189$ values.
Similarly, panel~(b) shows the distribution of the standard deviations, 
$\langle S_{k,k}\rangle$ (diagonal elements of $\mat{S}$; \refsec{sec:prior}; 
189 values).
Panel~(c) shows the distribution of the non-diagonal elements of the 
correlation matrix, $\langle R_{k,k^\prime}\rangle$ (\refsec{sec:prior}; 
$7(7-1)/2\times27=567$ values).
Panel~(d) shows the calibration offsets, $\langle\ln(1+\delta_j)\rangle$.
There are 11 photometric filters (\refsec{sec:refgridesc}), thus this 
distribution contains $11\times27=297$ values.

We can first note that the recovered values are tightly distributed around the 
true values.
Most of the values of the hyperparameters are within $3\sigma$ 
of their true values.
The quantity having the widest distribution is the standard deviation 
$S_{k,k}$.
Second, comparing the HB to the $\chi^2$ results, we can see that the HB method 
systematically improves the results.
In particular, the standard deviations, $S_{k,k}$, and correlation 
coefficients, $R_{k,k^\prime}$, are the quantities for which the improvement is  
the most notable.
The $\chi^2$ distribution of $S_{k,k}$ is clearly skewed towards positive 
values.
It is due to the fact that the $\chi^2$ method always leads to more dispersed 
parameter distributions (\eg\ \reffig{fig:dissect_correl}, panel~a).


\begin{table*}
  \centering
  \begin{tabular}{l*{5}{r}}
    \hline\hline
       (1) & (2) & (3) & (4) & (5) & (6) \\
       $y$ & $\langle D^\sms{HB}[y]\rangle$
           & $f(|D^\sms{HB}[y]|\le1)$
           & $f(|D^\sms{HB}[y]|\le3)$
           & ${\rm Max}(|D^\sms{HB}[y]|)$
           & ${\rm med}(|D_{\chi^2}/D^\sms{HB}[y]|)$ \\
    \hline
      $\mu_k$ & $-0.13$
      & $79.9\,\%$
      & $99.5\,\%$
      & $3.8$
      & $1.7$ \\
      $S_{k,k}$ & $-0.52$
      & $56.1\,\%$
      & $91.5\,\%$
      & $5.8$
      & $9.0$ \\
      $R_{k,k^\prime}$ & $0.0069$
      & $95.4\,\%$
      & $100.0\,\%$
      & $2.2$
      & $8.4$ \\
      $\ln(1+\delta_j)$ & $0.027$
      & $69.4\,\%$
      & $99.3\,\%$
      & $4.2$
      & $2.4$ \\
    \hline
      $\ln M_i$ & $0.16$
      & $76.6\,\%$
      & $99.8\,\%$
      & $4.8$
      & $2.3$ \\
      $\ln \Uavm_i$ & $-0.29$
      & $78.6\,\%$
      & $99.9\,\%$
      & $5.0$
      & $2.7$ \\
      $q^\sms{PAH}_i$ & $0.46$
      & $60.4\,\%$
      & $97.8\,\%$
      & $5.5$
      & $2.2$ \\
      $f^+_i$ & $-0.072$
      & $68.2\,\%$
      & $99.2\,\%$
      & $5.0$
      & $3.6$ \\
    \hline
      \multicolumn{6}{c}{Common Probability Distributions} \\
    \hline
      Gaussian
      & 0
      &  $68.3\,\%$
      &  $99.7\,\%$
      & \ldots & \ldots \\
      Student's $t$ ($f=3$)
      & 0
      &  $60.9\,\%$
      &  $94.2\,\%$
      & \ldots & \ldots \\
    \hline
  \end{tabular}
  \caption{{\sl Statistics of the recovery performances.}
            For each parameter and hyperparameter, $y$, we
            quote the following properties of the histograms
            presented in \reffigs{fig:refgrid_hyp_all} and
            \ref{fig:refgrid_par_all}.
            (2) $\langle D^\sms{HB}[y]\rangle$ is the average
            of the distribution.
            (3,4) $f(|D^\sms{HB}[y]|\le N)$ is the fraction of 
            absolute values below $N$.
            (5) ${\rm Max}(|D^\sms{HB}[y]|)$ is the maximum 
            deviation, in number of $\sigma$.
            (6) finally, ${\rm med}(|D_{\chi^2}/D^\sms{HB}[y]|)$
            is the median of the ratio between the $\chi^2$ and
            HB absolute deviations.
            The latter quantifies by how much the parameter
            recovery has been improved.
            The last two lines show the corresponding confidence
            levels of two common probability distributions:
            a gaussian and a Student's $t$ distribution with 
            $f=3$ degrees of freedom.}
  \label{tab:refgrid_stat}
\end{table*}


\reftab{tab:refgrid_stat} quantifies the main properties of these histograms.
The fraction of outliers ($>3\sigma$; column~4) is consistent with a gaussian 
distribution, except for the elements of $\mat{S}$.
The $S_{k,k}$ values are notably more spread out.
They also have the most skewed distribution (largest absolute 
$\langle D^\sms{HB}[y]\rangle$; column~2).
Inspecting these histograms, we notice that most of the outliers correspond to 
the PAH charge fraction, $f^+$ \refeqp{eq:deltaU}.
It is indeed the most degenerate parameter, with the collection of photometric 
filters we have chosen.
This parameter controls mainly the 3.3, 11.2 and 12.7~\mic\ PAH 
features. 
In our simulation grid, the only constraint on these bands is provided by 
the \IRACi\ band, as can be seen on \reffig{fig:SEDshapes}.
However, this photometric band also constrains the stellar continuum, 
$L^\star$, and the calibration offset, $\delta_\sms{\IRACi}$.
In addition, the $3.3\mmic$-to-continuum ratio is often very low (\eg\ 
panel~b of \reffig{fig:dissect_sed}).
The value of $f^+$ is thus very poorly constrained.
All things considered, it is remarkable that the values of this 
parameter are properly recovered, in most of the cases.

Column (6) of \reftab{tab:refgrid_stat} shows that the standard deviations and 
correlation coeffcients are recovered a factor $\simeq8$ times better for the HB 
method than with the $\chi^2$.
Quoting the median instead of the mean of the ratio in column (6) is 
conservative, as there are more outliers with the $\chi^2$ method.

\subsubsection{Performances for the parameters}
\label{sec:refpar}

\begin{figure*} 
  \includegraphics[width=\linewidth]{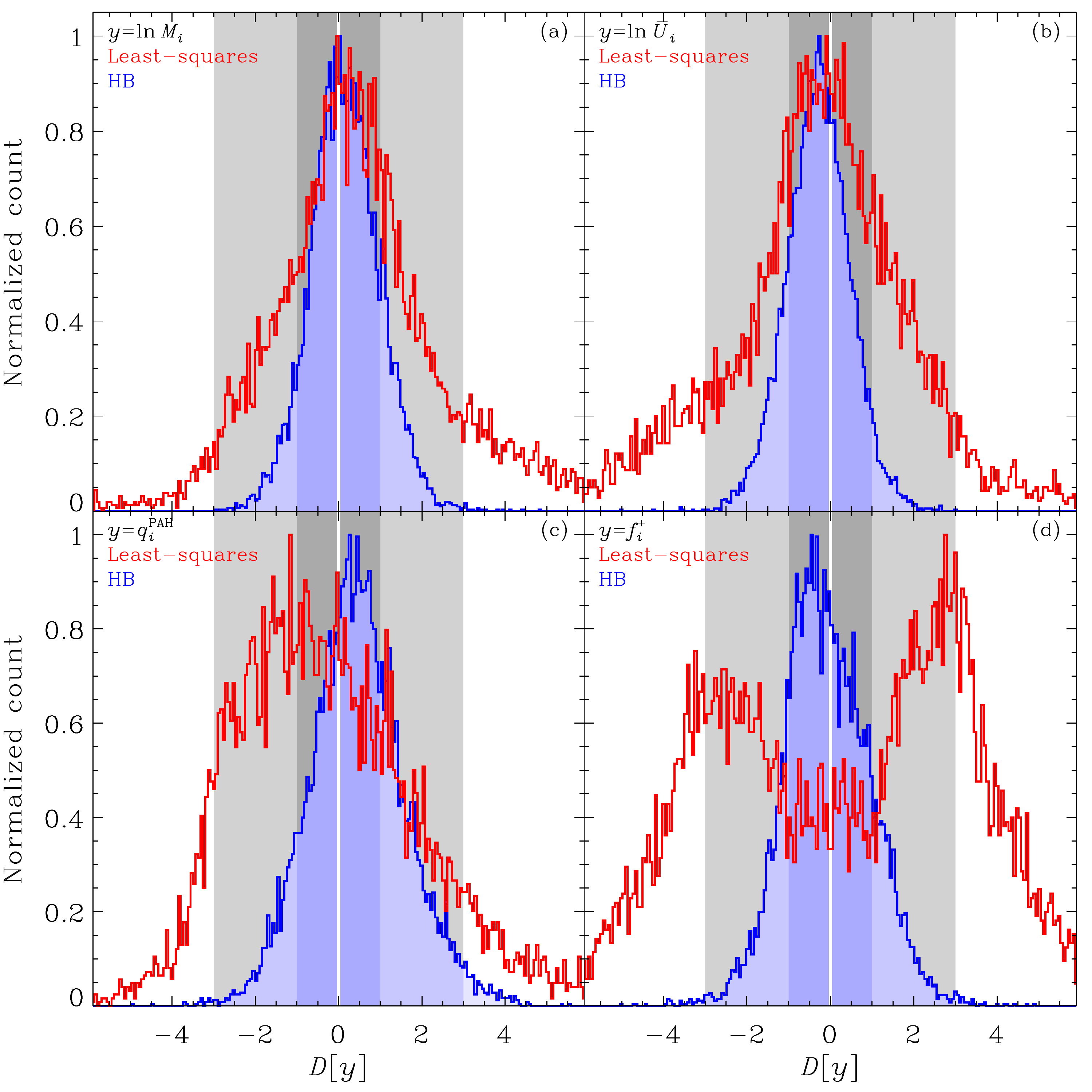}
  \caption{{\sl Recovery performances for the main parameters.}
           The four panels show the distribution of the
           relative deviation $D[y]$ \refeqp{eq:D}, for the main parameters
           of each source in all the simulations of \refsec{sec:refgridesc}.
           We focus here on the most important parameters: panel (a)
           shows the distribution of the dust mass per source, 
           $\ln M_i$;
           panel (b), the average starlight intensity, $\Uavm_i$
           \refeqp{eq:avU};
           panel (c), PAH mass fraction, $q^\sms{PAH}_i$;
           and panel (d), the charge fraction of PAHs, $f^+_i$.
           The blue histograms correspond to the HB values, while the
           red histograms represent the $\chi^2$ results.
           We have highlighted the $1\sigma$ (in dark grey) and 
           the $3\sigma$ (in light grey) ranges.}
  \label{fig:refgrid_par_all}
\end{figure*}  
\reffig{fig:refgrid_par_all} shows the distributions of $D[y]$ for the four 
most physically relevant parameters ($M$, \Uav, $q^\sms{PAH}$ and $f^+$).
Each of these panels contains the values of the parameter for each region in 
the 27 simulations (\refsec{sec:refgridesc}).
Since, there are 9 simulations with 10 regions, 9 with 100 and 9 with 1000, the 
total number of points in each of these distributions is $9990$.
The HB distributions are tightly centered around 0. 
The two best constrained parameters, $M$ and \Uav, have a $3\sigma$ degree of 
confidence (column 4 of \reftab{tab:refgrid_stat}) close to a gaussian.
The two other parameters, $q_\sms{PAH}$ and $f^+$, which are more degenerate, 
are still well constrained.
They are, however, slightly more spread out, with a fraction of outliers between
a Gaussian and a Student's $t$ distribution.
On average, there is no clear bias.
After inspection, we do not find any clear trend of these residuals with signal-to-noise ratio, SED shape or sample size.

The comparison to the $\chi^2$ distribution demonstrates that the parameters 
are recovered $\simeq2-3$ times better with the HB method (column 6 of 
\reftab{tab:refgrid_stat}).
In addition, the outliers are more numerous with the $\chi^2$ method, as the
$\chi^2$ uncertainties are less reliable (\refsec{sec:dissect_par}).

 \subsection{The Integrated Autocorrelation Times}
  \label{sec:tint}

\begin{figure*} 
  \includegraphics[width=\linewidth]{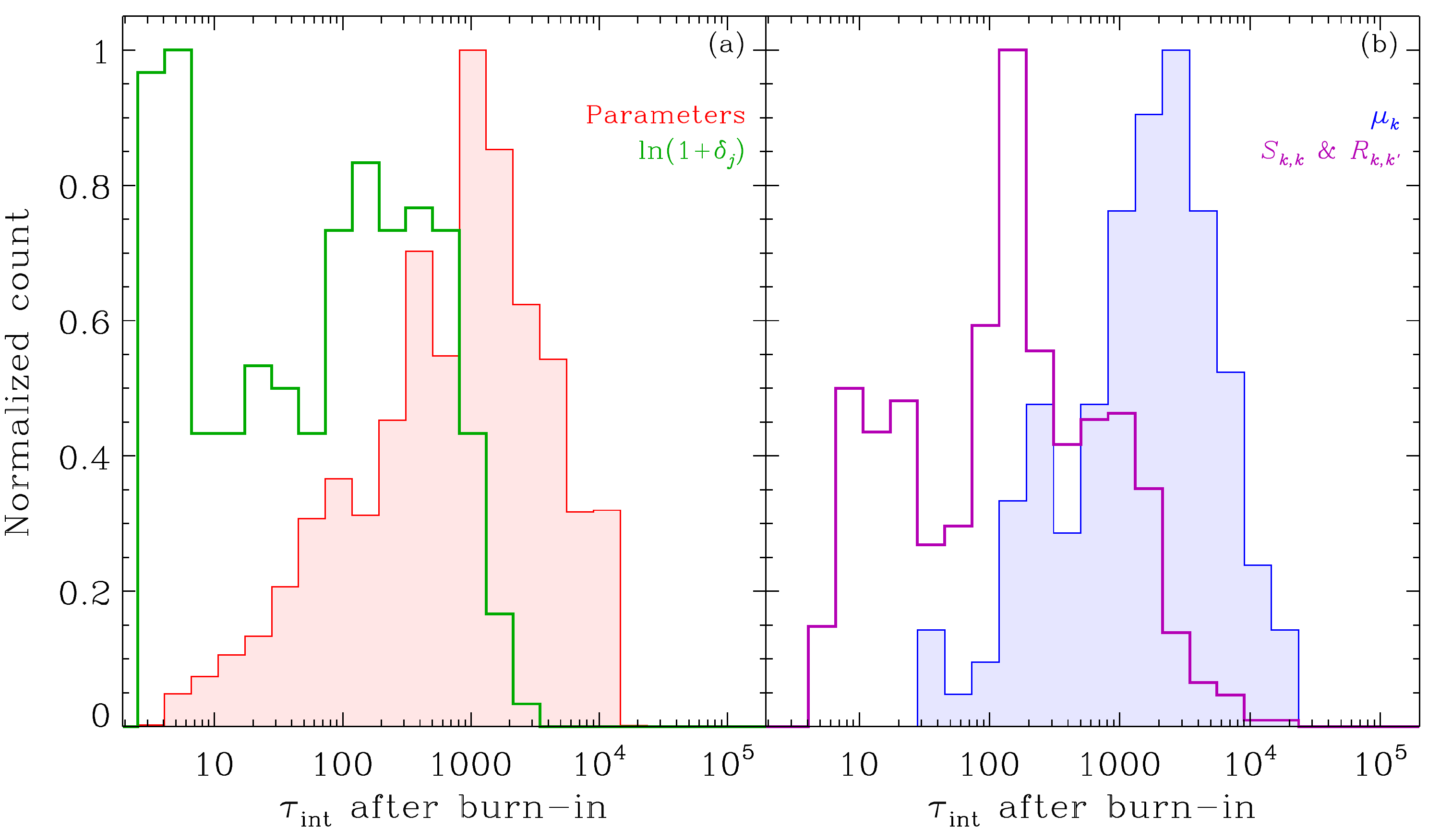}
  \caption{{\sl Integrated autocorrelation times for the simulation grid.}
           This figure shows the distribution of $\tau_\sms{int}$ 
           \refeqp{eq:tint} for: 
           the distribution of individual parameters (panel a, red);
           the calibration offsets, $\ln(1+\delta_j)$ (panel a, green);
           the averages, $\mu_k$ (panel b, blue);
           and the elements of the covariance matrix, $S_{k,k}$ and 
           $R_{k,k^\prime}$ (panel b, purple).
           To build the histogram of the parameters (panel a), we have
           normalized the number of points by the number of 
           sources, $n$, in order to give the same weight to each simulation.}
  \label{fig:refgrid_tint}
\end{figure*}  
The integrated autocorrelation times, $\tau_\sms{int}$ \refeqp{eq:tint}, for 
each parameter and hyperparameters of the simulation grid are represented in 
\reffig{fig:refgrid_tint}.
We can see that the calibration offsets and the elements of the covariance 
matrix converge faster than the averages and the parameters.
The inspection of these distributions does not show any obvious trend of 
$\tau_\sms{int}$ with signal-to-noise ratio, sample size or SED shape.
Overall, the maximum time in the whole grid is $\tau_\sms{int}\simeq 3\E{4}$.
It means that, to make sure the MCMC has converged towards the stationary 
posterior, a chain of length of $N_\sms{MCMC}\simeq10^6$, after the burn-in 
phase, is adequate.

Knowing where the burn-in phase ends is however more difficult. 
One can, for example, run several parallel chains, with different initial conditions.
In our case, since we are applying our code to simulated data, we know the value towards which each parameter should converge.
It is thus possible to estimate if the burn-in phase is passed.
Inspecting our results, we did not find any parameter where the burn-in phase lasts longer than a few $\tau_\sms{int}$.
This reassuring feature seems to be the consequence of properly chosen initial
conditions: the least-squares values (\refsec{sec:sampler}).
Indeed, starting the chain with random initial conditions can lead to burn-in phases longer than $10^6$.
In this paper, we have excluded the first $10^5$ steps of each MCMC to account for burn-in.

  \section{Variations on the Simulation Grid}
  \label{sec:vargrid}

In this section, we demonstrate the performances of our HB code on additional 
effects, that were not covered by the simulation grid of 
\refsec{sec:refgridesc}.

  \subsection{The Presence of intrinsic correlations}
  \label{sec:priorcorr}

The model grid studied in \refsec{sec:refgrid} was simulated with no intrinsic
correlation between parameters ($R[{\rm all}]=0$; \reftab{tab:prioref}).
The purpose was to stay as general as possible.
However, our HB code is designed to efficiently recover correlation coefficients between parameters.

To demonstrate its efficiency, we have simulated two samples, with $n=100$ sources, and a median signal-to-noise ratio, $f_\sms{S/N}=3$.
The distribution of hyperparameters is based on the warm SED shape (\reftab{tab:prioref}), with the following modifications:
$S[\ln U_-]=0.6$, $S[\ln\Delta U]=0.7$, $S[q^\sms{PAH}]=0.2$,
$R[\ln M,\ln U_-]=\varsigma\,0.5$, $R[\ln M,\ln\Delta U]=\varsigma\,0.5$,
$R[\ln M,q^\sms{PAH}]=-\varsigma\,0.5$, $R[\ln M,f^+]=\varsigma\,0.5$,
$R[\ln U_-,\ln\Delta U]=0.5$, $R[\ln U_-,q^\sms{PAH}]=-0.5$,
$R[\ln U_-,f^+]=0.5$, $R[\ln\Delta U,q^\sms{PAH}]=-0.5$,
$R[\ln\Delta U,f^+]=0.5$ and $R[q^\sms{PAH},f^+]=-0.5$.
The parameter $\varsigma$ is set to $\varsigma=-1$, to induce a negative 
correlation between $M$ and \Uav, and to $\varsigma=1$, to induce a positive
correlation.

\begin{figure*} 
  \includegraphics[width=\textwidth]{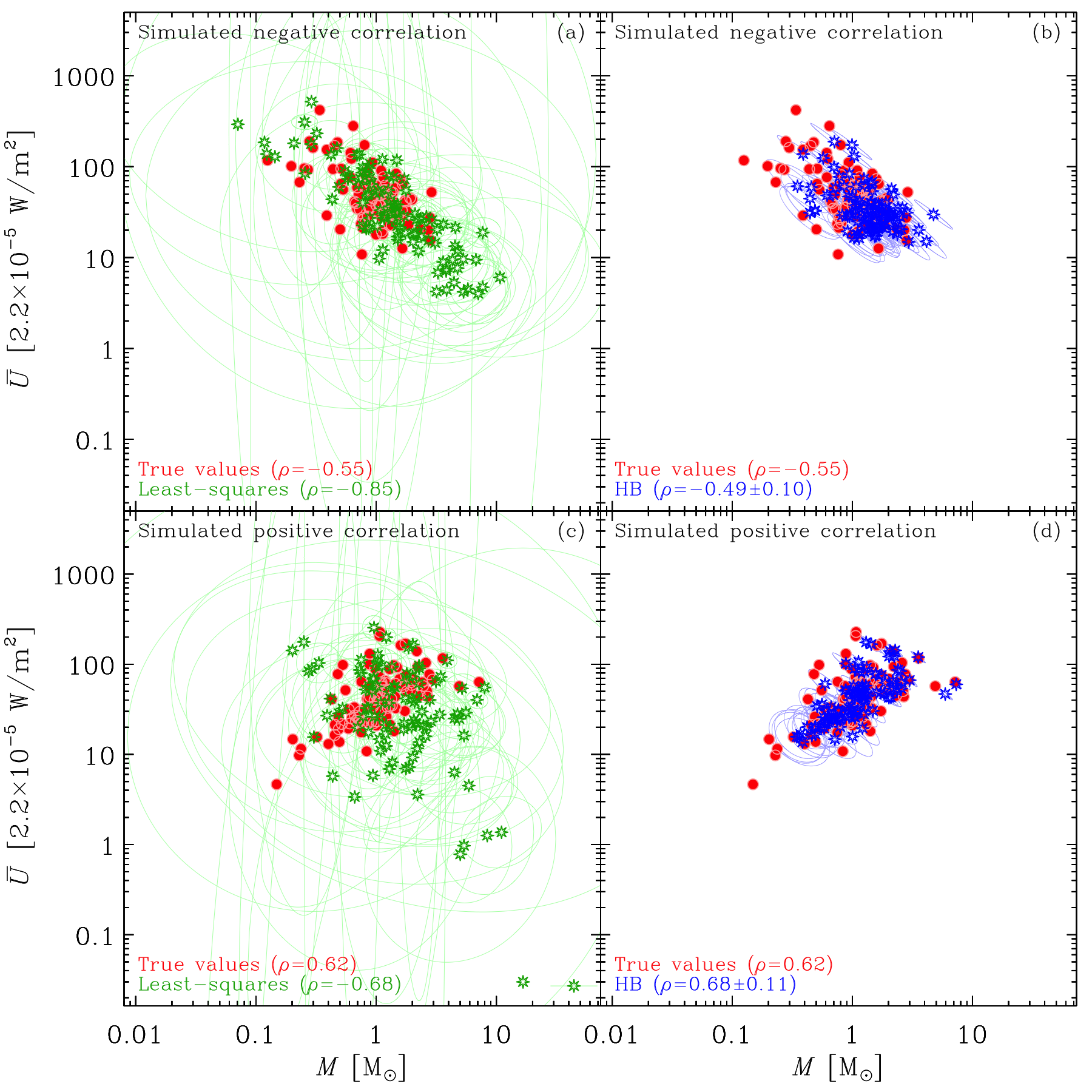}
  \caption{{\sl Efficiency of the method in presence of intrinsic 
           correlations.}
           Each panel shows the relation between the dust mass, $M$, 
           and the average starlight intensity, \Uav.
           The true values are the red dots.
           Panels (a) and (b) show the results applied to simulations with an
           intrinsic negative correlation.
           Panels (c) and (d) show the results with a positive correlation.
           The green stars and uncertainty ellipses, in panels (a) and (c),
           are the least-squares results.
           The blue stars and uncertainty ellipses, in panels (b) and (d),
           are the HB results.
           We quote, in each panel, the true and inferred values of the 
           correlation coefficient, $\rho$, between $\ln M$ and $\ln\Uavm$.}
  \label{fig:vargrid_prior}
\end{figure*}
\reffig{fig:vargrid_prior} shows the results of the $\chi^2$ and HB codes on these two simulations.
As noted on the previous examples, the $\chi^2$ method leads to a more dispersed distribution of values (panels a and c).
More interestingly here, we see that:
\begin{itemize}
  \item when the true correlation is negative (panel a; $\rho=-0.55$), the 
    $\chi^2$ method finds a tighter correlation ($\rho=-0.85$);
  \item when the true correlation is positive (panel c; $\rho=0.62$), the 
    $\chi^2$ method finds a negative correlation ($\rho=-0.68$).
\end{itemize}
In contrast, the HB method is able to consistently recover the correlation coefficients, whether the true correlation is negative (panel b) or positive
(panel d).
The quoted correlation coefficients, in panels b and d, for the HB method, have been derived directly from the MCMC, as explained in \refsec{sec:derived}.

  \subsection{Effect of the wavelength coverage}
  \label{sec:wavcov}
 
It is obvious that the wavelength coverage has an impact on the recovery of the 
parameters.
However, the nature of this impact on the HB results is not trivial.
We have generated two simulations to illustrate this effect.
We start from the central simulation of \refsec{sec:refgridesc} (warm SED, with 
$n=100$ and $f_\sms{S/N}=3$) and make the following modifications.
\begin{description}
  \item[Far-IR coverage:]we remove the three SPIRE bands from the set of 
    constraints. 
    The longest wavelength constraint is thus moved from $\lambda=500\mmic$ to
    $\lambda=160\mmic$.
  \item[Mid-IR coverage:]we add the following \wise\ and \akari/IRC bands:
    \wisei, \wiseii, \wiseiii, \wiseiv,
    \akariii, \akariiii, \akariiv, \akariv, \akarivi, \akarivii\ and 
    \akariviii.
\end{description}

\begin{figure*} 
  \includegraphics[width=\textwidth]{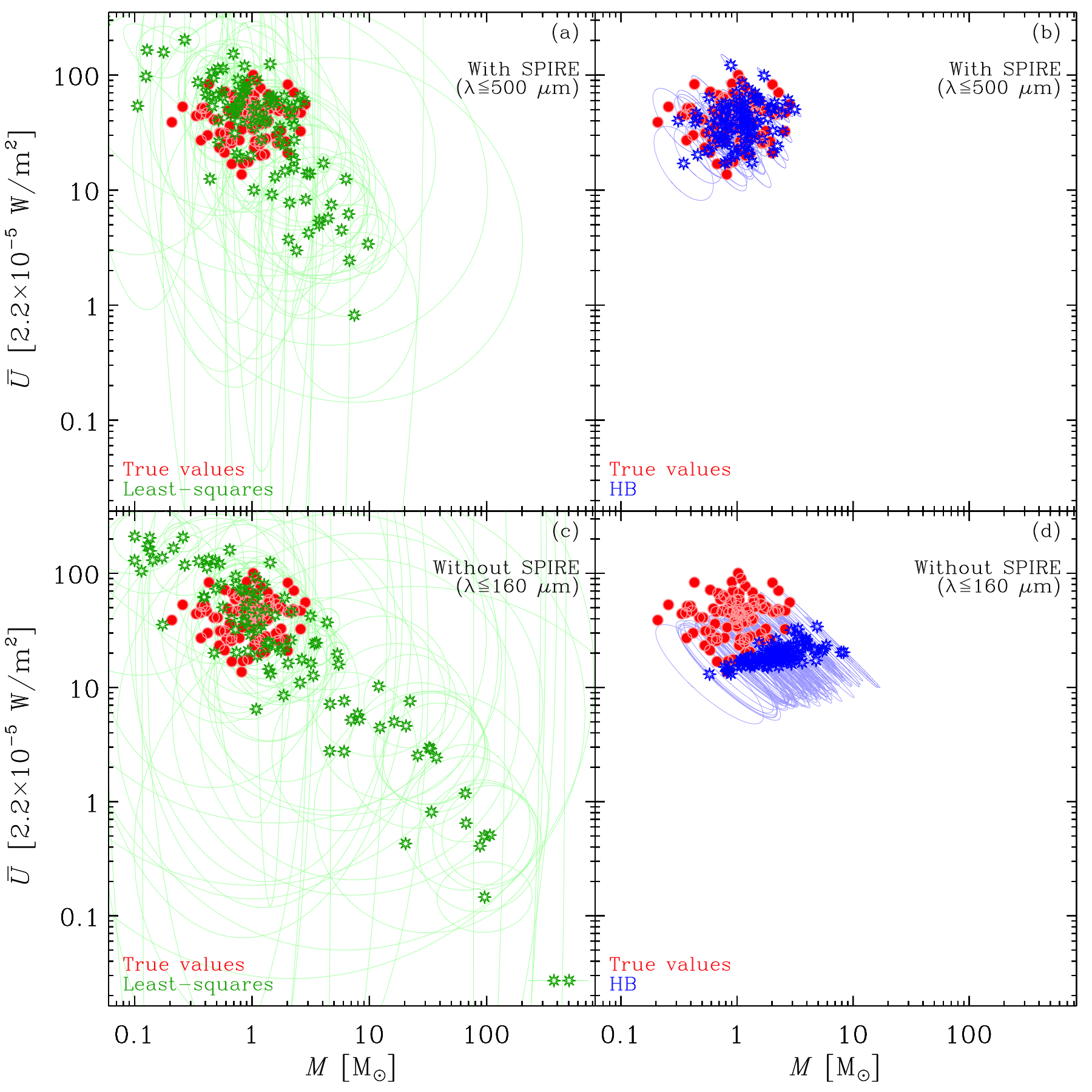}
  \caption{{\sl Effect of the far-IR wavelength coverage.}
           The four panels show the results of different methods, applied on 
           the central simulation of \refsec{sec:refgridesc} 
           (warm SED, with $n=100$ and $f_\sms{S/N}=3$).
           In each panel, the red dots show the true values of the dust mass, 
           $M$, and average starlight intensity, \Uav\ \refeq{eq:avU}.
           Panels (a) and (b) correspond to the case where the whole wavelength
           coverage is used.
           Panels (c) and (d) show the results obtained, excluding the three
           SPIRE bands, from the set of constraints.
           The green stars and uncertainty ellipses, in panels (a) and (c),
           are the least-squares results.
           The blue stars and uncertainty ellipses, in panels (b) and (d),
           are the HB results.}
  \label{fig:vargrid_FIR}
\end{figure*}
\reffig{fig:vargrid_FIR} shows the results of the HB code, with reduced far-IR
coverage.
The parameters which are the most affected by this modification are those 
controling the shape of the far-IR part of the SED, mainly $M$ and \Uav.
Comparing panels (a) and (c), we see that, without long wavelength constraints,
the $\chi^2$ results are significantly more dispersed.
The noise induced correlation, discussed in \refsecs{sec:dissect_par} and 
\ref{sec:priorcorr}, is enhanced in panel (c).
On the contrary, comparing panels (b) and (d), we see that the HB inference 
does not result in an increase of the dispersion, nor any false correlation.
Most true values are within $1\sigma$ of their HB results.
We note that decreasing the wavelength coverage has the same qualitative 
effect as decreasing the signal-to-noise ratio (see \refsec{sec:prioref}):
the prior becomes dominant in the posterior distribution.

\begin{figure} 
  \includegraphics[width=\linewidth]{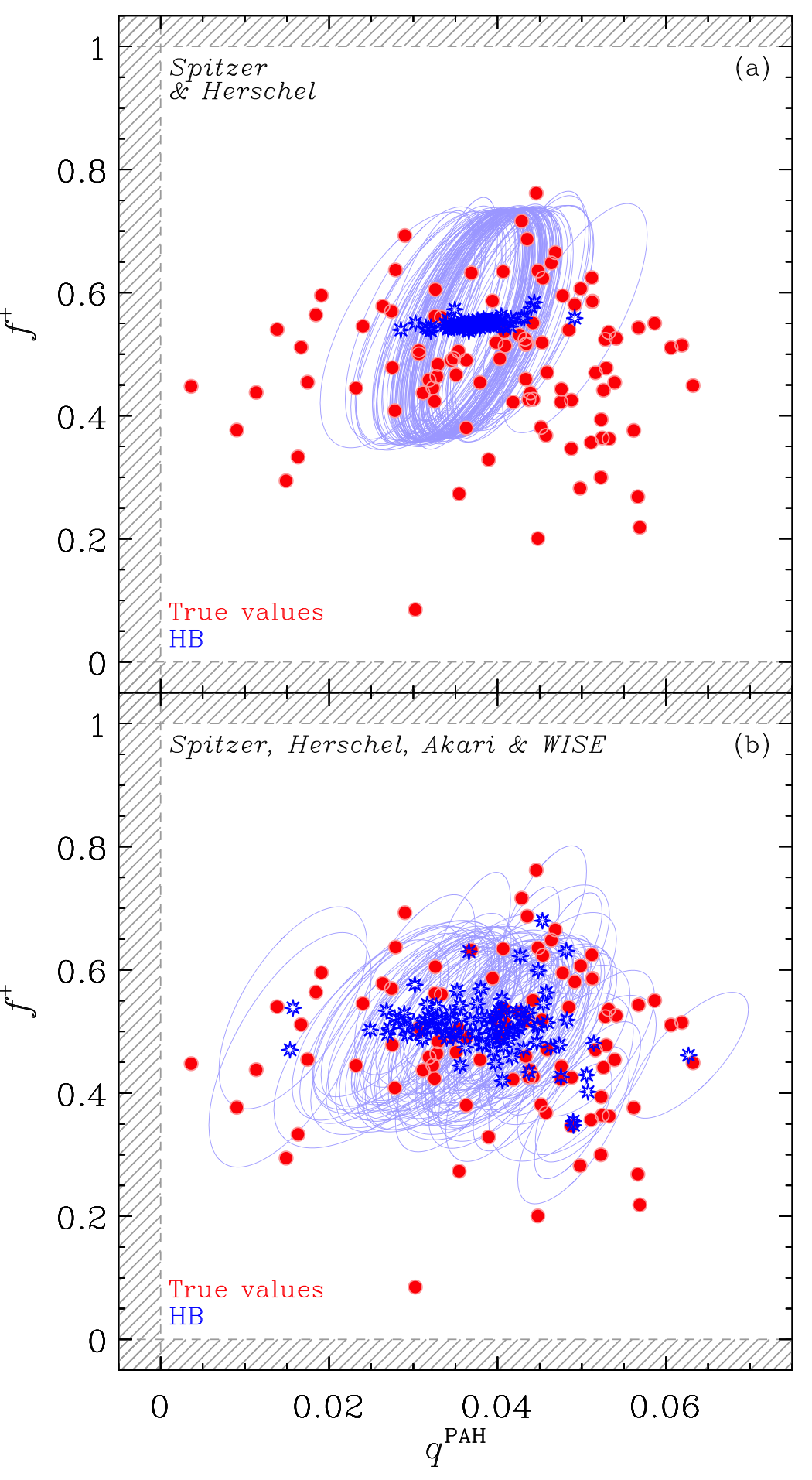}
  \caption{{\sl Effect of the mid-IR wavelength coverage.}
           The two panels show the HB results, applied on 
           the central simulation of \refsec{sec:refgridesc} 
           (warm SED, with $n=100$ and $f_\sms{S/N}=3$).
           In each panel, the red dots show the true values of the PAH mass
           fraction, $q^\sms{PAH}$, and PAH charge fraction, $f^+$.
           The blue stars and uncertainty ellipses are the HB results.
           Panel (a) corresponds to the original coverage with \spitz\ and
           \hersc\ photometry, only.
           Panel (b) shows the results increasing the mid-IR coverage, by 
           adding \wise\ and \akari\ bands.}
  \label{fig:vargrid_MIR}
\end{figure}
The mid-IR coverage has an effect, mainly on the PAH mass and charge fractions,
$q^\sms{PAH}$ and $f^+$.
We discussed in \refsec{sec:refhyp} the fact that $f^+$ was particularly poorly 
constrained.
This can be seen in panel (a) of \reffig{fig:vargrid_MIR}.
Increasing the mid-IR wavelength coverage (panel b), this problem is solved and 
the distribution of parameters is now more accurately recovered.

  \subsection{The Introduction of an external parameter}
  \label{sec:varext}

As we start to see here, the HB method is 
particularly efficient at recovering intrinsic correlations between parameters 
of the SED model.
However, the correlation of these SED parameters with other quantities
(\eg\ gas mass, metallicity, \etc) is also relevant.
In order to treat these external parameters in the HB framework, we need to 
extend the prior distribution to them (see \refapp{app:extra}).
Although these external parameters are not free parameters,
their mean values are going to be modified consistently with their 
uncertainties.
To demonstrate this process, we have drawn an SED sample and an 
associated distribution of gas mass $M_\sms{gas}$ (assuming a typical Galactic
$M_\sms{gas}/M_\sms{dust}\simeq100$).
This sample has $n=300$ sources and a median signal-to-noise ratio of
$f_\sms{S/N}=0.3$.
We adopt the warm SED distribution (\refsec{sec:refgridesc}), with the 
following modifications: 
$S[\ln U_-]=0.6$, $S[\ln\Delta U]=0.7$, 
$\mu[\ln M_\sms{gas}]=\ln(100\,\rm M_\odot)$, 
$S[\ln M_\sms{gas}] = 0.5$ and $R[\ln M_\sms{dust},\ln M_\sms{gas}]=0.9$,
where we have noted the dust mass $M_\sms{dust}$ to avoid confusion.
We attribute a $10\,\%$ uncertainty to the gas mass.

\begin{figure} 
  \includegraphics[width=\linewidth]{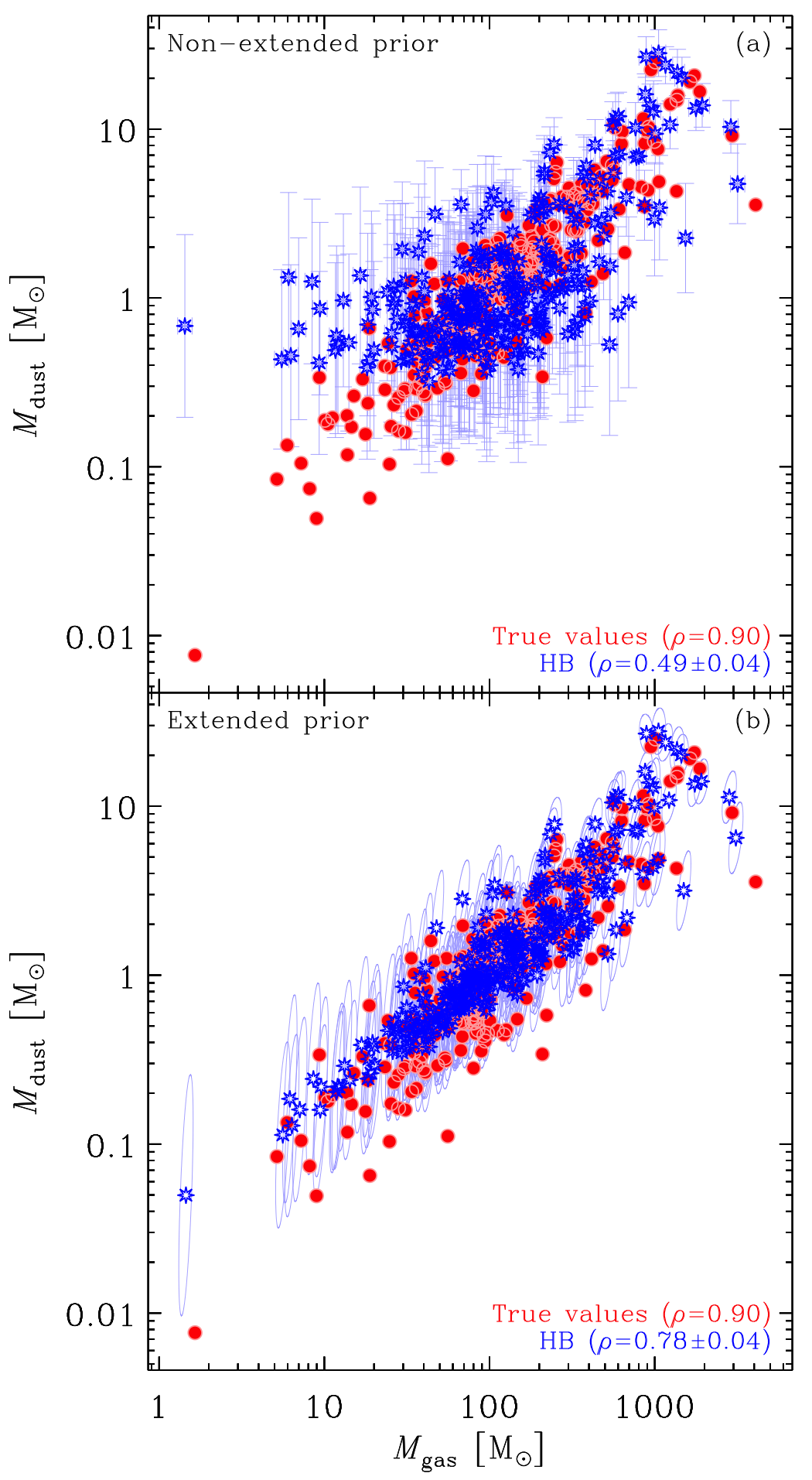}
  \caption{{\sl Extension of the prior to an external parameter.}
           The two panels show the relation between the dust mass, noted 
           $M_\sms{dust}$, for clarity here, and the gas mass, $M_\sms{gas}$, 
           a parameter not controlling the SED model.
           The red dots are the true values in both panels.
           The blue stars and uncertainty bars/ellipses are the HB inference.
           Panel (a) shows the standard result, with the prior of 
           \refeq{eq:HB}: $M_\sms{dust}$ is derived from the HB code and
           $M_\sms{gas}$ is plotted as is.
           Panel (b) shows the result, introducing the gas mass as an extra 
           parameter in the prior (see \refapp{app:extra}).
           We show error bars rather than ellipses in panel (a), because the
           two plotted quantities do not come from the same joint PDF.}
  \label{fig:vargrid_extra}
\end{figure}
\reffig{fig:vargrid_extra} compares the results obtained with and without prior 
extension.
Panel (a) shows that, when $M_\sms{gas}$ is not in the prior, the
correlation found is significantly weaker ($\rho=0.49\pm0.04$) than the 
intrinsic one ($\rho=0.9$).
Notice however that the inferred dust masses are consistent with the true 
values.
The largest discrepancies ($2-3\sigma$) happen at low column density, where 
signal-to-noise ratio is low.
Panel (b) presents the results with extension of the prior.
The correlation between the two parameters is better recovered
($\rho=0.78\pm0.04$).
The efficiency of extending the prior to external parameters is increased if 
the number of sources is larger.

  \section{Application to Other Physical Models}
  \label{sec:othergrid}
  
\subsection{Modified black bodies}
\label{sec:MBBfit}

As discussed in \refsec{sec:MBB}, \MBB\ is the most widely used dust model, due 
to its simplicity.
The HB dust codes that have been previously presented in the literature all 
implement it \citep{kelly12,veneziani13,juvela13,marsh15}.
In this section, we apply our HB code to one \MBB\ simulation, in order to 
confirm its ability to correct the noise induced negative correlation between 
$T$ and $\beta$ \citep{shetty09}.
This simulation is designed to mimic typical \planck\ observations of the diffuse Galactic ISM \citep[\eg][]{planck-collaboration14b,ysard15}.
We assume: $\mu[\ln M]=0$, $\mu[ln T]=\ln(20.5\,{\rm K})$, $\mu[\beta]=1.65$,
$S[\ln M]=0.05$, $S[\ln T]=0.05$, $S[\beta]=0.01$ and $R[{\rm all}]=0$.
We draw $n=1000$ sources, observed through the following \cobe\ and \planck\ photometric bands: \DIRBEviii, \DIRBEix, \DIRBEx, \HFIi, \HFIii\ and \HFIiii.
We set the median signal-to-noise ratio to $f_\sms{S/N}=10$.

\begin{figure} 
  \includegraphics[width=\linewidth]{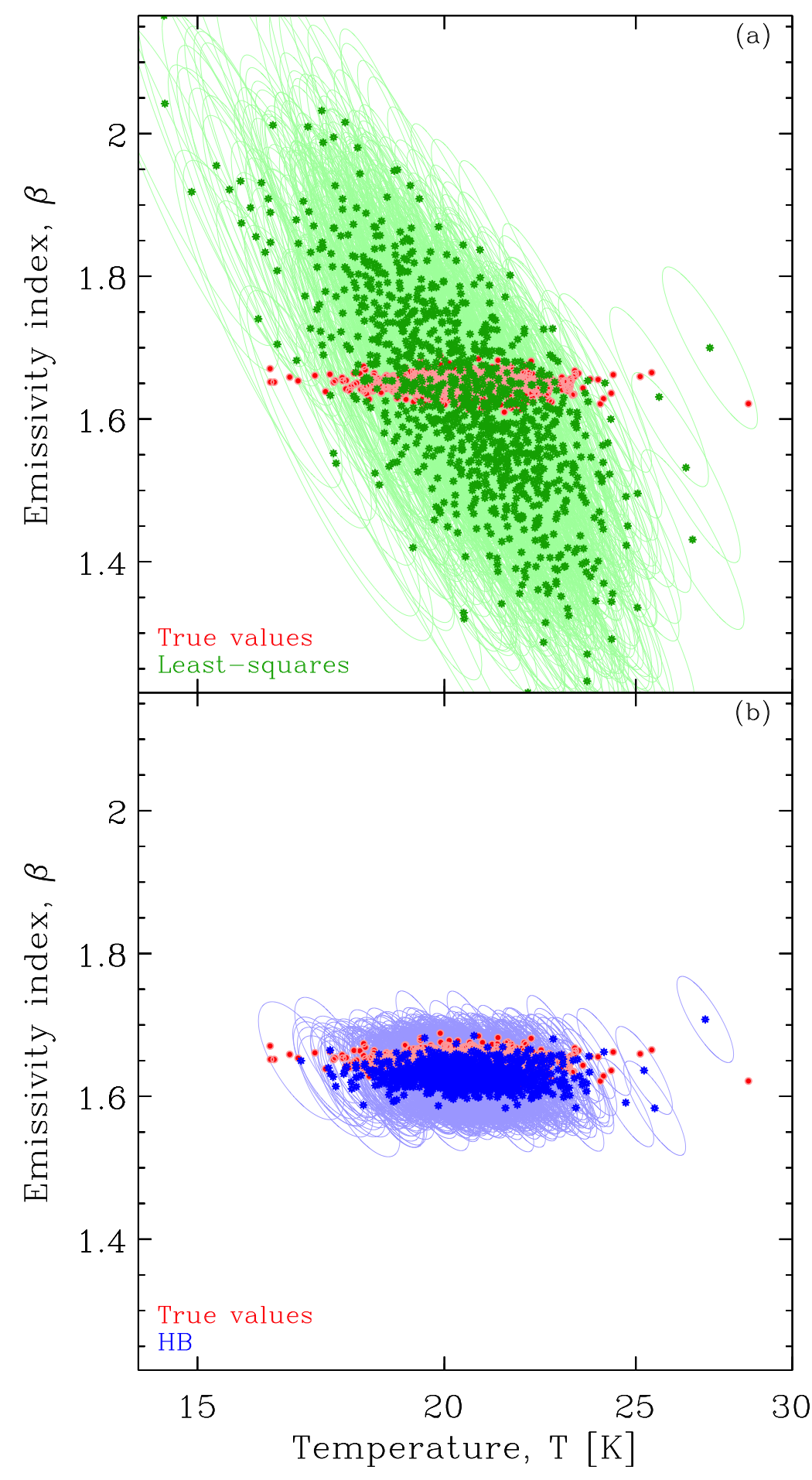}
  \caption{{\sl Application to \MBB\ models.}
           The two panels represent the temperature and emissivity index
           of a simulation with $n=1000$ sources having an \MBB\ SED 
           (\refsec{sec:MBB}).
           The red dots, in both panels, are the true values.
           The green stars and uncertainty ellipses, in panel (a), are the
           $\chi^2$ values.
           The blue stars and uncertainty ellipses, in panel (b), are the HB
           results.}
  \label{fig:other_MBB}
\end{figure}
The results are presented in \reffig{fig:other_MBB}.
Panel (a) displays the least-squares values.
Notice the well-kown false $\beta-T$ correlation.
Panel (b) demonstrates that the HB method can indeed accurately correct this 
false correlation, as was also shown by \citet{kelly12}.
\citetalias{galliano18b} will present more \MBB\ simulations to discuss 
robustness.

\subsection{Broken emissivity modified black bodies}
\label{sec:BEMBBfit}

\BEMBB\ models (\refsec{sec:BEMBB}) have been used by \citet{gordon14} and \citet{roman-duval14} to model the Magellanic clouds.
To roughly mimic these data, we simulate a sample of $n=300$ sources, 
observed through: \PACSii, \PACSiii, \SPIREi, \SPIREii\ and \SPIREiii.
The parameters are distributed as: $\mu[\ln M]=0$, $S[\ln M]=0.05$, 
$\mu[\ln T]=\ln(20.5\,\rm K)$, $S[\ln T]=0.05$,
$\mu[\beta_1]=1.65$, $S[\beta_1]=0.01$, $\mu[\beta_2]=1.65$, $S[\beta_2]=0.01$,
$\mu[\ln \nu_b]=\ln(c/350\mmic)$, $S[\ln\nu_b]=0.01$ and $R[{\rm all}]=0$.
We assume that the median signal-to-noise ratio is $f_\sms{S/N}=10$.
One of the derived quantities discussed by \citet{gordon14} is the {\it submm 
excess}, $e_{500}$, defined as the relative difference between the actual 
emission at 500~\mic\ and the emission of an \MBB\ with the same temperature 
and $\beta=\beta_1$:
\begin{equation}
  e_{500} 
  = \frac{1-\nu_{500}^{\beta_1}}{\nu_{500}^{\beta_2}\nu_b^{\beta_1-\beta_2}},
\end{equation}
where $\nu_{500}=c/500\mmic$.
Since $\mu[\beta_1]=\mu[\beta_2]$, our simulation exhibits on average a zero 
excess.

\begin{figure} 
  \includegraphics[width=\linewidth]{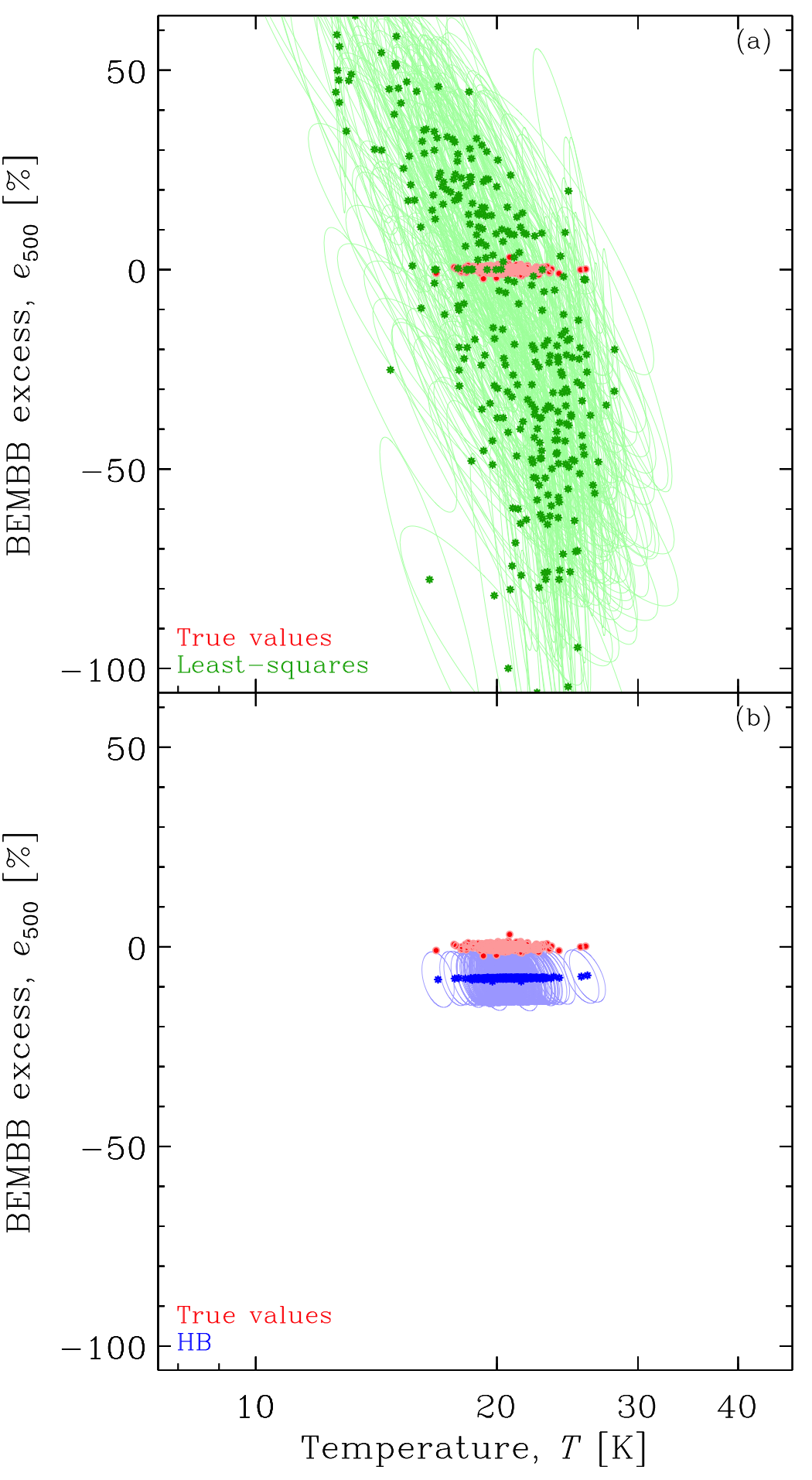}
  \caption{{\sl Application to \BEMBB\ models.}
           The two panels represent the temperature and 500~\mic\ excess, 
           $e_{500}$, of a simulation with $n=300$ sources having a \BEMBB\ SED 
           (\refsec{sec:BEMBB}).
           The red dots, in both panels, are the true values.
           The green stars and uncertainty ellipses, in panel (a), are the
           $\chi^2$ values.
           The blue stars and uncertainty ellipses, in panel (b), are the HB
           results.}
  \label{fig:other_BEMBB}
\end{figure}
\reffig{fig:other_BEMBB} presents the results.
Panel (a) shows that the $\chi^2$ method is unable to recover any meaningful 
excess.
It is due to the fact that the parameters of this model are extremely 
degenerate.
On the contrary the HB approach is able to provide a tight distribution, consistent with the true values (panel b).

\subsection{Multi-component dust SED models}
\label{sec:DL07}

The number of combinations of dust models is almost infinite.
We end this section with a widely used one, introduced by
\citet{draine07}.
It consists in the linear combination of a \powU\ (\refsec{sec:powerU}), a \deltaU\ (\refsec{sec:deltaU}) and a \Star\ (\refsec{sec:starBB}) component.
It is aimed at approximating the multi-phase nature of the ISM of galaxies, the
\deltaU\ component being attributed to diffuse ISM and the \powU\ component to hotter PDRs.
Several parameters are tied: $U_\sms{\deltaU} = U_-^\sms{\powU}$,
$q^\sms{PAH}_\sms{\deltaU}=q^\sms{PAH}_\sms{\powU}$ and
$f^+_\sms{\deltaU}=f^+_\sms{\powU}$.
In addition, some parameters are fixed: $\Delta U_\sms{\powU}=10^6$ and
$\alpha_\sms{\powU}=2$.
This way, the \powU\ component represents dust hotter than the \deltaU\ component and can account for the contribution of star forming regions to the mid-IR emission of galaxies.
The two PAH parameters are tied, as they would otherwise be degenerate.
The mass fraction of the \powU\ component is:
\begin{equation}
  \gamma = \frac{M_\sms{\powU}}{M_\sms{\powU}+M_\sms{\deltaU}}.
  \label{eq:DL07}
\end{equation}
We have performed one simulation to illustrate the performances of the HB method with this model.
We have drawn parameters from the warm SED distribution 
(\refsec{sec:refgridesc}), with $n=100$ sources, and a median signal-to-noise 
ratio, $f_\sms{S/N}=1$, with the following modifications: 
$\mu[\ln M_\sms{\powU}]=\ln0.01$ and $\mu[\ln U_-^\sms{\powU}]=\ln0$.
We have added the \deltaU\ component with $\mu[\ln M_\sms{\deltaU}]=\ln 1$.

\begin{figure} 
  \includegraphics[width=\linewidth]{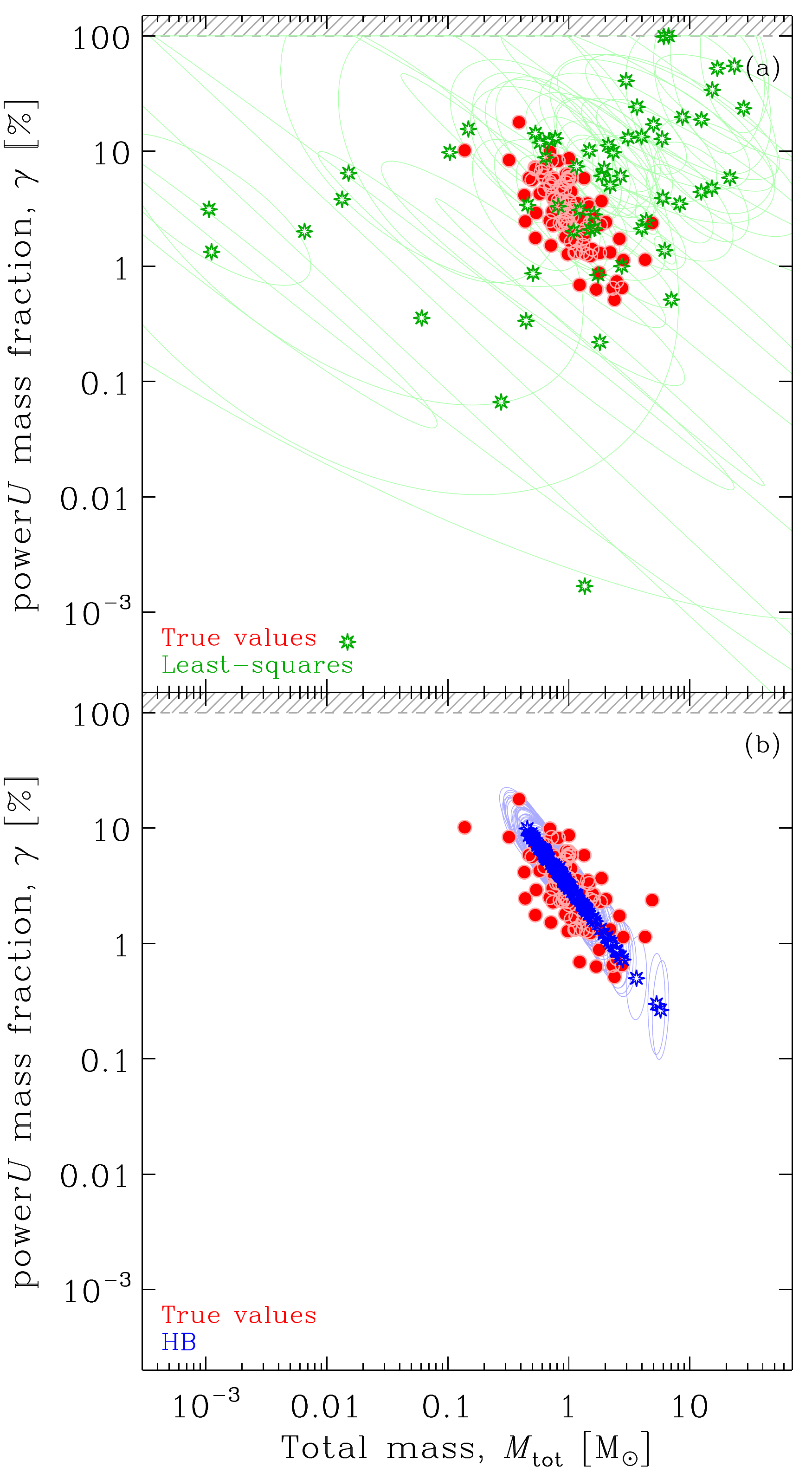}
  \caption{{\sl Application to a multi-phase dust model.}
           The two panels represent the total dust mass, $M_\sms{tot}$, and
           the mass fraction of non-uniformly illuminated dust, $\gamma$ 
           \refeqp{eq:DL07}, of a simulation with $n$ sources.
           The red dots, in both panels, are the true values.
           The green stars and uncertainty ellipses, in panel (a), are the
           $\chi^2$ values.
           The blue stars and uncertainty ellipses, in panel (b), are the HB
           results.}
  \label{fig:other_DL07}
\end{figure}
The results are presented in \reffig{fig:other_DL07}.
It shows the relation between the mass of the two components, $M_\sms{tot}=M_\sms{\powU}+M_\sms{\deltaU}$, and the parameter $\gamma$.
As can be seen in panel (a), with the $\chi^2$ method, $\gamma$ is completely degenerate.
It spreads the whole range of values between 0 and 100$\,\%$.
On the contrary, the HB method is able to significantly reduce the uncertainties and provide a relevant distribution of parameters, consistent with their true values.


  \section{Summary \&\ Conclusion}
  \label{sec:concl}
    
This is the first article in a series of two papers presenting a new 
model, \HB, aimed at deriving dust parameters from infrared observations.
The main originality of this model is to apply hierarchical Bayesian (HB) 
inference to full dust models, taking into account realistic optical 
properties, size distribution, stochastic heating and distribution of starlight 
intensities, as well as color correction and correlated calibration 
uncertainties.
Simply put, the HB method consists of sampling the probability distribution of 
the physical parameters of an ensemble of sources.
This distribution of parameters is modelled with a prior distribution 
controlled by hyperparameters.
The inferred prior distribution does not significantly modify the PDF of 
sources for which the likelihood is much
narrower than the dispersion of the distribution of parameters
(\refsec{sec:dissect_par} and \ref{sec:prioref}).
However, the prior has an important effect on sources with low 
signal-to-noise ratios.

We have described the formalism of our model and its numerical implementation 
(\refsecs{sec:phys}-\ref{sec:HB}).
We have subsequently applied our model to synthetic observations, in order to 
quantify its performances (\refsecs{sec:refgrid}-\ref{sec:othergrid}).
The main conclusions are the following.
\begin{enumerate}
  \item We have compared the performances of least-squares, non-hierarchical 
    and hierarchical Bayesian methods (\refsec{sec:dissect_par}).
    We have shown that, although the non-hierarchical Bayesian approach is 
    better than least-squares at estimating the uncertainties on derived 
    parameters, it is not able to correct noise-induced correlations and 
    scatter.
    On the contrary, the HB approach is particularly efficient at 
    recovering true correlations between parameters and their intrinsic scatter
    (\refsecs{sec:anaref} and \ref{sec:priorcorr}).
    The HB inferred values are also consistently closer to their true values
    than the least-squares values (\refsec{sec:anaref}).
  \item We have systematically studied the performances of our model, varying 
    signal-to-noise ratio, SED shape and sample size (\refsec{sec:anaref}).
    We have shown that the recovered parameters are distributed symmetrically 
    around their true values.
    We did not find any clear bias.
    The scatter of the best constrained parameters from their true values are 
    close to a normal distribution, with almost $100\,\%$ of the distribution
    within $3\sigma$ (\refsec{sec:refpar}).
    Poorly constrained, degenerate, parameters are also consistently recovered, 
    but with a fraction of outliers closer to a Student's $t$ distribution.
  \item The HB method can reasonably compensate for a partial lack of spectral 
    coverage (\refsec{sec:wavcov}).
  \item We have demonstrated that one can easily include external parameters
    in the prior distribution in order to improve the recovery of a potential
    correlation between these external parameters and the dust properties
    (\refsec{sec:varext}).
  \item We have applied our HB code with other physical models 
    (different types of modified black bodies and multi-component dust SED 
    models; \refsec{sec:othergrid}).
    We have demonstrated that it works well in these cases, too, and that it
    can help obtain better constraints on degenerate parameters.
  \item We have discussed the issues of convergence towards the stationary 
    posterior.
    Our choice of starting the MCMC at the least-squares values appears 
    to lead to relatively short burn-in phases.
    The integrated autocorrelation times of the simulations in this paper
    are all shorter than a few $10^4$ (\refsec{sec:tint}).
    Having implemented ancillarity-sufficiency interweaving strategy 
    to all parameters, we have demonstrated that it significantly reduces
    the autocorrelation of the Markov chain (\refsec{sec:MCMC}).
\end{enumerate}

The HB technique is quite general \citep[\eg][]{shahmoradi17}.
It has been successfully applied to several fields in astrophysics, among others:
luminosity distribution of $\gamma$-ray bursts \citep{loredo98};
supernova studies \citep{mandel09};
exoplanet eccentricities \citep{hogg10};
galaxy clusters \citep{andreon10};
Milky Way satellites \citep{martinez15};
exoplanet mass-radius relationship \citep{wolfgang15}\footnote{List found on \href{http://astrostatistics.psu.edu/\-RLectures/\-hierar\-chical\-.pdf}{astrostatistics.psu.edu/RLectures/hierarchical.pdf}}.
We have demonstrated in this paper that it remains accurate over a wide range of sample sizes, source 
properties, signal-to-noise ratios and number of model parameters.
Such a method can in principle be applied to any field.
However, as discussed in \refsec{sec:prioref}, it is relevant mainly when the typical uncertainty on
a model parameter is comparable or larger than the typical dispersion of this parameter through the 
sample.
If, on the contrary, the uncertainties on the parameters are much smaller than the dispersion of the 
sample, this method does not provide significant improvements compared to simpler techniques.

The second article of this series \citepalias{galliano18b} will address 
the robustness of our code.
For that purpose, we will apply it to data simulated with different hypotheses 
from the model (noise distribution, physical components and distribution of 
parameters; see \refsec{sec:refgrid}).
In parallel, several up-coming papers are presenting the application of this 
model to real astrophysical data: the Magellanic clouds 
\citep[][\citeprep{galliano17c}]{galliano17a}, 
the Carina nebula \citep[][\citeprep{wu18}]{wu18b}, the dwarf galaxy IC$\,$10 
\citepprep{lianou17}, and the DustPedia galaxy sample \citep{davies17}.

Overall, observational dust studies need to emancipate themselves from dated 
techniques.
Although a modified black body least-squares fit can be useful in some cases,
the development of modern instrumentation renders it more and more obsolete.
More precise observations mean that we can access the complexity of the ISM
conditions, beyond the simple isothermal approximation.
In addition, more accurate fluxes call for an optimal way to extract relevant 
information from the data.
With the model presented here, we have pursued this goal, hoping it will contribute to refining our understanding of dust evolution.
Adopting such an approach represents a significant investment in terms of 
computation time and storage (\refapp{app:CPU}).
However, this will naturally be facilitated in the near future by the increase 
of computational performances.
Indeed, nowadays, the industry favors the development of large multi-core 
platforms and Gibbs sampling can be reasonably well parallelized.
Finally, this type of study would benefit from a better 
synergy between observers and instrumentalists.
Ideally, the development of new instruments and their reduction pipelines 
should account for the progress in data analysis techniques, and reciprocally.
In particular, the various sources of uncertainties, their correlations, their 
biases, \etc\ should be considered as important as the measured fluxes.
Their accurate knowledge should be planned in the design of the new detectors,
and thoroughly documented.
This will be particularly relevant for future IR missions like \spica\ \citep[\eg][]{van-der-tak17}.

  \section*{Acknowledgements}
    I thank Jean-Philippe Bernard, Karl Gordon, Mika Juvela, Martin 
    Kilbinger and Kirill Tchernyshyov for fruitful discussions about Bayesian  
    techniques and SED modelling, 
    Anthony Jones for sharing his expertise on dust evolution, 
    and S\'ebastien Fromang and Pierre-Fran\c{c}ois Lavall\'ee for coding 
    advice.
    Sophia Lianou provided me with several bug reports, running earlier 
    versions of the code.
    Finally, I thank Sacha Hony, Suzanne Madden and Ronin Wu, as well as an 
    anonymous referee, for providing me 
    with useful comments that helped clarify the manuscript.
    I acknowledge support from the EU FP7 project DustPedia (Grant No.\ 
    606847), from the Agence Nationale pour la Recherche (ANR) through the 
    program SYMPATICO (Programme Blanc, Projet ANR-11-BS56-0023) and 
    through the PRC program 1311 between CNRS (France) and JSPS (Japan).
	This work was supported by the Programme National "Physique et Chimie du 
	Milieu Interstellaire" (PCMI) of CNRS/INSU with INC/INP co-funded by CEA 
	and CNES.
 
  \bibliographystyle{$HOME/Astro/TeXstyle/bib_notes}
  \bibliography{$HOME/Astro/TeXstyle/references}

\begin{thebibliography}{89}
\expandafter\ifx\csname natexlab\endcsname\relax\def\natexlab#1{#1}\fi

\bibitem[{{Allamandola} {et~al.}(1999){Allamandola}, {Hudgins}, \&
  {Sandford}}]{allamandola99}
{Allamandola}, L.~J., {Hudgins}, D.~M., \& {Sandford}, S.~A.
  1999\href{http://cdsads.u-strasbg.fr/abs/1999ApJ...511L.115A}{, \apjl, 511,
  L115}

\bibitem[{{Andreon} \& {Hurn}(2010)}]{andreon10}
{Andreon}, S. \& {Hurn}, M.~A.
  2010\href{http://cdsads.u-strasbg.fr/abs/2010MNRAS.404.1922A}{, \mnras, 404,
  1922}

\bibitem[{{Asano} {et~al.}(2013){Asano}, {Takeuchi}, {Hirashita}, \&
  {Inoue}}]{asano13}
{Asano}, R.~S., {Takeuchi}, T.~T., {Hirashita}, H., \& {Inoue}, A.~K.
  2013\href{http://cdsads.u-strasbg.fr/abs/2013EP%26S...65..213A}{, Earth,
  Planets, and Space, 65, 213}

\bibitem[{{Bakes} \& {Tielens}(1994)}]{bakes94}
{Bakes}, E.~L.~O. \& {Tielens}, A.~G.~G.~M.
  1994\href{http://adsabs.harvard.edu/cgi-bin/nph-bib_query?bibcode=1994ApJ...427..822B&db_key=AST}{,
  \apj, 427, 822}

\bibitem[{Barnard {et~al.}(2000)Barnard, McCulloch, \& Meng}]{barnard00}
Barnard, J., McCulloch, R., \& Meng, X.-L.
  2000\href{http://www3.stat.sinica.edu.tw/statistica/j10n4/j10n416/j10n416.htm}{,
  Statistica Sinica, 10, 1281}

\bibitem[{{Bron} {et~al.}(2014){Bron}, {Le Bourlot}, \& {Le Petit}}]{bron14}
{Bron}, E., {Le Bourlot}, J., \& {Le Petit}, F.
  2014\href{http://cdsads.u-strasbg.fr/abs/2014A%26A...569A.100B}{, \aap, 569,
  A100}

\bibitem[{{Chastenet} {et~al.}(2017){Chastenet}, {Bot}, {Gordon}, {Bocchio},
  {Roman-Duval}, {Jones}, \& {Ysard}}]{chastenet17}
{Chastenet}, J., {Bot}, C., {Gordon}, K.~D., {et~al.}
  2017\href{http://cdsads.u-strasbg.fr/abs/2017A%26A...601A..55C}{, \aap, 601,
  A55}

\bibitem[{{Chevance} {et~al.}(2016){Chevance}, {Madden}, {Lebouteiller},
  {Godard}, {Cormier}, {Galliano}, {Hony}, {Indebetouw}, {Le Bourlot}, {Lee},
  {Le Petit}, {Pellegrini}, {Roueff}, \& {Wu}}]{chevance16}
{Chevance}, M., {Madden}, S.~C., {Lebouteiller}, V., {et~al.}
  2016\href{http://cdsads.u-strasbg.fr/abs/2016A%26A...590A..36C}{, \aap, 590,
  A36}

\bibitem[{{Compi{\`e}gne} {et~al.}(2011){Compi{\`e}gne}, {Verstraete}, {Jones},
  {Bernard}, {Boulanger}, {Flagey}, {Le Bourlot}, {Paradis}, \&
  {Ysard}}]{compiegne11}
{Compi{\`e}gne}, M., {Verstraete}, L., {Jones}, A., {et~al.}
  2011\href{http://cdsads.u-strasbg.fr/abs/2011A%26A...525A.103C}{, \aap, 525,
  A103+}

\bibitem[{{Coupeaud} {et~al.}(2011){Coupeaud}, {Demyk}, {Meny}, {Nayral},
  {Delpech}, {Leroux}, {Depecker}, {Creff}, {Brubach}, \& {Roy}}]{coupeaud11}
{Coupeaud}, A., {Demyk}, K., {Meny}, C., {et~al.}
  2011\href{http://cdsads.u-strasbg.fr/abs/2011A%26A...535A.124C}{, \aap, 535,
  A124}

\bibitem[{{da Cunha} {et~al.}(2008){da Cunha}, {Charlot}, \&
  {Elbaz}}]{da-cunha08}
{da Cunha}, E., {Charlot}, S., \& {Elbaz}, D.
  2008\href{http://adsabs.harvard.edu/abs/2008MNRAS.388.1595D}{, \mnras, 388,
  1595}

\bibitem[{{Dale} {et~al.}(2001){Dale}, {Helou}, {Contursi}, {Silbermann}, \&
  {Kolhatkar}}]{dale01}
{Dale}, D.~A., {Helou}, G., {Contursi}, A., {Silbermann}, N.~A., \&
  {Kolhatkar}, S.
  2001\href{http://adsabs.harvard.edu/cgi-bin/nph-bib_query?bibcode=2001ApJ...549..215D&db_key=AST}{,
  \apj, 549, 215}

\bibitem[{{Davies} {et~al.}(2017){Davies}, {Baes}, {Bianchi}, {Jones},
  {Madden}, {Xilouris}, {Bocchio}, {Casasola}, {Cassara}, {Clark}, {De Looze},
  {Evans}, {Fritz}, {Galametz}, {Galliano}, {Lianou}, {Mosenkov}, {Smith},
  {Verstocken}, {Viaene}, {Vika}, {Wagle}, \& {Ysard}}]{davies17}
{Davies}, J.~I., {Baes}, M., {Bianchi}, S., {et~al.}
  2017\href{http://cdsads.u-strasbg.fr/abs/2017PASP..129d4102D}{, \pasp, 129,
  044102}

\bibitem[{{De Cia} {et~al.}(2013){De Cia}, {Ledoux}, {Savaglio}, {Schady}, \&
  {Vreeswijk}}]{de-cia13}
{De Cia}, A., {Ledoux}, C., {Savaglio}, S., {Schady}, P., \& {Vreeswijk}, P.~M.
  2013\href{http://adsabs.harvard.edu/abs/2013A%26A...560A..88D}{, \aap, 560,
  A88}

\bibitem[{{De Vis} {et~al.}(2017){De Vis}, {Gomez}, {Schofield}, {Maddox},
  {Dunne}, {Baes}, {Cigan}, {Clark}, {Gomez}, {Lara-L{\'o}pez}, \&
  {Owers}}]{de-vis17}
{De Vis}, P., {Gomez}, H.~L., {Schofield}, S.~P., {et~al.}
  2017\href{http://cdsads.u-strasbg.fr/abs/2017MNRAS.471.1743D}{, \mnras, 471,
  1743}

\bibitem[{{Draine}(2003)}]{draine03c}
{Draine}, B.~T.
  2003\href{http://cdsads.u-strasbg.fr/abs/2003ARA%26A..41..241D}{, \araa, 41,
  241}

\bibitem[{{Draine} {et~al.}(2007){Draine}, {Dale}, {Bendo}, {Gordon}, {Smith},
  {Armus}, {Engelbracht}, {Helou}, {Kennicutt}, {Li}, {Roussel}, {Walter},
  {Calzetti}, {Moustakas}, {Murphy}, {Rieke}, {Bot}, {Hollenbach}, {Sheth}, \&
  {Teplitz}}]{draine07b}
{Draine}, B.~T., {Dale}, D.~A., {Bendo}, G., {et~al.}
  2007\href{http://adsabs.harvard.edu/abs/2007ApJ...663..866D}{, \apj, 663,
  866}

\bibitem[{{Draine} \& {Li}(2007)}]{draine07}
{Draine}, B.~T. \& {Li}, A.
  2007\href{http://adsabs.harvard.edu/abs/2007ApJ...657..810D}{, \apj, 657,
  810}

\bibitem[{{Fitzpatrick} \& {Massa}(2007)}]{fitzpatrick07}
{Fitzpatrick}, E.~L. \& {Massa}, D.
  2007\href{http://cdsads.u-strasbg.fr/abs/2007ApJ...663..320F}{, \apj, 663,
  320}

\bibitem[{{Foreman-Mackey} {et~al.}(2013){Foreman-Mackey}, {Conley},
  {Meierjurgen Farr}, {Hogg}, {Long}, {Marshall}, {Price-Whelan}, {Sanders}, \&
  {Zuntz}}]{foreman-mackey13}
{Foreman-Mackey}, D., {Conley}, A., {Meierjurgen Farr}, W., {et~al.} 2013,
  {emcee: The MCMC Hammer}, Astrophysics Source Code Library

\bibitem[{{Galametz} {et~al.}(2016){Galametz}, {Hony}, {Albrecht}, {Galliano},
  {Cormier}, {Lebouteiller}, {Lee}, {Madden}, {Bolatto}, {Bot}, {Hughes},
  {Israel}, {Meixner}, {Oliviera}, {Paradis}, {Pellegrini}, {Roman-Duval},
  {Rubio}, {Sewi{\l}o}, {Fukui}, {Kawamura}, \& {Onishi}}]{galametz16}
{Galametz}, M., {Hony}, S., {Albrecht}, M., {et~al.}
  2016\href{http://cdsads.u-strasbg.fr/abs/2016MNRAS.456.1767G}{, \mnras, 456,
  1767}

\bibitem[{{Galametz} {et~al.}(2011){Galametz}, {Madden}, {Galliano}, {Hony},
  {Bendo}, \& {Sauvage}}]{galametz11}
{Galametz}, M., {Madden}, S.~C., {Galliano}, F., {et~al.}
  2011\href{http://cdsads.u-strasbg.fr/abs/2011A%26A...532A..56G}{, \aap, 532,
  A56}

\bibitem[{{Galliano}(????)}]{galliano18b}
{Galliano}, F., {\it in prep.}

\bibitem[{{Galliano}(2017)}]{galliano17a}
{Galliano}, F.
  2017\href{http://cdsads.u-strasbg.fr/abs/2017P%26SS..149...38G}{, \planss,
  149, 38}

\bibitem[{{Galliano} {et~al.}(2008{\natexlab{a}}){Galliano}, {Dwek}, \&
  {Chanial}}]{galliano08a}
{Galliano}, F., {Dwek}, E., \& {Chanial}, P.
  2008{\natexlab{a}}\href{http://adsabs.harvard.edu/abs/2008ApJ...672..214G}{,
  \apj, 672, 214}

\bibitem[{{Galliano} {et~al.}(????){Galliano}, {Galametz}, {Hony}, {Madden},
  {the HERITAGE consortium}, \& consoritum}]{galliano17c}
{Galliano}, F., {Galametz}, M., {Hony}, S., {et~al.}, {\it in prep.}

\bibitem[{{Galliano} {et~al.}(2011){Galliano}, {Hony}, {Bernard}, {Bot},
  {Madden}, {Roman-Duval}, {Galametz}, {Li}, {Meixner}, {Engelbracht},
  {Lebouteiller}, {Misselt}, {Montiel}, {Panuzzo}, {Reach}, \&
  {Skibba}}]{galliano11}
{Galliano}, F., {Hony}, S., {Bernard}, J.-P., {et~al.}
  2011\href{http://cdsads.u-strasbg.fr/abs/2011A%26A...536A..88G}{, \aap, 536,
  A88}

\bibitem[{{Galliano} {et~al.}(2008{\natexlab{b}}){Galliano}, {Madden},
  {Tielens}, {Peeters}, \& {Jones}}]{galliano08b}
{Galliano}, F., {Madden}, S.~C., {Tielens}, A.~G.~G.~M., {Peeters}, E., \&
  {Jones}, A.~P.
  2008{\natexlab{b}}\href{http://adsabs.harvard.edu/abs/2008ApJ...679..310G}{,
  \apj, 679, 310}

\bibitem[{{Gelman} {et~al.}(2004){Gelman}, {Carlin}, {Stern}, \&
  {Rubin}}]{gelman04}
{Gelman}, A., {Carlin}, J., {Stern}, H., \& {Rubin}, D. 2004, Bayesian Data
  Analysis (Chapman \&\ Hall)

\bibitem[{Geman \& Geman(1984)}]{geman84}
Geman, S. \& Geman, D. 1984\href{http://dl.acm.org/citation.cfm?id=2286617}{,
  IEEE Trans. Pattern Anal. Mach. Intell., 6, 721}

\bibitem[{{Gordon} {et~al.}(2003){Gordon}, {Clayton}, {Misselt}, {Landolt}, \&
  {Wolff}}]{gordon03}
{Gordon}, K.~D., {Clayton}, G.~C., {Misselt}, K.~A., {Landolt}, A.~U., \&
  {Wolff}, M.~J.
  2003\href{http://adsabs.harvard.edu/cgi-bin/nph-bib_query?bibcode=2003ApJ...594..279G&db_key=AST}{,
  \apj, 594, 279}

\bibitem[{{Gordon} {et~al.}(2014){Gordon}, {Roman-Duval}, {Bot}, {Meixner},
  {Babler}, {Bernard}, {Bolatto}, {Boyer}, {Clayton}, {Engelbracht}, {Fukui},
  {Galametz}, {Galliano}, {Hony}, {Hughes}, {Indebetouw}, {Israel}, {Jameson},
  {Kawamura}, {Lebouteiller}, {Li}, {Madden}, {Matsuura}, {Misselt}, {Montiel},
  {Okumura}, {Onishi}, {Panuzzo}, {Paradis}, {Rubio}, {Sandstrom}, {Sauvage},
  {Seale}, {Sewi{\l}o}, {Tchernyshyov}, \& {Skibba}}]{gordon14}
{Gordon}, K.~D., {Roman-Duval}, J., {Bot}, C., {et~al.}
  2014\href{http://cdsads.u-strasbg.fr/abs/2014ApJ...797...85G}{, \apj, 797,
  85}

\bibitem[{{Grenier} {et~al.}(2005){Grenier}, {Casandjian}, \&
  {Terrier}}]{grenier05}
{Grenier}, I.~A., {Casandjian}, J.-M., \& {Terrier}, R.
  2005\href{http://cdsads.u-strasbg.fr/abs/2005Sci...307.1292G}{, Science, 307,
  1292}

\bibitem[{{Guhathakurta} \& {Draine}(1989)}]{guhathakurta89}
{Guhathakurta}, P. \& {Draine}, B.~T.
  1989\href{http://adsabs.harvard.edu/cgi-bin/nph-bib_query?bibcode=1989ApJ...345..230G&db_key=AST}{,
  \apj, 345, 230}

\bibitem[{{Guillet} {et~al.}(2017){Guillet}, {Fanciullo}, {Verstraete},
  {Boulanger}, {Jones}, {Miville-Desch{\^e}nes}, {Ysard}, {Levrier}, \&
  {Alves}}]{guillet17}
{Guillet}, V., {Fanciullo}, L., {Verstraete}, L., {et~al.}
  2017\href{http://adsabs.harvard.edu/abs/2017arXiv171004598G}{,
  ArXiv:1710.04598}

\bibitem[{{Hildebrand}(1983)}]{hildebrand83}
{Hildebrand}, R.~H.
  1983\href{http://cdsads.u-strasbg.fr/abs/1983QJRAS..24..267H}{, \qjras, 24,
  267}

\bibitem[{{Hogg} {et~al.}(2010){Hogg}, {Myers}, \& {Bovy}}]{hogg10}
{Hogg}, D.~W., {Myers}, A.~D., \& {Bovy}, J.
  2010\href{http://cdsads.u-strasbg.fr/abs/2010ApJ...725.2166H}{, \apj, 725,
  2166}

\bibitem[{{Hollenbach} \& {Tielens}(1997)}]{hollenbach97}
{Hollenbach}, D.~J. \& {Tielens}, A.~G.~G.~M.
  1997\href{http://adsabs.harvard.edu/cgi-bin/nph-bib_query?bibcode=1997ARA%26A..35..179H&db_key=AST}{,
  \araa, 35, 179}

\bibitem[{{Jones} {et~al.}(2017){Jones}, {K{\"o}hler}, {Ysard}, {Bocchio}, \&
  {Verstraete}}]{jones17}
{Jones}, A.~P., {K{\"o}hler}, M., {Ysard}, N., {Bocchio}, M., \& {Verstraete},
  L. 2017\href{http://cdsads.u-strasbg.fr/abs/2017A%26A...602A..46J}{, \aap,
  602, A46}

\bibitem[{{Juvela} {et~al.}(2013){Juvela}, {Montillaud}, {Ysard}, \&
  {Lunttila}}]{juvela13}
{Juvela}, M., {Montillaud}, J., {Ysard}, N., \& {Lunttila}, T.
  2013\href{http://cdsads.u-strasbg.fr/abs/2013A%26A...556A..63J}{, \aap, 556,
  A63}

\bibitem[{{Kelly}(2011)}]{kelly11}
{Kelly}, B.~C.
  2011\href{http://amstat.tandfonline.com/doi/abs/10.1198/jcgs.2011.203c}{,
  JCGS, 20, 584}

\bibitem[{{Kelly} {et~al.}(2012){Kelly}, {Shetty}, {Stutz}, {Kauffmann},
  {Goodman}, \& {Launhardt}}]{kelly12}
{Kelly}, B.~C., {Shetty}, R., {Stutz}, A.~M., {et~al.}
  2012\href{http://cdsads.u-strasbg.fr/abs/2012ApJ...752...55K}{, \apj, 752,
  55}

\bibitem[{{Kimura}(2016)}]{kimura16}
{Kimura}, H. 2016\href{http://cdsads.u-strasbg.fr/abs/2016MNRAS.459.2751K}{,
  \mnras, 459, 2751}

\bibitem[{{K{\"o}hler} {et~al.}(2012){K{\"o}hler}, {Stepnik}, {Jones},
  {Guillet}, {Abergel}, {Ristorcelli}, \& {Bernard}}]{kohler12}
{K{\"o}hler}, M., {Stepnik}, B., {Jones}, A.~P., {et~al.}
  2012\href{http://cdsads.u-strasbg.fr/abs/2012A%26A...548A..61K}{, \aap, 548,
  A61}

\bibitem[{{K{\"o}hler} {et~al.}(2015){K{\"o}hler}, {Ysard}, \&
  {Jones}}]{kohler15}
{K{\"o}hler}, M., {Ysard}, N., \& {Jones}, A.~P.
  2015\href{http://cdsads.u-strasbg.fr/abs/2015A%26A...579A..15K}{, \aap, 579,
  A15}

\bibitem[{{Le Bourlot} {et~al.}(2012){Le Bourlot}, {Le Petit}, {Pinto},
  {Roueff}, \& {Roy}}]{le-bourlot12}
{Le Bourlot}, J., {Le Petit}, F., {Pinto}, C., {Roueff}, E., \& {Roy}, F.
  2012\href{http://cdsads.u-strasbg.fr/abs/2012A%26A...541A..76L}{, \aap, 541,
  A76}

\bibitem[{{Lebouteiller} {et~al.}(2017){Lebouteiller}, {P{\'e}quignot},
  {Cormier}, {Madden}, {Pakull}, {Kunth}, {Galliano}, {Chevance}, {Heap},
  {Lee}, \& {Polles}}]{lebouteiller17}
{Lebouteiller}, V., {P{\'e}quignot}, D., {Cormier}, D., {et~al.}
  2017\href{http://cdsads.u-strasbg.fr/abs/2017A%26A...602A..45L}{, \aap, 602,
  A45}

\bibitem[{{Leroy} {et~al.}(2011){Leroy}, {Bolatto}, {Gordon}, {Sandstrom},
  {Gratier}, {Rosolowsky}, {Engelbracht}, {Mizuno}, {Corbelli}, {Fukui}, \&
  {Kawamura}}]{leroy11}
{Leroy}, A.~K., {Bolatto}, A., {Gordon}, K., {et~al.}
  2011\href{http://cdsads.u-strasbg.fr/abs/2011ApJ...737...12L}{, \apj, 737,
  12}

\bibitem[{{Li} \& {Draine}(2001)}]{li01}
{Li}, A. \& {Draine}, B.~T.
  2001\href{http://adsabs.harvard.edu/cgi-bin/nph-bib_query?bibcode=2001ApJ...554..778L&db_key=AST}{,
  \apj, 554, 778}

\bibitem[{{Lianou} {et~al.}(????){Lianou}, {Madden}, \& {Galliano}}]{lianou17}
{Lianou}, S., {Madden}, S.~C., \& {Galliano}, F., {\it in prep.}

\bibitem[{{Lisenfeld} \& {Ferrara}(1998)}]{lisenfeld98}
{Lisenfeld}, U. \& {Ferrara}, A.
  1998\href{http://adsabs.harvard.edu/cgi-bin/nph-bib_query?bibcode=1998ApJ...496..145L&db_key=AST}{,
  \apj, 496, 145}

\bibitem[{{Loredo} \& {Wasserman}(1998)}]{loredo98}
{Loredo}, T.~J. \& {Wasserman}, I.~M.
  1998\href{http://cdsads.u-strasbg.fr/abs/1998ApJ...502...75L}{, \apj, 502,
  75}

\bibitem[{{Mandel} {et~al.}(2009){Mandel}, {Wood-Vasey}, {Friedman}, \&
  {Kirshner}}]{mandel09}
{Mandel}, K.~S., {Wood-Vasey}, W.~M., {Friedman}, A.~S., \& {Kirshner}, R.~P.
  2009\href{http://cdsads.u-strasbg.fr/abs/2009ApJ...704..629M}{, \apj, 704,
  629}

\bibitem[{{Markwardt}(2009)}]{markwardt09}
{Markwardt}, C.~B. 2009, in
  \href{http://adsabs.harvard.edu/abs/2009ASPC..411..251M}{Astronomical Society
  of the Pacific Conference Series, Vol. 411, Astronomical Data Analysis
  Software and Systems XVIII, ed. D.~A. {Bohlender}, D.~{Durand}, \&
  P.~{Dowler}}, 251

\bibitem[{{Marsh} {et~al.}(2015){Marsh}, {Whitworth}, \& {Lomax}}]{marsh15}
{Marsh}, K.~A., {Whitworth}, A.~P., \& {Lomax}, O.
  2015\href{http://cdsads.u-strasbg.fr/abs/2015MNRAS.454.4282M}{, \mnras, 454,
  4282}

\bibitem[{{Martinez}(2015)}]{martinez15}
{Martinez}, G.~D.
  2015\href{http://cdsads.u-strasbg.fr/abs/2015MNRAS.451.2524M}{, \mnras, 451,
  2524}

\bibitem[{{Mathis} {et~al.}(1983){Mathis}, {Mezger}, \& {Panagia}}]{mathis83}
{Mathis}, J.~S., {Mezger}, P.~G., \& {Panagia}, N.
  1983\href{http://adsabs.harvard.edu/cgi-bin/nph-bib_query?bibcode=1983A%26A...128..212M&db_key=AST}{,
  \aap, 128, 212}

\bibitem[{{Mor\'e} {et~al.}(1984){Mor\'e}, {Sorensen}, {Hillstrom}, \&
  {Garbow}}]{more84}
{Mor\'e}, J.~J., {Sorensen}, D.~C., {Hillstrom}, K.~E., \& {Garbow}, B.~S.
  1984, in \href{http://www.mcs.anl.gov/~more/}{Sources and Development of
  Mathematical Software, ed. W.~J. {Cowell}}, 88--111

\bibitem[{{Paradis} {et~al.}(2010){Paradis}, {Veneziani}, {Noriega-Crespo},
  {Paladini}, {Piacentini}, {Bernard}, {de Bernardis}, {Calzoletti},
  {Faustini}, {Martin}, {Masi}, {Montier}, {Natoli}, {Ristorcelli}, {Thompson},
  {Traficante}, \& {Molinari}}]{paradis10}
{Paradis}, D., {Veneziani}, M., {Noriega-Crespo}, A., {et~al.}
  2010\href{http://cdsads.u-strasbg.fr/abs/2010A%26A...520L...8P}{, \aap, 520,
  L8}

\bibitem[{{Planck Collaboration} {et~al.}(2014{\natexlab{a}}){Planck
  Collaboration}, {Abergel}, {Ade}, {Aghanim}, {Alves}, {Aniano}, {Arnaud},
  {Ashdown}, {Aumont}, {Baccigalupi}, {Banday}, {Barreiro}, {Bartlett},
  {Battaner}, {Benabed}, {Benoit-L{\'e}vy}, {Bernard}, {Bersanelli},
  {Bielewicz}, {Bobin}, {Bonaldi}, {Bond}, {Bouchet}, {Boulanger}, {Burigana},
  {Cardoso}, {Catalano}, {Chamballu}, {Chiang}, {Christensen}, {Clements},
  {Colombi}, {Colombo}, {Couchot}, {Crill}, {Cuttaia}, {Danese}, {Davis}, {de
  Bernardis}, {de Rosa}, {de Zotti}, {Delabrouille}, {D{\'e}sert}, {Dickinson},
  {Diego}, {Dole}, {Donzelli}, {Dor{\'e}}, {Douspis}, {Dupac}, {Efstathiou},
  {En{\ss}lin}, {Eriksen}, {Falgarone}, {Finelli}, {Forni}, {Frailis},
  {Franceschi}, {Galeotta}, {Ganga}, {Ghosh}, {Giard}, {Giraud-H{\'e}raud},
  {Gonz{\'a}lez-Nuevo}, {G{\'o}rski}, {Gregorio}, {Gruppuso}, {Guillet},
  {Hansen}, {Harrison}, {Helou}, {Henrot-Versill{\'e}},
  {Hern{\'a}ndez-Monteagudo}, {Herranz}, {Hildebrandt}, {Hivon}, {Hobson},
  {Holmes}, {Hornstrup}, {Hovest}, {Huffenberger}, {Jaffe}, {Jaffe}, {Joncas},
  {Jones}, {Jones}, {Juvela}, {Kalberla}, {Keih{\"a}nen}, {Kerp}, {Keskitalo},
  {Kisner}, {Kneissl}, {Knoche}, {Kunz}, {Kurki-Suonio}, {Lagache},
  {L{\"a}hteenm{\"a}ki}, {Lamarre}, {Lasenby}, {Lawrence}, {Leonardi},
  {Levrier}, {Liguori}, {Lilje}, {Linden-V{\o}rnle}, {L{\'o}pez-Caniego},
  {Lubin}, {Mac{\'{\i}}as-P{\'e}rez}, {Maffei}, {Maino}, {Mandolesi}, {Maris},
  {Marshall}, {Martin}, {Mart{\'{\i}}nez-Gonz{\'a}lez}, {Masi}, {Massardi},
  {Matarrese}, {Mazzotta}, {Melchiorri}, {Mendes}, {Mennella}, {Migliaccio},
  {Mitra}, {Miville-Desch{\^e}nes}, {Moneti}, {Montier}, {Morgante},
  {Mortlock}, {Munshi}, {Murphy}, {Naselsky}, {Nati}, {Natoli}, {Noviello},
  {Novikov}, {Novikov}, {Oxborrow}, {Pagano}, {Pajot}, {Paoletti}, {Pasian},
  {Perdereau}, {Perotto}, {Perrotta}, {Piacentini}, {Piat}, {Pierpaoli},
  {Pietrobon}, {Plaszczynski}, {Pointecouteau}, {Polenta}, {Ponthieu}, {Popa},
  {Pratt}, {Prunet}, {Puget}, {Rachen}, {Reach}, {Rebolo}, {Reinecke},
  {Remazeilles}, {Renault}, {Ricciardi}, {Riller}, {Ristorcelli}, {Rocha},
  {Rosset}, {Roudier}, {Rusholme}, {Sandri}, {Savini}, {Spencer}, {Starck},
  {Sureau}, {Sutton}, {Suur-Uski}, {Sygnet}, {Tauber}, {Terenzi}, {Toffolatti},
  {Tomasi}, {Tristram}, {Tucci}, {Umana}, {Valenziano}, {Valiviita}, {Van
  Tent}, {Verstraete}, {Vielva}, {Villa}, {Wade}, {Wandelt}, {Winkel}, {Yvon},
  {Zacchei}, \& {Zonca}}]{planck-collaboration14b}
{Planck Collaboration}, {Abergel}, A., {Ade}, P.~A.~R., {et~al.}
  2014{\natexlab{a}}\href{http://adsabs.harvard.edu/abs/2014A%26A...566A..55P}{,
  \aap, 566, A55}

\bibitem[{{Planck Collaboration} {et~al.}(2011{\natexlab{a}}){Planck
  Collaboration}, {Abergel}, {Ade}, {Aghanim}, {Arnaud}, {Ashdown}, {Aumont},
  {Baccigalupi}, {Balbi}, {Banday}, \& et~al.}]{planck-collaboration11c}
{Planck Collaboration}, {Abergel}, A., {Ade}, P.~A.~R., {et~al.}
  2011{\natexlab{a}}\href{http://cdsads.u-strasbg.fr/abs/2011A%26A...536A..24P}{,
  \aap, 536, A24}

\bibitem[{{Planck Collaboration} {et~al.}(2014{\natexlab{b}}){Planck
  Collaboration}, {Ade}, {Aghanim}, {Armitage-Caplan}, {Arnaud}, {Ashdown},
  {Atrio-Barandela}, {Aumont}, {Baccigalupi}, {Banday}, \&
  et~al.}]{planck-collaboration14a}
{Planck Collaboration}, {Ade}, P.~A.~R., {Aghanim}, N., {et~al.}
  2014{\natexlab{b}}\href{http://adsabs.harvard.edu/abs/2014A%26A...571A...8P}{,
  \aap, 571, A8}

\bibitem[{{Planck Collaboration} {et~al.}(2011{\natexlab{b}}){Planck
  Collaboration}, {Ade}, {Aghanim}, {Arnaud}, {Ashdown}, {Aumont},
  {Baccigalupi}, {Balbi}, {Banday}, {Barreiro}, {Bartlett}, {Battaner},
  {Benabed}, {Beno{\^\i}t}, {Bernard}, {Bersanelli}, {Bhatia}, {Bock},
  {Bonaldi}, {Bond}, {Borrill}, {Bouchet}, {Bucher}, {Burigana}, {Cabella},
  {Cardoso}, {Catalano}, {Cay{\'o}n}, {Challinor}, {Chamballu}, {Chary},
  {Chiang}, {Christensen}, {Clements}, {Colombi}, {Couchot}, {Coulais},
  {Crill}, {Cuttaia}, {Danese}, {Davies}, {Davis}, {de Bernardis}, {de
  Gasperis}, {de Rosa}, {de Zotti}, {Delabrouille}, {Delouis}, {D{\'e}sert},
  {Dickinson}, {Dole}, {Donzelli}, {Dor{\'e}}, {D{\"o}rl}, {Douspis}, {Dupac},
  {Efstathiou}, {En{\ss}lin}, {Finelli}, {Forni}, {Frailis}, {Franceschi},
  {Galeotta}, {Ganga}, {Giard}, {Giardino}, {Giraud-H{\'e}raud},
  {Gonz{\'a}lez-Nuevo}, {G{\'o}rski}, {Gratton}, {Gregorio}, {Gruppuso},
  {Hansen}, {Harrison}, {Helou}, {Henrot-Versill{\'e}}, {Herranz},
  {Hildebrandt}, {Hivon}, {Hobson}, {Holmes}, {Hovest}, {Hoyland},
  {Huffenberger}, {Jaffe}, {Jones}, {Juvela}, {Keih{\"a}nen}, {Keskitalo},
  {Kisner}, {Kneissl}, {Knox}, {Kurki-Suonio}, {Lagache},
  {L{\"a}hteenm{\"a}ki}, {Lamarre}, {Lasenby}, {Laureijs}, {Lawrence}, {Leach},
  {Leonardi}, {Linden-V{\o}rnle}, {L{\'o}pez-Caniego}, {Lubin},
  {Mac{\'{\i}}as-P{\'e}rez}, {MacTavish}, {Madden}, {Maffei}, {Maino},
  {Mandolesi}, {Mann}, {Maris}, {Mart{\'{\i}}nez-Gonz{\'a}lez}, {Masi},
  {Matarrese}, {Matthai}, {Mazzotta}, {Melchiorri}, {Mendes}, {Mennella},
  {Miville-Desch{\^e}nes}, {Moneti}, {Montier}, {Morgante}, {Mortlock},
  {Munshi}, {Murphy}, {Naselsky}, {Natoli}, {Netterfield},
  {N{\o}rgaard-Nielsen}, {Noviello}, {Novikov}, {Novikov}, {Osborne}, {Pajot},
  {Partridge}, {Pasian}, {Patanchon}, {Peel}, {Perdereau}, {Perotto},
  {Perrotta}, {Piacentini}, {Piat}, {Plaszczynski}, {Pointecouteau}, {Polenta},
  {Ponthieu}, {Poutanen}, {Pr{\'e}zeau}, {Prunet}, {Puget}, {Reach}, {Rebolo},
  {Reinecke}, {Renault}, {Ricciardi}, {Riller}, {Ristorcelli}, {Rocha},
  {Rosset}, {Rowan-Robinson}, {Rubi{\~n}o-Mart{\'{\i}}n}, {Rusholme}, {Sandri},
  {Savini}, {Scott}, {Seiffert}, {Shellard}, {Smoot}, {Starck}, {Stivoli},
  {Stolyarov}, {Sudiwala}, {Sygnet}, {Tauber}, {Terenzi}, {Toffolatti},
  {Tomasi}, {Torre}, {Tristram}, {Tuovinen}, {T{\"u}rler}, {Umana},
  {Valenziano}, {Varis}, {Vielva}, {Villa}, {Vittorio}, {Wade}, {Wandelt},
  {Yvon}, {Zacchei}, \& {Zonca}}]{planck-collaboration11d}
{Planck Collaboration}, {Ade}, P.~A.~R., {Aghanim}, N., {et~al.}
  2011{\natexlab{b}}\href{http://adsabs.harvard.edu/abs/2011A%26A...536A..16P}{,
  \aap, 536, A16}

\bibitem[{{Planck Collaboration} {et~al.}(2011{\natexlab{c}}){Planck
  Collaboration}, {Ade}, {Aghanim}, {Arnaud}, {Ashdown}, {Aumont},
  {Baccigalupi}, {Balbi}, {Banday}, {Barreiro}, \&
  et~al.}]{planck-collaboration11b}
{Planck Collaboration}, {Ade}, P.~A.~R., {Aghanim}, N., {et~al.}
  2011{\natexlab{c}}\href{http://adsabs.harvard.edu/abs/2011A%26A...536A..19P}{,
  \aap, 536, A19}

\bibitem[{Press {et~al.}(2007)Press, Teukolsky, Vetterling, \&
  Flannery}]{press07}
Press, W.~H., Teukolsky, S.~A., Vetterling, W.~T., \& Flannery, B.~P. 2007,
  Numerical Recipes 3rd Edition: The Art of Scientific Computing (Cambridge
  University Press)

\bibitem[{{R{\'e}my-Ruyer} {et~al.}(2014){R{\'e}my-Ruyer}, {Madden},
  {Galliano}, {Galametz}, {Takeuchi}, {Asano}, {Zhukovska}, {Lebouteiller},
  {Cormier}, {Jones}, {Bocchio}, {Baes}, {Bendo}, {Boquien}, {Boselli},
  {DeLooze}, {Doublier-Pritchard}, {Hughes}, {Karczewski}, \&
  {Spinoglio}}]{remy-ruyer14}
{R{\'e}my-Ruyer}, A., {Madden}, S.~C., {Galliano}, F., {et~al.}
  2014\href{http://cdsads.u-strasbg.fr/abs/2014A%26A...563A..31R}{, \aap, 563,
  A31}

\bibitem[{{R{\"o}llig} {et~al.}(2007){R{\"o}llig}, {Abel}, {Bell}, {Bensch},
  {Black}, {Ferland}, {Jonkheid}, {Kamp}, {Kaufman}, {Le Bourlot}, {Le Petit},
  {Meijerink}, {Morata}, {Ossenkopf}, {Roueff}, {Shaw}, {Spaans}, {Sternberg},
  {Stutzki}, {Thi}, {van Dishoeck}, {van Hoof}, {Viti}, \&
  {Wolfire}}]{rollig07}
{R{\"o}llig}, M., {Abel}, N.~P., {Bell}, T., {et~al.}
  2007\href{http://cdsads.u-strasbg.fr/abs/2007A%26A...467..187R}{, \aap, 467,
  187}

\bibitem[{{Roman-Duval} {et~al.}(2014){Roman-Duval}, {Gordon}, {Meixner},
  {Bot}, {Bolatto}, {Hughes}, {Wong}, {Babler}, {Bernard}, {Clayton}, {Fukui},
  {Galametz}, {Galliano}, {Glover}, {Hony}, {Israel}, {Jameson},
  {Lebouteiller}, {Lee}, {Li}, {Madden}, {Misselt}, {Montiel}, {Okumura},
  {Onishi}, {Panuzzo}, {Reach}, {Remy-Ruyer}, {Robitaille}, {Rubio}, {Sauvage},
  {Seale}, {Sewilo}, {Staveley-Smith}, \& {Zhukovska}}]{roman-duval14}
{Roman-Duval}, J., {Gordon}, K.~D., {Meixner}, M., {et~al.}
  2014\href{http://cdsads.u-strasbg.fr/abs/2014ApJ...797...86R}{, \apj, 797,
  86}

\bibitem[{{Roy} {et~al.}(2013){Roy}, {Martin}, {Polychroni}, {Bontemps},
  {Abergel}, {Andr{\'e}}, {Arzoumanian}, {Di Francesco}, {Hill}, {Konyves},
  {Nguyen-Luong}, {Pezzuto}, {Schneider}, {Testi}, \& {White}}]{roy13}
{Roy}, A., {Martin}, P.~G., {Polychroni}, D., {et~al.}
  2013\href{http://cdsads.u-strasbg.fr/abs/2013ApJ...763...55R}{, \apj, 763,
  55}

\bibitem[{{Serra} {et~al.}(2011){Serra}, {Amblard}, {Temi}, {Burgarella},
  {Giovannoli}, {Buat}, {Noll}, \& {Im}}]{serra11}
{Serra}, P., {Amblard}, A., {Temi}, P., {et~al.}
  2011\href{http://cdsads.u-strasbg.fr/abs/2011ApJ...740...22S}{, \apj, 740,
  22}

\bibitem[{{Shahmoradi}(2017)}]{shahmoradi17}
{Shahmoradi}, A.
  2017\href{http://cdsads.u-strasbg.fr/abs/2017arXiv171110599S}{, ArXiv
  e-prints}

\bibitem[{{Shetty} {et~al.}(2009){Shetty}, {Kauffmann}, {Schnee}, \&
  {Goodman}}]{shetty09}
{Shetty}, R., {Kauffmann}, J., {Schnee}, S., \& {Goodman}, A.~A.
  2009\href{http://cdsads.u-strasbg.fr/abs/2009ApJ...696..676S}{, \apj, 696,
  676}

\bibitem[{{Siebenmorgen} {et~al.}(2014){Siebenmorgen}, {Voshchinnikov}, \&
  {Bagnulo}}]{siebenmorgen14}
{Siebenmorgen}, R., {Voshchinnikov}, N.~V., \& {Bagnulo}, S.
  2014\href{http://adsabs.harvard.edu/abs/2014A%26A...561A..82S}{, \aap, 561,
  A82}

\bibitem[{{Sokal}(1996)}]{sokal96}
{Sokal}, A. 1996, in
  \href{http://www.stat.unc.edu/faculty/cji/Sokal.pdf}{Lectures at the
  Carg\`ese Summer School on "Functional Integration: Basics and Applications"}

\bibitem[{{Stepnik} {et~al.}(2003){Stepnik}, {Abergel}, {Bernard}, {Boulanger},
  {Cambr{\'e}sy}, {Giard}, {Jones}, {Lagache}, {Lamarre}, {Meny}, {Pajot}, {Le
  Peintre}, {Ristorcelli}, {Serra}, \& {Torre}}]{stepnik03}
{Stepnik}, B., {Abergel}, A., {Bernard}, J., {et~al.}
  2003\href{http://cdsads.u-strasbg.fr/abs/2003A%26A...398..551S}{, \aap, 398,
  551}

\bibitem[{{van der Tak} {et~al.}(2017){van der Tak}, {Madden}, {Roelfsema}, b,
  c, \& d}]{van-der-tak17}
{van der Tak}, F.~F.~S., {Madden}, S.~C., {Roelfsema}, P.~R., {et~al.}
  2017\href{http://aanda.org}{, \pasp, {\it in prep.}}

\bibitem[{{Veneziani} {et~al.}(2013){Veneziani}, {Piacentini},
  {Noriega-Crespo}, {Carey}, {Paladini}, \& {Paradis}}]{veneziani13}
{Veneziani}, M., {Piacentini}, F., {Noriega-Crespo}, A., {et~al.}
  2013\href{http://cdsads.u-strasbg.fr/abs/2013ApJ...772...56V}{, \apj, 772,
  56}

\bibitem[{Villani \& Larsson(2006)}]{villani06}
Villani, M. \& Larsson, R.
  2006\href{http://www.ingentaconnect.com/content/tandf/lsta/2006/00000035/00000006/art00015}{,
  Communications in Statistics-Theory and Methods, 35, 1123}

\bibitem[{{Weingartner} \& {Draine}(2001)}]{weingartner01b}
{Weingartner}, J.~C. \& {Draine}, B.~T.
  2001\href{http://cdsads.u-strasbg.fr/abs/2001ApJS..134..263W}{, \apjs, 134,
  263}

\bibitem[{{Wolfgang} \& {Lopez}(2015)}]{wolfgang15}
{Wolfgang}, A. \& {Lopez}, E.
  2015\href{http://cdsads.u-strasbg.fr/abs/2015ApJ...806..183W}{, \apj, 806,
  183}

\bibitem[{{Wraith} {et~al.}(2009){Wraith}, {Kilbinger}, {Benabed}, {Capp{\'e}},
  {Cardoso}, {Fort}, {Prunet}, \& {Robert}}]{wraith09}
{Wraith}, D., {Kilbinger}, M., {Benabed}, K., {et~al.}
  2009\href{http://adsabs.harvard.edu/abs/2009PhRvD..80b3507W}{, \prd, 80,
  023507}

\bibitem[{{Wu} {et~al.}(2018){Wu}, {Bron}, {Onaka}, {Le Petit}, {Galliano},
  {Languignon}, {Nakamura}, \& {Okada}}]{wu18b}
{Wu}, R., {Bron}, E., {Onaka}, T., {et~al.}
  2018\href{http://cdsads.u-strasbg.fr/abs/2018arXiv180101643W}{, ArXiv
  e-prints}

\bibitem[{{Wu} {et~al.}(????){Wu}, {Galliano}, {Onaka}, {Bron}, \& {Le
  Petit}}]{wu18}
{Wu}, R., {Galliano}, F., {Onaka}, T., {Bron}, E., \& {Le Petit}, F., {\it in
  prep.}

\bibitem[{{Wu} {et~al.}(2015){Wu}, {Madden}, {Galliano}, {Wilson},
  {Kamenetzky}, {Lee}, {Schirm}, {Hony}, {Lebouteiller}, {Spinoglio},
  {Cormier}, {Glenn}, {Maloney}, {Pereira-Santaella}, {R{\'e}my-Ruyer}, {Baes},
  {Boselli}, {Bournaud}, {De Looze}, {Hughes}, {Panuzzo}, \& {Rangwala}}]{wu15}
{Wu}, R., {Madden}, S.~C., {Galliano}, F., {et~al.}
  2015\href{http://cdsads.u-strasbg.fr/abs/2015A%26A...575A..88W}{, \aap, 575,
  A88}

\bibitem[{{Ysard} {et~al.}(2015){Ysard}, {K{\"o}hler}, {Jones},
  {Miville-Desch{\^e}nes}, {Abergel}, \& {Fanciullo}}]{ysard15}
{Ysard}, N., {K{\"o}hler}, M., {Jones}, A., {et~al.}
  2015\href{http://cdsads.u-strasbg.fr/abs/2015A%26A...577A.110Y}{, \aap, 577,
  A110}

\bibitem[{{Yu} \& {Meng}(2011)}]{yu11}
{Yu}, Y. \& {Meng}, X.-L.
  2011\href{http://amstat.tandfonline.com/doi/abs/10.1198/jcgs.2011.203main}{,
  JCGS, 20, 531}

\bibitem[{{Zafar} \& {Watson}(2013)}]{zafar13}
{Zafar}, T. \& {Watson}, D.
  2013\href{http://adsabs.harvard.edu/abs/2013A%26A...560A..26Z}{, \aap, 560,
  A26}

\bibitem[{{Zhukovska}(2014)}]{zhukovska14}
{Zhukovska}, S.
  2014\href{http://cdsads.u-strasbg.fr/abs/2014A%26A...562A..76Z}{, \aap, 562,
  A76}

\bibitem[{{Zubko} {et~al.}(2004){Zubko}, {Dwek}, \& {Arendt}}]{zubko04}
{Zubko}, V., {Dwek}, E., \& {Arendt}, R.~G.
  2004\href{http://adsabs.harvard.edu/cgi-bin/nph-bib_query?bibcode=2004ApJS..152..211Z&db_key=AST}{,
  \apjs, 152, 211}

\end{thebibliography}

\appendix

  \section{Model Computation and Inversion}
  \label{app:template}
  
Calculating the full SED model at each iteration of the MCMC would be 
prohibitive, in terms of computing time.
As discussed in \refsec{sec:SEDcomb}, we rather 
interpolate a finely sampled grid of pre-computed templates.

  \subsection{The Pre-computed model grid}
  \label{app:tempgrid}

For each model component presented in \refsec{sec:SEDcomp}, we generate a grid 
of specific luminosities, $l_\nu^\sms{mod}(\vect{x},\nu_j)$, as a function of 
each parameter $x_k$.
We separate components that can be linearly combined. 
For instance, for the \deltaU\ and \powU\ components, we compute separate grids 
for each grain sub-species: neutral and ionized PAHS, small and large carbon grains, small and large silicates.
The frequencies $\nu_j$ belong to a list of the most widely used infrared
photometric filters.
Integration of the SED model into the filter bandpass and color corrections are 
thus included in $l_\nu^\sms{mod}$.

To build the templates, we start from a coarse parameter grid, and add a mid-point 
if the linear interpolation of $\ln l_\nu^\sms{mod}$ is less accurate than 
$10^{-3}$.
We repeat this process until no more mid-points need to be added.

  \subsection{SED model evaluation}
  \label{app:tempev}

At each iteration of the MCMC, we evaluate the template by performing a 
multidimensional linear interpolation of $\ln l_\nu^\sms{mod}(\vect{x},\nu_j)$, 
as a function of $\vect{x}$.
By construction of the grid, we know this interpolation will be more accurate 
than $10^{-3}$.

  \subsection{SED model inversion}
  \label{app:tempinv}  

The model inversion used by ASIS (\refsec{sec:sampler}) is also performed with a multidimensional linear interpolation.
The only difference here is that we now interpolate the parameter we are looking for, $x_{k^\prime}$, as a function of $\ln l_\nu^\sms{mod}(\vect{x},\nu_j)$ and of the other parameters, $x_{k\ne k^\prime}$.

  \section{Additional Elements on the HB Method}
  
  \subsection{The Split-Normal Distribution}
  \label{app:split}

The split-normal distribution \citep[][used in \refsec{sec:noise}]{villani06} is defined as:
\begin{equation}
  p(x) =
  \left\{
  \begin{array}{ll}
    \displaystyle
    A\times\exp\left[-\frac{1}{2}\left(\frac{x-x_0}{\lambda}\right)^2\right] 
      & \mbox{if } x \le x_0 \\
      & \\
    \displaystyle
    A\times\exp\left[-\frac{1}{2}\left(\frac{x-x_0}{\lambda\tau}\right)^2\right] 
      & \mbox{if } x > x_0,
  \end{array}
  \right.
\end{equation}
where:
\begin{equation}
  A = \sqrt{\frac{2}{\pi}}\frac{1}{\lambda(1+\tau)},
\end{equation}
and $x_0$ is a position parameter, $\lambda$, a scale paramter, and $\tau$, 
a shape parameter.
These parameters are linked to the mean, $\mu$, standard devitation, $\sigma$, and skewness, $\gamma_1$, through the following set of equations:
\begin{eqnarray}
  b & = & \frac{\pi-2}{\pi} (\tau-1)^2+\tau \\
  \mu    & = & x_0 + \sqrt{\frac{2}{\pi}}\lambda(\tau-1) \label{eq:split1}\\
  \sigma & = & \sqrt{b}\lambda^2 \label{eq:split2}\\
  \gamma_1 & = & b^{-3/2}\sqrt{\frac{2}{\pi}}(\tau-1)\times
                 \left[\left(\frac{4}{\pi}-1\right)(\tau-1)^2+\tau\right].
                 \label{eq:split3}
\end{eqnarray}
There is no simple inversion.
We therefore solve $\tau$ numerically from \refeq{eq:split3}, and then inverse \refeqs{eq:split1} and (\ref{eq:split2}).

  \subsection{The Gibbs Sampling}
  \label{app:gibbs}

Gibbs sampling consists in drawing each parameter, $x_k$, one by one, from its
conditional distribution, fixing the other parameters to their current value in the MCMC, $p(x_k|x_{k^\prime\ne k})$.
In practice, this distribution is a complex function, depending on the SED model.
We numerically perform this drawing as follows.
Let's assume that parameter $x_k$ is bound to the inerval $[x_k^\sms{min},x_k^\sms{max}]$.
We first compute the cumulative distribution function (CDF) as:
\begin{equation}
  F(x_k|x_{k^\prime\ne k}) 
  = \frac{\displaystyle\int_{x_k^\sms{min}}^{x_k} p(y|x_{k^\prime\ne k})
          \ddiff y}
         {F(x_k^\sms{max}|x_{k^\prime\ne k})}.
  \label{eq:CDF}
\end{equation}
We then draw a random variable $\theta$, uniformly distributed between 0 and 1. 
The updated value of $x_k$ is derived by solving 
$F(x_k|x_{k^\prime\ne k})=\theta$, by linear interpolation of the $x_k$ grid as a function of $\ln F$.

Gibbs sampling is more adapted to our problem than the rejection strategy of the Metropolis-Hastings method \citep[\eg][]{foreman-mackey13} or than importance sampling \citep[\eg][]{wraith09}.
Indeed, the latter methods are not efficient when the number of dimensions, $N_\sms{dim}$ \refeqp{eq:Ndim}, becomes large. 
Several runs presented in this paper lead to $N_\sms{dim}\simeq 7000$.

To provide an accurate integration of \refeq{eq:CDF} in a reasonable computing 
time, we implement the following adaptative grid.
We start from a regular grid of $N^{(0)}=10$ values, $x_k^{(t)}$, 
with $t=1,\ldots,N^{(0)}$, spanning the whole range of the interval $[x_k^\sms{min},x_k^\sms{max}]$.
We compute the normalization of \refeq{eq:CDF}, $F^{(0)}(x_k^\sms{max}|x_{k^\prime\ne k})$, as the sum of elemental trapeziums, $\Delta F^{(0)}(x^{(t)})$.
We then add a mid-point in each interval and compute the new trapeziums 
$\Delta F^{(1)}(x_k^{(t)})$.
We stop the refinement of element $t$ if:
\begin{equation}
  \begin{multlined}
    |\Delta F^{(1)}(x_k^{(t)})
    +\Delta F^{(1)}\left(\frac{x_k^{(t)}+x_k^{(t+1)}}{2}\right)
    -\Delta F^{(0)}(x_k^{(t)})|
  \\
    \le 10^{-3}\times\frac{\displaystyle
                           \sum_{t^\prime=1}^{N^{(1)}}
                            \Delta F^{(1)}(x_k^{(t^\prime)})}{\sqrt{N^{(1)}}}.
  \end{multlined}
\end{equation}
We iterate until no more refinement is needed.
Most of the drawings require between 50 an 200 points.

  \subsection{Consistent Treatment of External Parameters}
  \label{app:extra}

There are several non-dusty parameters that we may want to correlate with the grain properties, like the gas column density, the metallicity, \etc\
In principle, we could study these extra correlations afterwards, using the final results of the MCMC.
However, including these parameters within the MCMC sampler, as part of our hierarchical model, will help improve the correlation, as they will share a common prior distribution with the model parameters.
Poorly constrained grain parameters could benefit from using these additional observations.

Let's decompose the parameters of a source $s_i$ into $q_1$ {\it SED model parameters} and $q_2$ {\it external parameters}:
\begin{equation}
  \vect{x}_i = \left(\left\{x_{i,k}^\sms{model}\right\}_{k=1,q_1},
                     \left\{x_{i,k}^\sms{extra}\right\}_{k=1,q_2}\right)
                     \;\;\;\mbox{ with }q=q_1+q_2.
\end{equation}
For instance, $\{x_{i,k}^\sms{model}\}_{k=1,q_1}$ could be $\{\ln M_i,\ln T_i,\beta_i\}$, ($q_1=3$) and $\{x_{i,k}^\sms{extra}\}_{k=1,q_2}$
could be $\{\ln M^\sms{\hi}_i,Z_i\}$ ($q_2=2$).
The SED parameters are constrained by the observed SED.
The extra parameters are fixed in principle, as they have been estimated from an independent method.
However, including them in the MCMC sampler will lead to a modification of their value consistent with their uncertainties.
Let's note their observed value $Q_{i,k}$, with an uncertainty 
$\sigma_{i,k}^\sms{extra}$.
The likelihood of the extra parameter $k$, of a source $s_i$, is simply:
\begin{equation}
  p\left(x_{i,k}^\sms{extra}|Q_{i,k}\right) = \mathcal{P}\left(Q_{i,k},\left(\sigma_{i,k}^\sms{extra}\right)^2\right),
\end{equation}
where $\mathcal{P}(\bar{y},\sigma_y^2)$ is one of the noise distributions of \refsec{sec:noise}, with mean $\bar{y}$ and variance $\sigma_y^2$.
The posterior distribution of this extra parameter is then:
\begin{equation}
  \begin{multlined}
  p\left(x_{i,k}^\sms{extra}|Q_{i,k},\vect{x}_i^\sms{model},
      x^\sms{extra}_{i,k^\prime\ne k},\vect{\mu},\mat{\Sigma}\right) 
    \\
    \propto p\left(x_{i,k}^\sms{extra}|Q_{i,k}\right)
            \times
            p\left(\vect{x}_i|\vect{\mu},\mat{\Sigma}\right)
            \times
            p\left(\vect{\mu},\mat{\Sigma}\right).
  \end{multlined}
  \label{eq:LHextra}
\end{equation}
We sample these extra parameters, at each iteration of the MCMC, from 
\refeq{eq:LHextra}.

  \section{Generalized least-squares with correlated errors}
  \label{app:chi2}

We have developped a chi-squared minimization SED fitter, having the same interface as our HB code.
It can fit any linear combination of the individual SED components of \refsec{sec:SEDcomp} to an observed SED.
It is based on the \minpack\ \citep{more84} Levenberg-Marquardt method, that we have converted in Fortran 90.
We have added the possibility to limit, fix and tie parameters, similarly to what \citet{markwardt09} did in IDL.

  \subsection{Total covariance matrix and chi-squared}

In most of our cases, the uncertainties on the observed fluxes, $\vect{L}_\nu^\sms{obs}$, come from two 
terms:\textlist{\thetextlist~the noise, $\vect{\sigma}_\nu^\sms{noise}$, which is uncorrelated between frequencies, with diagonal covariance matrix $\mat{V}_\sms{noise}$; and \thetextlist~correlated systematic calibration uncertainties, $\vect{\sigma}_\nu^\sms{syst}$, with non-diagonal covariance matrix $\mat{V}_\sms{syst}$.}The total covariance matrix is simply:
\begin{equation}
  \mat{V}_\sms{obs} = \mat{V}_\sms{noise} + \mat{V}_\sms{syst}.
\end{equation}

Properly taking into account the different sources of uncertainties requires to account for correlated terms by minimizing the \citengl{generalized least-squares} problem:
\begin{equation}
  \chi^2 = \vect{r}^T\,\mat{V}_\sms{obs}^{-1}\,\vect{r},
  \label{eq:chi2}
\end{equation}
where $\vect{r} = \vect{L}_\nu^\sms{obs}-\vect{L}_\nu^\sms{mod}$, is the residual between the observations and the model $\vect{L}_\nu^\sms{mod}$.

  \subsection{Adapting the Levenberg-Marquardt algorithm}
  
The Levenberg-Marquardt algorithm \citep[{\eg}][hereafter LM]{press07,markwardt09} is aimed at finding the best parameter vector $\vect{x}$ minimizing the $\chi^2$, starting from an initial estimate $\vect{x}^{(0)}$.
Computing, for each new value of $\vect{x}^{(n+1)}=\vect{x}^{(n)}+\vect{\delta x}^{(n+1)}$, the Jacobian matrix of the model, at $\vect{x}^{(n)}$:
\begin{equation}
  J_{k,j} = \frac{\partial L_{\nu,j}^\sms{mod}(x_k)}{\partial x_k},
          = - \frac{\partial w_j}{\partial x_k},
\end{equation}
the LM method solves:
\begin{equation}
  \left(\mat{J}^T\,\mat{J}\right)\,\vect{\delta x} = \mat{J}^T\,\vect{w},
\end{equation}
where the normalized residual is $\vect{w}=\vect{r}/\vect{\sigma}_\nu^\sms{obs}$, noting the total (uncorrelated) uncertainty:
\begin{equation}
  \vect{\sigma}_\nu^\sms{obs} = \sqrt{\left(\vect{\sigma}_\nu^\sms{noise}\right)^2 
                            + \left(\vect{\sigma}_\nu^\sms{syst}\right)^2}.
\end{equation}
This vector does not account for correlated terms.
The classical $\chi^2$, for non-correlated uncertainties, is 
$\chi^2=|\vect{w}|^2$.

When introducing correlated errors, the new problem to solve becomes:
\begin{equation}
  \left(\mat{J}^T\,\mat{V}_\sms{obs}^{-1}\,\mat{J}\right)\,\vect{\delta x} = \mat{J}^T\,\left(\mat{V}_\sms{obs}^{-1}\,\vect{r}\right).
\end{equation}
The Cholesky decomposition \citep[{\eg}][]{press07} of the covariance matrix results in a lower triangle matrix, $\mat{L}$, such that $\mat{V}_\sms{obs} = \mat{L}\,\mat{L}^T$.
This decomposition is always possible as the covariance matrix is positive-definite.
We also have $\mat{V}_\sms{obs}^{-1} = \left(\mat{L}^{-1}\right)^T\,\mat{L}^{-1}$.
The new system to be solved is:
\begin{equation}
  \mat{\mathcal{J}}^T\,\mat{\mathcal{J}}\,\vect{x} 
    = \mat{\mathcal{J}}^T\,\vect{t},
\end{equation}
where $\mat{\mathcal{J}} = L^{-1}\,\mat{J}$ and $\vect{t}=L^{-1}\vect{r}$.

In practice, to account for correlated uncertainties in the LM method, one just 
needs to perform the following steps.
\begin{enumerate}
  \item Compute once and for all, at the beginning (before the iterations), the 
    Cholesky decomposition of the covariance matrix, to obtain $\mat{L}$;
  \item Compute once and for all the inverse $\mat{L}^{-1}$.
  \item Set the output of the user-defined function in \minpack, returning the 
    weighted residual, to $\vect{t}=\mat{L}^{-1}\,\vect{r}$ instead of 
    $\vect{w}$.
\end{enumerate}

  \subsection{Uncertainties on derived quantities}
  
One of the product of the Levenberg-Marquardt method is the covariance matrix $\mat{V}_\sms{par}$ of the parameters $\vect{x}$.
We may want to also compute the uncertainties of quantities derived from these parameters, $f(\vect{x})$.
In the case of SED fitting, such quantities can be the total luminosity, the average starlight intensity, \etc

The error on any function of the parameters is:
\begin{equation}
  \sigma^2_{f(\vect{x})} = \left(\vect{\nabla}f(\vect{x})\right)^T
    \mat{V}_\sms{par}\,\vect{\nabla}f(\vect{x}).
  \label{eq:deriverr}
\end{equation}
For example if we have only two parameters, $\vect{x}=(a\;b)$, then:
\begin{eqnarray}
  \vect{\nabla}f=\left(
  \begin{array}{c}
    \displaystyle\frac{\partial f}{\partial a} \\ \\
    \displaystyle\frac{\partial f}{\partial b} 
  \end{array}
  \right)
& \mbox{\&} &
  \mat{V}_\sms{par}=\left(
  \begin{array}{cc}
    \displaystyle\sigma_a^2 & \rho\sigma_a\sigma_b \\
    \displaystyle\rho\sigma_a\sigma_b & \sigma_b^2
  \end{array}
  \right),
\end{eqnarray}
and \refeq{eq:deriverr} gives the usual expression:
\begin{equation}
  \sigma^2_{f(\vect{x})} 
  = \left(\frac{\partial f}{\partial a}\right)^2\sigma_a^2 
      + \left(\frac{\partial f}{\partial b}\right)^2\sigma_b^2 
      + 2\left(\frac{\partial f}{\partial a}\right)
         \left(\frac{\partial f}{\partial b}\right)\rho\sigma_a\sigma_b.
\end{equation}
Similarly, the covariance of two functions of the parameters $f(\vect{x})$ and
$g(\vect{x})$ is:
\begin{equation}
  {\rm cov}(f(\vect{x}),g(\vect{x})) 
    = \left(\vect{\nabla}f(\vect{x})\right)^T\mat{V}_\sms{par}
       \vect{\nabla}g(\vect{x}).
\end{equation}

  \subsection{Monte-Carlo intial conditions}
  \label{app:iniMC}
  
To avoid the LM algorithm to converge toward a local minimum, we generate
$N_\sms{ini}=30$ random initial conditions, drawn from a uniform distribution 
over the whole range of parameters.
We then run the LM code from these initial conditions and keep only the 
parameters corresponding to the lowest of these $N_\sms{ini}$ minimum chi-squared.

  \section{Computing Times for the Least-Squares and Bayesian Methods}
  \label{app:CPU}

Let's assume that we have a large number of sources, $n\gg1$, and that the 
bottleneck in terms of CPU is the calculation of one SED model at all the 
photometric filters, $\vect{L}_\nu^\sms{mod}(\vect{x};\vect{\nu})$.
The LM method (\refapp{app:chi2}) typically performs $N_\sms{iter}\simeq50-150$ 
iterations.
At each iteration, the model is evaluated $n\times q$ times to compute the 
gradient, by finite differences on each one of the $q$ parameters.
Since we run $N\sms{ini}=30$ initial conditions (\refapp{app:iniMC}), the 
number of model calculations is:
\begin{equation}
  N_{\chi2} \simeq N_\sms{ini}\times N_\sms{iter}\times n \times q.
  \label{eq:Nchi2}
\end{equation}

The parameter space for the HB method \refeqp{eq:Ndim} becomes
$N_\sms{dim} \simeq n\times q$, when $n\gg1$.
This is the number of Gibbs samplings of the parameters, per iteration.
Each Gibbs sampling requires $N_\sms{samp}\simeq 50-200$ points 
(\refapp{app:gibbs}).
For a chain of length $N_\sms{MCMC}$, the number of model calculations is thus:
\begin{equation}
  N_\sms{HB} \simeq N_\sms{MCMC}\times n\times q\times N_\sms{samp}.
  \label{eq:NHB}
\end{equation}
Noting that $N_\sms{iter}\simeq N_\sms{samp}$, the ratio of $\chi^2$ and HB 
CPU times is:
\begin{equation}
  \frac{CPU_\sms{HB}}{CPU_{\chi^2}}
  \simeq \frac{N_\sms{MCMC}}{N_\sms{ini}}\simeq{3\E{4}}.
  \label{eq:CPU}
\end{equation}
This approximation gives the correct order of magnitude.
For instance, the central simulation of \refsec{sec:refgridesc} had a ratio:
\begin{equation}
  \frac{CPU_\sms{HB}}{CPU_{\chi^2}}\simeq \frac{3\;\mbox{weeks}}{90\;\mbox{s}}
    \simeq 2\E{4}.
\end{equation}

The simulations presented in this paper add up to a total of 
$\simeq 10^5$~hours of CPU.
It points out that the use of the HB method represents an important
investment in terms of computing time.
In terms of storage, the MCMCs of all the simulations in this study add up to $\simeq1$~Tb of HDF5\footnote{\href{https://support.hdfgroup.org/HDF5/}{https://support.hdfgroup.org/HDF5/}.} files.

  \section{Round-Off Errors}

Round-off errors could be a potential problem for such a code, as it requires a large number of 
operations \refeqp{eq:Ndim}.
Round-off errors are however extremely difficult to track.
We note that all the tests we performed in this manuscript lead to a good agreement with the true 
values of the parameters.
Thus, round-off errors, if any, do not significantly affect the results. 
The reason is likely due to the stochasticity of the HB code.
Indeed, each step of the MCMC is only weakly dependent on the accuracy of the previous step.

\begin{figure}
  \includegraphics[width=\linewidth]{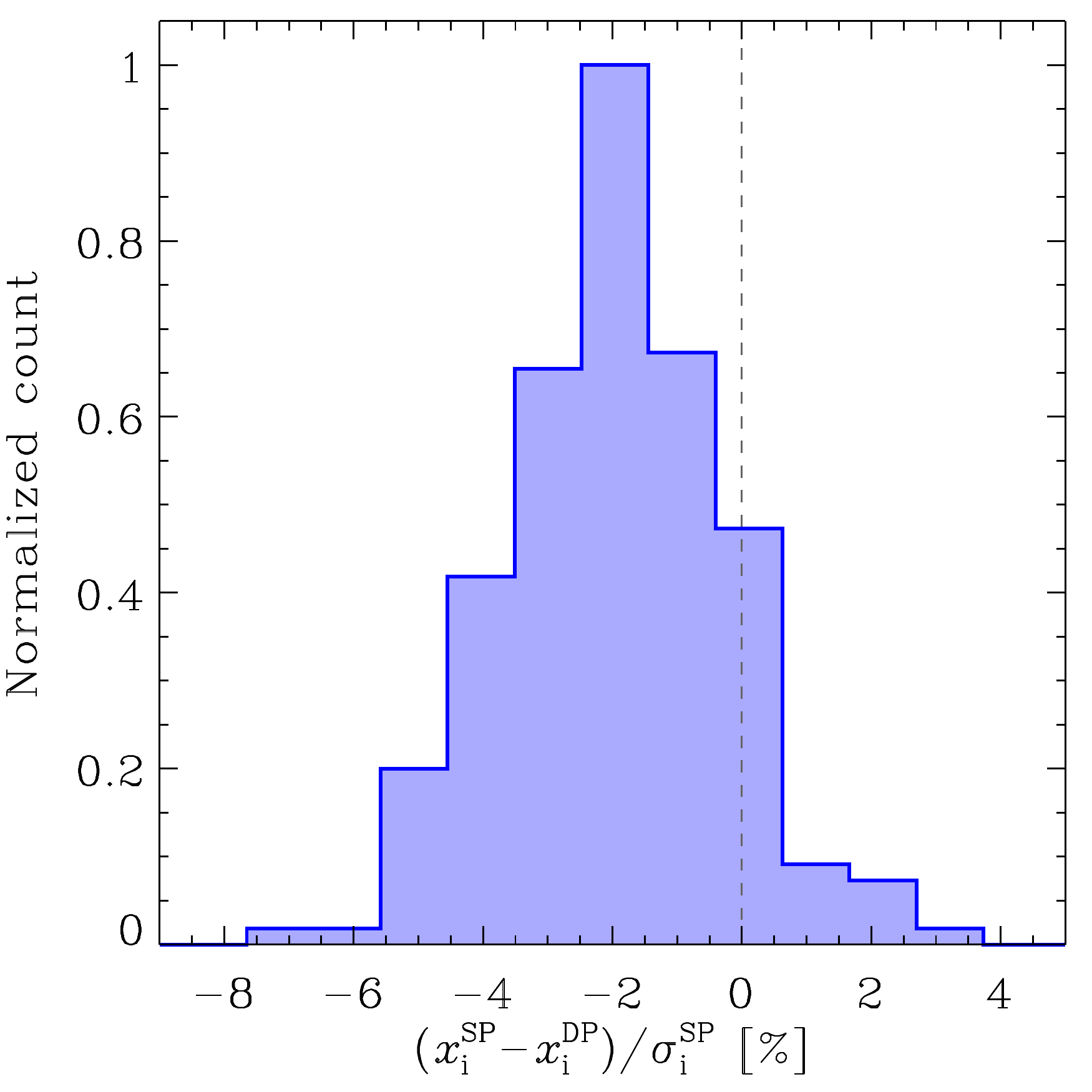}
  \caption{
           {\sl Test on rounding errors.}
                  This is the distribution of the difference between the inferred model parameters of 
                  each 
                  source $i$, $x_i$, in single precision (SP) and in double precision (DP), divided
                  by the uncertainty on the parameters, $\sigma_i$, in SP.
                  This test is performed on the central simulation of \refsec{sec:refgridesc}
                  (warm SED, $n=100$, $f_\sms{S/N}=3$).
           }
  \label{fig:roundoff}
\end{figure}
In order to quantify the typical order of magnitude of rounding errors we have re-run the central 
simulation of 
\refsec{sec:refgridesc} (warm SED, $n=100$, $f_\sms{S/N}=3$) in single precision (SP), while the rest
of the runs in this paper were all performed in double precision (DP).
If round-off errors were important, we would notice a significant difference between the results with 
the two precisions.
\reffig{fig:roundoff} shows the distribution of the difference of the model parameters of every source, 
$x_i$, in both precisions, divided by the uncertainty on these parameters, $\sigma_i$, in SP.
It shows that for any parameter, reducing the floating number precision, from DP to SP, results in an 
absolute difference of only a few percents of its uncertainty.
The actual round-off errors of the DP parameters are expected to be smaller than this value, as the 
precision is higher than SP.
Thus these errors can reasonably be considered negligible.

\bsp	
\label{lastpage}
\end{document}